\documentclass[12pt]{article}
\usepackage{url,color}

\usepackage[margin=1.15in]{geometry}
\usepackage{amsfonts,bm}
\usepackage{graphicx}
\usepackage{amsmath}
\usepackage{setspace}
\usepackage{float}
\usepackage{subfigure,color,epsfig,multirow,bbm,graphicx,graphics, sidecap}
\usepackage{tabularx,ulem}
\usepackage{algorithm} 
\usepackage{algpseudocode}
\usepackage{xcolor}
\usepackage[numbers]{natbib}
\usepackage[T1]{fontenc}
\usepackage{authblk}

\newcommand\Algphase[1]{%
\vspace*{-.2\baselineskip}\Statex\hspace*{\dimexpr-\algorithmicindent-2pt\relax}\rule{\textwidth}{0.4pt}%
\Statex\hspace*{-\algorithmicindent}\textbf{#1}%
\vspace*{-.2\baselineskip}\Statex\hspace*{\dimexpr-\algorithmicindent-2pt\relax}\rule{\textwidth}{0.4pt}%
}

\newcommand{\J}{\mathcal{J}}
\newcommand{\M}{\mathcal{M}}
\renewcommand{\H}{\mathcal{H}}
\newcommand{\x}{{\mathbf x}}
\newcommand{\xr}{{\bf \tilde x}}
\newcommand{\Ua}{\mathbf{U}_\textnormal{a}}
\newcommand{\Uf}{\mathbf{U}_\textnormal{f}}
\newcommand{\Wa}{\mathbf{W}_\textnormal{a}}
\newcommand{\Wf}{\mathbf{W}_\textnormal{f}}
\newcommand{\Wg}{\mathbf{W}_\textnormal{g}}
\newcommand{\Ug}{\mathbf{U}_\textnormal{g}}
\newcommand{\Id}{\mathbf{I}}
\newcommand{\la}{{\bm \lambda}}
\newcommand{\lar}{{\bm \tilde \lambda}}
\newcommand{\Lpod}{\mathcal{L}^\textsc{pod}}
\renewcommand{\L}{\mathcal{L}}
\newcommand{\Ns}{{N_\textnormal{state}}}

\newtheorem{theorem}{Theorem}[section]

\newtheorem{definition}[theorem]{Definition}

\makeatletter
\newcommand{\mathleft}{\@fleqntrue\@mathmargin\parindent}
\newcommand{\mathcenter}{\@fleqnfalse}
\makeatother

\begin{document}

\thispagestyle{empty}
\setcounter{page}{0}

\begin{Huge}
\begin{center}
Computer Science Technical Report CSTR-{17/2015} \\
\today
\end{center}
\end{Huge}
\vfil
\begin{huge}
\begin{center}
{\tt R. \c{S}tef\u{a}nescu, A. Sandu, I.M. Navon }
\end{center}
\end{huge}

\vfil
\begin{huge}
\begin{it}
\begin{center}
``{\tt POD/DEIM Reduced-Order Strategies for Efficient Four Dimensional Variational Data Assimilation}''
\end{center}
\end{it}
\end{huge}
\vfil

\begin{large}
\begin{center}
Computational Science Laboratory \\
Computer Science Department \\
Virginia Polytechnic Institute and State University \\
Blacksburg, VA 24060 \\
Phone: (540)-231-2193 \\
Fax: (540)-231-6075 \\
Email: \url{sandu@cs.vt.edu} \\
Web: \url{http://csl.cs.vt.edu}
\end{center}
\end{large}

\vspace*{1cm}

\begin{tabular}{ccc}
\includegraphics[width=2.5in]{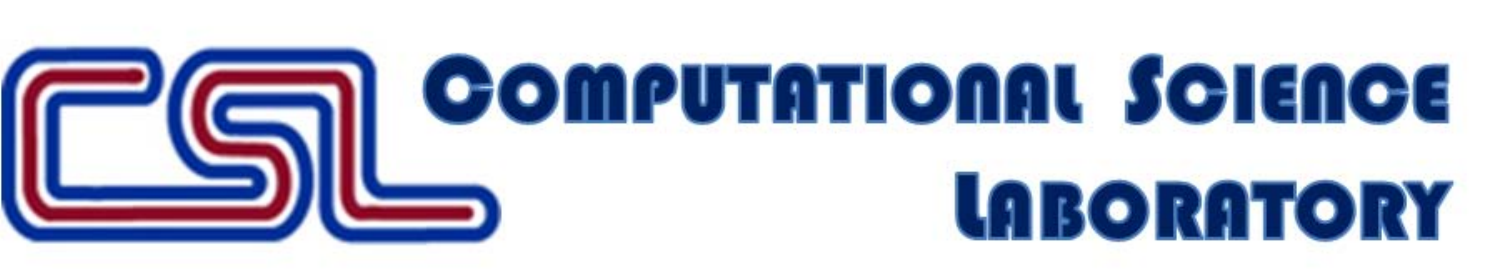}
&\hspace{2.5in}&
\includegraphics[width=2.5in]{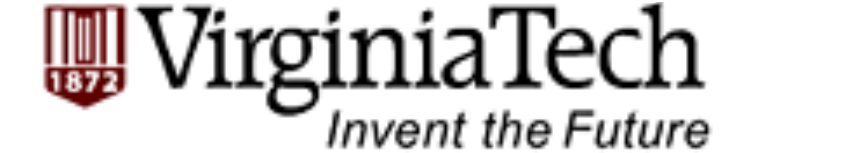} \\
{\bf\em Innovative Computational Solutions} &&\\
\end{tabular}

\newpage

\title{POD/DEIM Reduced-Order Strategies for Efficient Four Dimensional Variational Data Assimilation}
\author[1]{R\u{a}zvan \c{S}tef\u{a}nescu \thanks{rstefane@vt.edu}}
\author[1]{Adrian Sandu \thanks{sandu@cs.vt.edu}}
\author[2]{Ionel Michael Navon \thanks{inavon@fsu.edu}}
\affil[1]{Computational Science Laboratory, Department of Computer Science, Virginia Polytechnic Institute and State University, Blacksburg, Virginia, USA, 24060}
\affil[2]{Department of Scientific Computing,The Florida State University, Tallahassee, Florida, USA, 32306}

\date{}
\maketitle

\begin{abstract}
This work studies reduced order modeling (ROM) approaches to speed up the solution of variational data assimilation problems with large scale nonlinear dynamical models. It is shown that a key requirement for a successful reduced order solution  is that reduced order Karush-Kuhn-Tucker conditions accurately represent their full order counterparts. In particular, accurate reduced order approximations are needed for the forward and adjoint dynamical models, as well as for the reduced gradient. New strategies to construct reduced order based are developed for Proper Orthogonal Decomposition (POD) ROM data assimilation using both Galerkin and Petrov-Galerkin projections. For the first time POD, tensorial POD, and discrete empirical interpolation method (DEIM) are employed to develop reduced data assimilation systems for a geophysical flow model, namely, the two dimensional shallow water equations. Numerical  experiments confirm the theoretical framework for Galerkin projection. In the case of Petrov-Galerkin projection, stabilization strategies must be considered for the reduced order models. The new reduced order shallow water data assimilation system provides  analyses similar to those produced by the full resolution data assimilation system in one tenth of the computational time.
\end{abstract}

\begin{keyword}
inverse problems; proper orthogonal decomposition; discrete empirical interpolation method (DEIM); reduced-order models (ROMs); shallow water equations;  finite difference methods;
\end{keyword}

\section{Introduction}\label{Sec:Introduction}

Optimal control problems for nonlinear partial differential equations often require very large computational resources. Recently the reduced order approach applied to optimal control problems for partial differential equations has received increasing attention as a way of reducing the computational effort. The main idea is to project the dynamical system onto subspaces consisting of basis elements that represent the characteristics of the expected solution. These low order models serve as surrogates for the dynamical system in the optimization process and the resulting small optimization problems can be solved efficiently.

Application of Proper Orthogonal Decomposition (POD) to solve optimal control problems has proved to be successful as evidenced in the works of \citet{Kunisch_Volkwein_1999, Kunisch_Volkwein_2004, Ito_Kunisch_2006, Ito_Kunisch_2008, Kunisch_Xie_2005}. However this approach may suffer from the fact that the basis elements are computed from a reference trajectory containing features which are quite different from those of the optimally controlled trajectory. A priori it is not evident what is the optimal strategy to generate snapshots for the reduced POD control procedure. A successful POD based reduced optimization should represent correctly the dynamics of the flow that is altered by the controller. To overcome the problem of unmodelled dynamics in the  basis \citet{Afanasiev_Hinze_2001,Ravindran_2002,Kunisch_2008,Bergmann_Cordier_2005,Ravindran_2000,Yue_Meerbergen_2013,Zahr_Farhat_2014,Zahr_Amsallem_Farhat_2013} proposed to update the basis according to the current optimal control. In \citet{Arian_2000,bergmann2008optimal}, this updating technique was combined with a trust region (TR) strategy to determine whether after an optimization step an update of the POD-basis should be performed. Additional work on TR/POD proved its efficiency, see \citet{Bergmann_2007,Leibfritz_Volkwein_2006,Sachs_Volkwein_2010}. Other studies proposed to include time derivatives, nonlinear terms, and adjoint information (see \citet{Diwoky_Volkwein2001,Hinze_2005,Hinze_adapt_2011,Gubisch_Volkwein2013,Hay_Borggaard_Pelletier_2009,Carlberg_2011,Zahr_Farhat_2014}) into POD basis for reduced order optimization purposes.
A-posteriori analysis for POD applied to optimal control problems governed by parabolic and elliptic PDEs were developed in \citet{Hinze_2005, Hinze_Volkwein_2008, Tonna_Volkwein_2010, Troltzsch_Volkwein_2010, Kahlbacher_Volkwein_2012,Kammannetal2013}. Optimal snapshot location strategies for selecting additional snapshots at different time instances and for changing snapshots weights to represent more accurately the reduced order solutions were introduced in \citet{Kunisch_Volkwein2010}. Extension to parameterized nonlinear systems is available in \citet{Lass_Volkwein_adapt_2012}.  

POD was successfully applied to solve strong constraint four dimensional variational (4D-Var) data assimilation problems for oceanic problems (\citet{Cao_Zhu_2007,Fang_Pain_Navon2009}) and atmospheric models (\citet{Chen2011,
Chen2012,Daescu_Navon2007,Daescu_Navon2_2008,Du_parab2_2012}). A strategy that formulates first order optimality conditions starting from POD models has been implemented in 4D-Var systems in \citet{Vermeulen2006,Sava2012}, while hybrid methods using reduced adjoint models but optimizing in full space were introduced in \citet{Altaf2013} and \citet{Ambrozic2013}. POD/DEIM has been employed to reduce the CPU complexity of a 1D Burgers 4D-Var system in \citet{Baumann2013}. Recently \citet{Amsallem_et_al2013} used a two-step gappy POD procedure to decrease the computational complexity of the reduced nonlinear terms in the solution of shape optimization problems.
Reduced basis approximation \citep{BMN2004,grepl2005posteriori,patera2007reduced,rozza2008reduced,Dihlmann_2013}  is known to be very efficient for parameterized problems and has recently been applied in the context of reduced order optimization \citep{Gubisch_Volkwein2013,Lassila_Rozza_2010,Manzoni_Quarteroni_Rozza_2012,Rozza_Manzoni_2010}.

This paper develops a systematic approach to POD bases selection for Petrov-Galerkin and Galerkin based reduced order data assimilation systems with non-linear models.
The fundamental idea is to provide an order reduction strategy that ensures that the reduced Karush Kuhn Tucker (KKT) optimality conditions
accurately approximate the full KKT optimality conditions \citep{Karush1939,KT1951}. This property is guaranteed by constraining the reduced KKT conditions to coincide with the projected high-fidelity KKT equations. 
An error estimation result shows that smaller reduced order projection errors lead to more accurate reduced order optimal solutions. This result provides practical guidance for the construction of reduced order data assimilation systems, and for solving general reduced optimization problems. The research extends the results of \citet{Hinze_Volkwein_2008} by considering nonlinear models and Petrov-Galerkin projections. 

The proposed reduced order strategy is applied to solve a 4D-Var data assimilation problem with the two dimensional shallow water equations model. We compare three reduced 4D-Var data assimilation systems using three different POD Galerkin based reduced order methods namely standard POD, tensorial POD and standard POD/DEIM (see \citet{Stefanescu_etal_forwardPOD_2014}). For Petrov-Galerkin projection stabilization strategies have to be considered. To the best of our knowledge this is the first application of POD/DEIM to obtain suboptimal solutions of reduced data assimilation system governed by a geophysical 2D flow model. For the mesh size used in our experiments the hybrid POD/DEIM reduced data assimilation system is approximately ten times faster then the full space data assimilation system, and this ratio is found out to be directly proportional with the mesh size.

The remainder of the paper is organized as follows. Section \ref{sec:Full_4D_Var_DA} introduces the optimality condition for the standard 4D-Var data assimilation problem. Section \ref{Sec:ROM} reviews the reduced order modeling methodologies deployed in this work: 
standard, tensorial, and DEIM POD. Section \ref{sec:POD_bases} derives efficient POD bases selection  strategies for reduced POD 4D-Var data assimilation systems governed by nonlinear state models using both Petrov-Galerkin and Galerkin projections. An estimation error result is also derived.  Section \ref{SWE_ROMS_DA} discusses the swallow water equations model and the three reduced order 4D-Var data assimilation systems employed for comparisons in this study. Results of extensive numerical experiments are discussed in Section \ref{sec:Numerical_resuls} while conclusions are drawn in Section \ref{sec:Conclusions}. Finally an appendix describing the high-fidelity alternating direction fully implicit (ADI) forward, tangent linear and adjoint shallow water equations (SWE) discrete models and reduced order SWE tensors is presented.

\section{Strong constraint 4D-Var data assimilation system}\label{sec:Full_4D_Var_DA}

Variational data assimilation seeks the optimal parameter values that provide the best fit (in some well defined sense)
of model outputs with physical observations. In this presentation we focus on the traditional approach where the
model parameters are the initial conditions ${\bf x}_0$, however the discussion can be easily generalized to further include
other model parameter  as  control variables. In 4D-Var the objective function $\J:\mathbb{R}^{N_{\textnormal{state}}} \to \mathbb{R}$ that quantifies the model-data misfit and accounts for prior information is minimized:
\begin{subequations}
\label{eqn:4DVar_problem}
\begin{equation}
\label{eqn:cost_function_4D_Var}
 \J({\mathbf x}_0) = \frac{1}{2}\big({\mathbf x}^{\rm b}_0-{\mathbf x}_0\big)^T{\mathbf B}_0^{-1}\big({\mathbf x}^{\rm b}_0-{\mathbf x}_0\big) +\frac{1}{2}\sum_{i=0}^N\big({\mathbf y}_i-\H_i\left({\mathbf x}_i\right)\big)^T\mathbf{R}_i^{-1}\big({\mathbf y}_i-\H\left({\mathbf x}_i\right)\big),
\end{equation}
subject to the constraints posed by the nonlinear forward model dynamics
\begin{equation}\label{eqn:full_forward}
 {\mathbf x}_{i+1} = \M_{i,i+1}\left({\mathbf x}_i\right), \quad i=0,..,N-1.
\end{equation}
\end{subequations}

Here ${\mathbf x}^{\rm b}_0 \in \mathbb{R}^{N_{\textnormal{state}}}$ is the background state and represents the best estimate of the true state ${\mathbf x}^{\rm true}_0 \in \mathbb{R}^{N_{\textnormal{state}}}$ 
prior to any measurement being available, where $N_{\textnormal{state}}$ is the number of spatial discrete variables. The background errors are generally
assumed to have a Gaussian distribution, i.e. ${\mathbf x}^{\rm b}_0 - {\mathbf x}^{\rm true}_0 \in \mathcal{N}(0,{\bf B}_0)$, where ${\bf B}_0 \in \mathbb{R}^{N_\textnormal{state}\times N_\textnormal{state}}$
is the background error covariance matrix. The nonlinear model $\M_{i,i+1}: \mathbb{R}^{N_{\textnormal{state}}} \to \mathbb{R}^{N_{\textnormal{state}}} $, $i=0,..,N-1$ advances the state vector in time from $t_i$ to $t_{i+1}$.  The state variables are ${\mathbf x}_i \in \mathbb{R}^{N_{\textnormal{state}}}$ and the data (observation) values are ${\mathbf y}_i \in \mathbb{R}^{N_{\textnormal{obs}}}$ at times $t_i$, $i=0,..,N$.  These observations are corrupted by instruments and representativeness errors \cite{Cohn1997} which are assumed to have a normal distribution $\mathcal{N}(0,{\mathbf{R}}_i)$, where $\mathbf{R}_i\in \mathbb{R}^{N_{\textnormal{obs}}\times N_{\textnormal{obs}}}$ describes the observation error
covariance matrix at time $t_i$. The nonlinear observation operator $\H_i: \mathbb{R}^{N_{\textnormal{state}}} \to \mathbb{R}^{N_{\textnormal{obs}}}$ maps the model state space to the observation space.

Using the Lagrange multiplier technique the constrained optimization problem \eqref{eqn:4DVar_problem} is  replaced with the unconstrained
optimization of the following Lagrangian function, $\mathcal{L}: \mathbb{R}^{N_\textnormal{state}} \to \mathbb{R}$
\begin{equation}\label{eqn:Lagrangian_cost_function}
\begin{split}
&\mathcal{L}({\mathbf x}_0) = \frac{1}{2}\big({\mathbf x}^{\rm b}_0-{\mathbf x}_0\big)^T{\mathbf B}_0^{-1}\big({\mathbf x}^{\rm b}_0-{\mathbf x}_0\big) + \frac{1}{2}\sum_{i=0}^N\big({\mathbf y}_i-\H({\bf x}_i)\big)^T\mathbf{R}_i^{-1}\big({\mathbf y}_i-\H({\bf x}_i)\big) +\\
& \quad \quad \quad \quad \quad \quad \sum_{i=0}^{N-1}{\mathbf \lambda}_{i+1}^T\big({\mathbf x}_{i+1}-\M_{i,i+1}\left({\mathbf x}_i\right)\big),
\end{split}
\end{equation}
where ${\mathbf \lambda}_i \in \mathbb{R}^{N_\textnormal{state}}$ is the Lagrange multipliers vector at observation time $t_i$.

Next we derive the first order optimality conditions. An infinitesimal change in $\mathcal{L}$ due to an infinitesimal  change $\delta {\mathbf x}_0$  in  ${\mathbf x}_0$ is
\begin{equation}
\label{eqn:Lagrangian_cost_function_variationI}
\begin{split}
 \delta \mathcal{L}({\mathbf x}_0) &= -\delta {\mathbf x}_0^T {\mathbf B}_0^{-1}\big({\mathbf x}^{\rm b}_0-{\mathbf x}_0\big) - \sum_{i=0}^N \delta {\mathbf x}_i^T {\mathbf H}_i^T \mathbf{R}_{i}^{-1} \big({\mathbf y}_i - \H({\mathbf x}_i)\big) + \\
 &\sum_{i=0}^{N-1} \lambda_{i+1}^T\big(\delta {\mathbf x}_{i+1} - {\mathbf M}_{i,i+1}\delta {\mathbf x}_i\big) + \sum_{i=0}^{N-1} \delta {\mathbf \lambda}_{i+1}^T\big({\mathbf x}_{i+1} - \M_{i,i+1}\left({\mathbf x}_i\right)\big), \\
\end{split}
\end{equation} 
where $\delta {\mathbf x}_i = (\partial {\mathbf x}_i / \partial {\mathbf x}_0)\, \delta {\mathbf x}_0$, and ${\bf H}_i$ and ${\mathbf M}_{i,i+1}$ are the 
Jacobian matrices of $\H_i,~i=0,..,N$ and $\M_{i,i+1}$ for all time instances $t_i$, $i=0,\dots,N-1$, at ${\bf x}_i$,
\begin{equation*}\label{eqn:tangent_linear_operators}
 {\bf H}_i = \frac{\partial \H_i}{\partial {\mathbf x}_i}({\mathbf x}_i) \in \mathbb{R}^{N_\textnormal{obs} \times N_\textnormal{state}} ,\quad {\mathbf M}_{i,i+1} = \frac{\partial \M_{i,i+1}}{\partial {\mathbf x}_i}({\mathbf x}_i) \in \mathbb{R}^{N_\textnormal{state} \times N_\textnormal{state}}.
\end{equation*}
The corresponding adjoint operators are ${\bf H}_i^T \in \mathbb{R}^{N_\textnormal{state} \times N_\textnormal{obs}}$ and ${\bf M}_{i+1,i}^* \in \mathbb{R}^{N_\textnormal{state} \times N_\textnormal{state}}$, respectively, and satisfy
\begin{equation*}\label{eqn:adjoint_operators}
\begin{split}
 \langle {\bf H}_iz_1, y_1 \rangle_{\mathbb{R}^{N_\textnormal{obs}}} = \langle z_1, {\bf H}^T_iy_1 \rangle_{\mathbb{R}^{N_\textnormal{state}}},~\forall z_1 \in {\mathbb{R}^{N_\textnormal{state}}},~\forall y_1 \in {\mathbb{R}^{N_\textnormal{obs}}}, \\
 \langle {\mathbf M}_{i,i+1} z_1, z_2 \rangle_{\mathbb{R}^{N_{\textnormal{state}}}} = \langle z_1, {\mathbf M}^*_{i+1,i} z_2 \rangle_{\mathbb{R}^{N_{\textnormal{state}}}},\quad \forall z_1,~z_2 \in \mathcal {\mathbb{R}^{N_{\textnormal{state}}}},\\
\end{split}
\end{equation*}
where $\langle \cdot , \cdot \rangle_{\mathbb{R}^{N_{\textnormal{state}}}},~\langle \cdot , \cdot \rangle_{\mathbb{R}^{N_{\textnormal{obs}}}}$ are the corresponding Euclidian products. The adjoint operators of ${\bf H}_i$ and ${\mathbf M}_{i,i+1}$ are their transposes. We prefer to use the notation ${\bf M}_{i+1,i}^*= {\bf M}_{i,i+1}^T$ to show that the corresponding adjoint model runs backwards in time. After rearranging \eqref{eqn:Lagrangian_cost_function_variationI} and using the definition of adjoint operators one obtains:
\begin{eqnarray}
\label{eqn:Lagrangian_cost_function_variationII}
   \delta\mathcal{L}({\mathbf x}_0) & =& -\delta {\mathbf x}_0^T {\mathbf B}_0^{-1}\big({\mathbf x}^{\rm b}_0-{\bf x}_0\big) + \sum_{i=1}^{N-1}\delta {\mathbf x}_i^T \Bigl({\mathbf \lambda}_i - {\mathbf M}_{i+1,i}^*\mathbf{\lambda}_{i+1}-{\bf H}_i^T \mathbf{R}_{i}^{-1} \big({\mathbf y}_i - \H({\mathbf x}_i)\Bigr) \\
  \nonumber
 && + \delta {\mathbf x}_N^T\Bigl({\mathbf \lambda}_N- {\bf H}_N^T \mathbf{R}_{N}^{-1} \big({\mathbf y}_N - \H({\mathbf x}_N)\big)\Bigr) -\delta {\mathbf x}_0^T\Bigl({\bf H}_0^T \mathbf{R}_{0}^{-1}\big({\mathbf y}_0 - \H({\mathbf x}_0)\big) + {\mathbf M}_{1,0}^*\mathbf{\lambda}_{1}\Bigr) \\
\nonumber
 && + \sum_{i=0}^{N-1}\delta {\mathbf \lambda}_{i+1}^T \big({\mathbf x}_{i+1} - \M_{i,i+1}\left({\mathbf x}_i\right)\big).
\end{eqnarray}  
The first order necessary optimality conditions for the Full order 4D-Var are obtained by zeroing the variations of \eqref{eqn:Lagrangian_cost_function_variationII}:
\begin{subequations}
\label{eqn:KKT_Full}
\begin{eqnarray}
\label{eqn:KKT_Full_forward}
&& \textnormal{\it Full order forward model:} \\
\nonumber
&& \qquad {\mathbf x}_{i+1} = \M_{i,i+1}\left({\mathbf x}_i\right), \quad i=0,..,N-1; \\
\label{eqn:KKT_Full_adjoint}
&& \textnormal{\it Full order adjoint model:} \\
\nonumber
&& \qquad  \mathbf{\lambda}_N = {\mathbf H}_N^T \mathbf{R}_{N}^{-1} \big({\bf y}_N - \H({\mathbf x}_N)\big), \\
\nonumber
&& \qquad {\mathbf \lambda}_i = {\mathbf M}_{i+1,i}^*\mathbf{\lambda}_{i+1} + {\mathbf H}_i^T \mathbf{R}_{i}^{-1} \big({\bf y}_i - \H({\mathbf x}_i)\big),\quad i=N-1,..,0; \\
\label{eqn:KKT_Full_gradient}
&& \textnormal{\it Full order gradient of the cost function:} \\
\nonumber
&& \qquad \nabla_{{\bf x}_0}\mathcal{L} = 
-{\mathbf B}_0^{-1}\big({\mathbf x}^{\rm b}_0-{\mathbf x}_0\big) - {\mathbf \lambda}_0 = 0.
\end{eqnarray}
\end{subequations}

\section{Reduced order forward modeling}\label{Sec:ROM}

The most prevalent basis selection method for model reduction of nonlinear problems is the
proper orthogonal decomposition, also known as Karhunen-Lo\`{e}ve expansion
\cite{karhunen1946zss,loeve1955pt}, principal component analysis \cite{hotelling1939acs},
and empirical orthogonal functions \cite{lorenz1956eof}.

Three reduced order models will be considered in this paper: standard POD (SPOD), tensorial POD (TPOD), and standard POD/Discrete Empirical 
Interpolation Method (POD/DEIM), which were described in \citet{Stefanescu_etal_forwardPOD_2014} and \citet{Stefanescu2012}. 
The reduced Jacobians required by the ADI schemes are obtained analytically for all three ROMs via tensorial calculus
and their computational complexity depends only on $k$, the dimension of POD basis. 
The above mentioned methods differ in the way the nonlinear terms are treated. We illustrate the application of the methods to
reduce a polynomial quadratic nonlinearity $N(\mathbf{x}_i)={\bf x}_i^2,~i=0,..,N$, where vector powers are taken component-wise. Details regarding standard POD
approach including the snapshot procedure, POD basis computation and the corresponding reduced equations can be found in \citep{Stefanescu_etal_forwardPOD_2014}.
We assume a Petrov-Galerkin projection for constructing the reduced order models with the two biorthogonal projection matrices 
$U,W \in \mathbb{R}^{N_\textnormal{state}\times k}$, $W^TU = I_k$,
where $I_k$ is the identity matrix of order $k$, $k \ll N_{\textnormal{state}}$. $U$ denotes the POD basis (trial functions) and the test functions are stored in $W$. 
We assume a POD expansion of the state ${\mathbf x}\approx U{\tilde {\mathbf x}}$, and the reduced order quadratic term 
$\widetilde{N}({\bf \tilde x}) \approx N(\mathbf{x})$ is detailed bellow. For simplicity we removed the index $i$ from the state variable notation, thus 
$\mathbf{x} \in \mathbb{R}^{N_\textnormal{state}},~{\tilde {\mathbf x}} \in \mathbb{R}^k$.
\paragraph{Standard POD}
\begin{equation}\label{eqn:POD_standard_nonlinearity_general}
\widetilde{N}({\bf \tilde x}) = \underbrace{W^T}_{k \times N_\textnormal{state}}
\underbrace{\big(U{\bf \tilde x}\big)^2}_{N_\textnormal{state} \times 1}, \quad \widetilde{N}({\bf \tilde x}) \in \mathbb{R}^k,
\end{equation}
where vector powers are taken component-wise.
\paragraph{Tensorial POD}
\begin{equation}\label{eqn:POD_tensorial_nonlinearity_general}
\widetilde{N}({\bf \tilde x}) = \big[{\widetilde{N}}_i\big]_{i=1,..,k}\in \mathbb{R}^k;\quad \widetilde{N}_i=\sum_{j=1}^k\sum_{l=1}^kT_{i,j,l}{\tilde x}_j{\tilde x}_l,
\end{equation}
where the rank-three tensor $T$ is defined as
\[
T = \left(T_{i,j,l}\right)_{i,j,l=1,..,k}\in \mathbb{R}^{k \times k \times k},\quad T_{i,j,l} = \sum_{r=1}^{N_{\textnormal{state}}} W_{r,i}U_{r,j}U_{r,l}.
\]
\paragraph{Standard POD/DEIM}
\begin{equation}
\tilde N({\bf \tilde x}) \approx \underbrace{W^TV(P^TV)^{-1}}_{k\times m} \underbrace{\left(P^TU{\bf \tilde x}\right)^2}_{m \times 1},
\label{eqn:POD_DEIM_nonlinearity_general}
\end{equation}
where $m$ is the number of interpolation points, $V\in \mathbb{R}^{N_{\textnormal{state}} \times m}$ gathers the first
$m$ POD basis modes of the nonlinear term while $P \in \mathbb{R}^{{N_{\textnormal{state}}} \times m}$ is the DEIM interpolation selection matrix (\citet{Cha2008,ChaSor2010,ChaSor2012}).

The systematic application of these techniques in the Petrov-Galerkin projection framework to \eqref{eqn:full_forward} leads to the following reduced order forward model
\begin{equation}\label{eqn:reduced_forward_AR_init}
{\bf \tilde x}_{i+1} = \widetilde{\M}_{i,i+1}\left({\bf \tilde x}_i\right),\quad \widetilde{\M}_{i,i+1}\left({\bf \tilde x}_i\right)= W^T\M_{i,i+1}\left(U {\bf \tilde x}_i\right),\quad i=0,..,N-1.
\end{equation}
%

\section{Reduced order 4D-Var data assimilation}\label{sec:POD_bases}

Two major strategies for solving data assimilation problems with reduced order models have been discussed in the literature. The ``reduced adjoint'' (RA) approach \cite{Altaf2013} projects the first order optimality equation \eqref{eqn:KKT_Full_adjoint} of the full system onto some POD reduced space and solves the optimization problem in the full space, while the ``adjoint of reduced'' (AR) approach \cite{Pelc2012,Vermeulen2006} formulates the first order optimality conditions from the forward reduced order model \eqref{eqn:reduced_forward_AR} and searches for the optimal solution by solving a collection of reduced optimization problems in the reduced space. Reduced order formulation of the cost function is employed in the ``adjoint of reduced'' case \cite{Dimitriu_2010}. In the control literature these approaches are referred to as ``design-then-reduce'' and ``reduce-then-design'' methods \cite{Atwell_King2001,Atwell_King2004}.

The ``reduced of adjoint''  approach avoids the implementation of the adjoint of the tangent linear approximation 
of the original nonlinear model by replacing it with a reduced adjoint. This is obtained by projecting the full set of equations
onto a POD space build on the snapshots computed with the high-resolution forward model. Once the gradient is obtained in reduced space it is projected back
in full space and the minimization process is carried in full space. The drawbacks of this technique consist in the use of an inaccurate 
low-order adjoint model with respect to its full counterpart that may lead to erroneous gradients. Moreover the computational cost of the optimization system is still high since it requires the run of the full forward model to evaluate the cost function during the minimization procedure.

A major concern with the ``adjoint of reduced''  approach is the lack of accuracy in the optimality conditions with respect to the full system.
The reduced forward model is accurate, but its adjoint and optimality condition poorly approximate their full counterparts since the POD basis relies only on forward dynamics information. Consequently both the RA and the AR methods may lead to inaccurate suboptimal solutions.

Snapshots from both primal and dual systems have been used for balanced truncation of a linear state-system. \cite{Willcox02balancedmodel}. \citet{Hinze_Wolkwein2008} developed a-priori error estimates for linear quadratic optimal control problems using proper orthogonal decomposition. They state that error estimates for the adjoint state yield error estimates of the control suggesting that accurate reduced adjoint models with respect to the full adjoint model lead to more accurate suboptimal surrogate solutions. Numerical results confirmed that a POD manifold built on snapshots taken from both forward and adjoint trajectories provides more accurate reduced optimization systems.

This work develops a systematic approach to select POD bases for reduced order data assimilation with non-linear models. Accuracy of the reduced optimum is ensured by constraining the ``adjoint of reduced'' optimality conditions to coincide with a generalized ``reduced of adjoint'' KKT conditions obtained by projecting  all equations \eqref{eqn:KKT_Full} onto appropriate reduced manifolds. This leads to the concept of accurate reduced KKT conditions and provides practical guidance to construct reduced order manifolds for reduced order optimization. The new strategy, named ARRA,  provides a unified framework where the AR and RA approaches lead to the same solution for general reduced order optimization problems. In the ARRA approach the model reduction and adjoint differentiation operations commute.

\subsection{The ``adjoint of reduced forward model'' approach}

We first define the objective function. To this end we assume that the forward POD manifold $\Uf$ is computed using only snapshots of 
the full forward model solution (the subscript denotes that only forward dynamics information is used for POD basis construction). 
The Petrov-Galerkin (PG) test functions $\Wf$ are different than the trial functions $\Uf$. 
Assuming a POD expansion of ${\bf x}_i \approx \Uf\,{\bf \tilde x}_i,~i=0,..,N,$ the reduced data assimilation problem minimizes the following reduced  order cost function $\J^\textsc{pod} : \mathbb{R}^k \to \mathbb{R}$
\begin{subequations}
\label{eqn:reduced_optimization}
\begin{equation}\label{eqn:reduced_cost_func}
\begin{split}
 &\J^\textsc{pod}({\bf \tilde x}_0) = \frac{1}{2}\big({\mathbf x}^{\rm b}_0-\Uf\,{\bf \tilde x}_0\big)^T{\bf B}_0^{-1}\big({\mathbf x}^{\rm b}_0-\Uf \tilde {\bf x}_0\big) \\
 &\quad \quad \quad \quad \quad \quad +\frac{1}{2}\sum_{i=0}^N\big({\mathbf y}_i-\H_i(\Uf\,{\bf \tilde x}_i)\big)^T\mathbf{R}_i^{-1}\big({\mathbf y}_i-\H_i(\Uf\,{\bf \tilde x}_i)\big)^T,
\end{split}
\end{equation}
subject to the constraints posed by the reduced order model dynamics
\begin{equation}
\label{eqn:reduced_forward_AR}
{\bf \tilde x}_{i+1} = \widetilde{\M}_{i,i+1}\left({\bf \tilde x}_i\right),\quad \widetilde{\M}_{i,i+1}\left({\bf \tilde x}_i\right)= \Wf^T\M_{i,i+1}\left(\Uf\,{\bf \tilde x}_i\right),\quad i=0,..,N-1.
\end{equation}
\end{subequations}
An observation operator that maps directly from the reduced model space to observations space may be introduced. For clarity sake we will 
continue to use operator notation $\H_i,~i=0,..,N$ and $\widetilde{\M}_{i,i+1}$ denotes the PG reduced order forward model that propagates the reduced order state from $t_i$ to $t_{i+1}$ for $i=0,..,N-1$.

Next the constrained optimization problem \eqref{eqn:reduced_optimization} is replaced by an unconstrained one for the reduced Lagrangian function $\mathcal{L}^\textsc{pod} : \mathbb{R}^k \to \mathbb{R}$
\begin{eqnarray}
\label{eqn:Lagrangian_reduced_cost_function_variationI}
 \mathcal{L}^\textsc{pod}({\bf \tilde x}_0) &=& \frac{1}{2}\big({\mathbf x}^{\rm b}_0-\Uf\,{\bf \tilde x}_0\big)^T{\bf B}_0^{-1}\big({\mathbf x}^{\rm b}_0-\Uf\,{\bf \tilde x}_0\big)\\
 \nonumber
  && +\frac{1}{2}\sum_{i=0}^N\big({\mathbf y}_i-\H_i(\Uf\,{\bf \tilde x}_i)\big)^T\mathbf{R}_i^{-1}\big({\mathbf y}_i-\H_i(\Uf\,{\bf \tilde x}_i)\big) \\
  \nonumber
  && +\sum_{i=0}^{N-1} {\bf \tilde \lambda}_{i+1}^T \big({\bf \tilde x}_{i+1} - \widetilde{\M}_{i,i+1}\left({\bf \tilde x}_i\right)\big),
\end{eqnarray}
where ${\bf \tilde \lambda}_{i} \in \mathbb{R}^k \in,~i=1,..,N$.
The variation of $\mathcal{L}^\textsc{pod}$ is given by
\begin{equation}\label{eqn:Lagrangian_reduced_cost_function_variationII}
\begin{split}
 &  \delta \mathcal{L}^\textsc{pod}({\bf \tilde x}_0) = -\delta {\bf \tilde x}_0^T {\mathbf B}_0^{-1}\big({\mathbf x}^{\rm b}_0-\Uf\,{\bf \tilde x}_0\big)
 + \sum_{i=1}^{N-1}\delta {\bf \tilde x}_i^T \Bigl({\bf \tilde \lambda}_i - {\bf \widetilde{M}}_{i+1,i}^*{\bf \tilde \lambda}_{i+1}\\
 &\quad \quad -\Uf^T{\bf \widehat H}_i^T \mathbf{R}_{i}^{-1} \big({\mathbf y}_i - \H_i(\Uf\,{\bf \tilde x}_i)\Bigr) +
  \delta {\bf \tilde x}_N^T \Bigl({\bf \tilde \lambda}_N- \Uf^T{\bf \widehat H}_N^T \mathbf{R}_{N}^{-1} \big({\bf y}_N - \H_i(\Uf\,{\bf \tilde x}_N)\big)\Bigr) \\
  & \quad \quad - \delta {\bf \tilde x}_0^T\Bigl(\Uf^T{\bf H}_0^T \mathbf{R}_{0}^{-1} \big({\mathbf y}_0 - \H_0(\Uf\,{\bf \tilde x}_0)\big) + {\bf \widetilde{M}}_{1,0}^* {\bf \tilde \lambda}_1\Bigr) + \sum_{i=0}^{N-1}\delta {\bf \tilde \lambda}_{i+1}^T \big({\bf \tilde x}_{i+1} - \widetilde{\M}_{i,i+1}\left({\bf \tilde x}_i\right)\big), \\
\end{split}
\end{equation}
where
\[
{\bf \widetilde{M}}_{i+1,i}^* = \Uf^T{\bf \widehat M}_{i+1,i}^*\Wf,\quad i=0,..,N-1
\]
is the adjoint of the reduced order linearized forward model $\M_{i,i+1}$. The operators ${\bf \widehat H}_i^T,~i=0,..,N$ and ${\bf \widehat M}_{i+1,i}^*,~i=0,..,N-1$ are  the high-fidelity adjoint models evaluated at $\Uf\,{\bf \tilde x}_i$. 
The reduced KKT conditions are obtained by setting the variations of \eqref{eqn:Lagrangian_reduced_cost_function_variationII} to zero:
\begin{subequations}
\label{eqn:KKT_AR}
\begin{eqnarray}
\label{eqn:KKT_AR_forward}
&& \textnormal{\it AR reduced forward model:}\\
\nonumber
&& \qquad {\bf \tilde x}_{i+1} = \widetilde{\M}_{i,i+1}\left({\bf \tilde x}_i\right),
\quad \widetilde{\M}_{i,i+1}\left({\bf \tilde x}_i\right) = \Wf^T\,{\M}_{i,i+1}\left({\Uf\,\bf \tilde x}_i\right),\quad i=0,..,N-1; \\
\label{eqn:KKT_AR_adjoint}
&& \textnormal{{\it AR reduced adjoint model}:  }  \\
\nonumber
&& \qquad {\bf \tilde \lambda}_N = \Uf^T\,{\bf \widehat H}_N^T\,\mathbf{R}_{N}^{-1}\, \big({\bf y}_N - \H_i(\Uf\,{\bf \tilde x}_N)\big), \quad \quad \quad \quad \quad \quad \quad \\
\nonumber
&& \qquad {\bf \tilde \lambda}_i = \Uf^T\,{\bf \widehat M}_{i+1,i}^*\,\Wf\, {\bf \tilde \lambda}_{i+1} + \Uf^T\,{\bf \widehat H}_i^T \,\mathbf{R}_{i}^{-1} \,\big({\mathbf y}_i - \H_i(\Uf\,{\bf \tilde x}_i)\big),\quad i=N-1,..,0; \\
\label{eqn:KKT_AR_gradient}
&& \textnormal{{\it AR cost function gradient} }: \\
\nonumber
&& \qquad \nabla_{{\bf \tilde x}_0}\mathcal{L}^\textsc{pod} = -\Uf^T {\bf B}_0^{-1}\big({\mathbf x}^{\rm b}_0-\Uf\,{\bf \tilde x}_0\big) - {\bf \tilde \lambda}_0 = 0.
\end{eqnarray}
\end{subequations}
A comparison between the KKT systems of the reduced order optimization \eqref{eqn:KKT_AR} and of the full order problem  \eqref{eqn:KKT_Full} reveals that the reduced forward model \eqref{eqn:KKT_AR_forward} is, by construction, an accurate approximation of the full forward model \eqref{eqn:KKT_Full_forward}. However, the AR adjoint \eqref{eqn:KKT_AR_adjoint} and AR gradient \eqref{eqn:KKT_AR_gradient} equations are constructed using the forward model bases and test functions, and are not guaranteed to approximate well the corresponding full adjoint \eqref{eqn:KKT_Full_adjoint} and full gradient \eqref{eqn:KKT_Full_gradient} equations, respectively.

\subsection{The ``reduced order adjoint model'' approach} \label{subsec:red_adjoint}

In a general RA framework the optimality conditions \eqref{eqn:KKT_Full} are projected onto reduced order subspaces. The forward, adjoint, and gradient variables are reduced using forward, adjoint, and gradient bases 
$\Uf,~\Ua$ and $\Ug$ such that $\x_i \approx \Uf\, \xr_i,~\la_i \approx \Ua\, \lar_i,~i=0,..,N$, and $\nabla_{\x_0} \L \approx \Ug\, \nabla_{\xr_0} \Lpod$. 
The test functions for the forward, adjoint, and gradient equations are $\Wf$, $\Wa$, and $\Wg$, respectively. 
The objective function to be minimized is \eqref{eqn:reduced_cost_func}. The projected KKT conditions read:
\begin{subequations}
\label{eqn:KKT_RA}
\begin{eqnarray}
\label{eqn:KKT_RA_forward}
&& \textnormal{\it RA reduced forward model:} \\
\nonumber
&& \qquad {\bf \tilde x}_{i+1} = \widetilde{\M}_{i,i+1}\,\left({\bf \tilde x}_i\right),\quad i=0,..,N-1;\\
\label{eqn:KKT_RA_adjoint}
&& \textnormal{\it RA reduced adjoint model:}  \\
\nonumber
&& \qquad  {\bf \tilde \lambda}_N = \Wa^T\,{\bf H}_N^T\, \mathbf{R}_{N}^{-1}\, \big({\bf y}_N - \H_i(\Uf{\xr}_N)\big),  \\
\nonumber
&& \qquad  {\bf \tilde \lambda}_{i} = \Wa^T\, {\bf  M}_{i+1,i}^*\,\Ua\, {\bf \tilde \lambda_{i+1}} + \Wa^T\, {\bf H}_i^T\,\mathbf{R}_i^{-1}\,\big({\bf y}_i - \H_i({\bf x}_i)\big), \quad i=N-1,..,0; \\
\label{eqn:KKT_RA_gradient}
&& \textnormal{{\it RA cost function gradient:}}  \\
\nonumber
&& \qquad \nabla_{\xr_0}\Lpod = -\Wg^T\, \big[{\mathbf B}_0^{-1}\,\big({\mathbf x}^{\rm b}_0-\Uf\,\xr_0\big) - \Ua\, {\bf \tilde \lambda_{0}}\big] = 0.
\end{eqnarray}
\end{subequations}
With appropriately chosen basis and test functions the system \eqref{eqn:KKT_RA} can accurately approximate the full order optimality conditions \eqref{eqn:KKT_Full}.  {\it However, in general the projected system \eqref{eqn:KKT_RA} does not represent the KKT conditions of any optimization problem}, and therefore the RA approach does not automatically provide a consistent optimization framework in the sense given by the following definition.

\begin{definition}\label{def:consistent_reduced_KKT}
A reduced KKT system \eqref{eqn:KKT_RA} is said to be consistent if it represents the first order optimality conditions of some reduced order optimization problem. 
\end{definition}

By constraining the projected KKT \eqref{eqn:KKT_RA} to match the ``adjoint of reduced'' optimality 
conditions \eqref{eqn:KKT_AR}, one can obtain a consistent optimality framework where the reduced system accurately approximates the high-fidelity  optimality system  \eqref{eqn:KKT_Full}.  This new theoretical framework is discussed next.

\subsection{The ARRA approach: ensuring accuracy of the first order optimality system}

The proposed method constructs the reduced order optimization problem \eqref{eqn:reduced_optimization} such that the reduced KKT equations \eqref{eqn:KKT_AR}  accurately approximate the high-fidelity KKT conditions \eqref{eqn:KKT_Full}. 

\begin{enumerate}
\item In the AR approach the first KKT condition \eqref{eqn:KKT_AR_forward} is an accurate approximation of  \eqref{eqn:KKT_Full_forward} by construction, since $\Uf$ is constructed with  snapshots of the full forward trajectory, i.e. ${\bf x} \approx \Uf\,{\bf \tilde x}$. 
\item Next, we require that the AR reduced adjoint model \eqref{eqn:KKT_AR_adjoint} is a low-order accurate approximation of the full adjoint model \eqref{eqn:KKT_Full_adjoint}.  This is achieved by imposing that the AR adjoint \eqref{eqn:KKT_AR_adjoint} is a reduced-order model obtained by projecting the full adjoint equation into a reduced basis containing high fidelity adjoint information, i.e., has the RA form \eqref{eqn:KKT_RA_adjoint}. We obtain
\begin{equation}
\label{eqn:POD_basis_optimal_strategy}
 \Wf = \Ua \textnormal{ and } \Wa = \Uf.
\end{equation}
The forward test functions are the bases $\Wf = \Ua$ constructed with  snapshots of the full adjoint trajectory, i.e. ${\bf \lambda} \approx \Ua\,{\bf \tilde \lambda}$. Note that in \eqref{eqn:KKT_RA_adjoint} the model and observation operators are evaluated at the full forward solution ${\bf x}$ and then projected. In \eqref{eqn:KKT_AR_adjoint} the model and observation  operators are evaluated at the reduced forward solution $\Uf\,{\bf \tilde x}_i  \approx {\bf x}_i$. The adjoint test functions are the bases $\Wa = \Uf$ constructed with  snapshots of the full forward trajectory.


%
\item With \eqref{eqn:POD_basis_optimal_strategy} the reduced gradient equation \eqref{eqn:KKT_AR_gradient} becomes   
\begin{equation}
\label{eqn:POD_basis_optimal_gradient}
\nabla_{{\bf \tilde x}_0}\mathcal{L}^\textsc{pod} = \Uf^T \left( -{\bf B}_0^{-1}\big({\mathbf x}^{\rm b}_0-\Uf\,{\bf \tilde x}_0\big) - \Ua\lar_0 \right) = 0.
\end{equation}
We require that the gradient \eqref{eqn:POD_basis_optimal_gradient} is a low-order accurate approximation of the full gradient \eqref{eqn:KKT_Full_gradient} obtained by projecting it, i.e., has the RA gradient form \eqref{eqn:KKT_RA_gradient}. This leads to
\begin{equation}
\label{eqn:POD_basis_optimal_strategy2}
 \Wg = \Uf .
\end{equation}
From PG construction, $\Ug$ is orthogonal with $\Wg=\Uf$, thus a good choice for $\Ug$ is $\Ua$ where the gradient of the background term with respect to the initial conditions $-{\bf B}_0^{-1}\big({\mathbf x}_0^{{\rm b}}-{\bf x}_0\big)$ is used to enrich the high-fidelity adjoint snapshots matrix, i.e. $\Ug = \Ua$.
\end{enumerate}

We refer to the technique described above as ``adjoint of reduced = reduced of adjoint'' (ARRA) framework. The ARRA reduced order optimality system is:
\begin{subequations}
\label{eqn:KKT_ARRA}
\begin{eqnarray}
\label{eqn:forward_ARRA_reduced_model}
&& \textnormal{{\it ARRA reduced forward model}:  }\\
\nonumber
&& \qquad {\bf \tilde x}_{i+1} = \widetilde{\M}_{i,i+1}\left({\bf \tilde x}_i\right),\quad \widetilde{\M}_{i,i+1}\left({\bf \tilde x}_i\right) = \Ua^T\,{\M}_{i,i+1}\left({\Uf\,\bf \tilde x}_i\right),\quad i=0,..,N-1, \\
\label{eqn:adjoint_ARRA_reduced_model}
&& \textnormal{{\it ARRA reduced adjoint model}:  }  \\
\nonumber
&& \qquad {\bf \tilde \lambda}_N = \Uf^T{\bf \widehat H}_N^T \mathbf{R}_{N}^{-1} \big({\bf y}_N - \H(\Uf\,{\bf \tilde x}_N)\big), \quad \quad \quad \quad \quad \quad \quad \\
\nonumber
&& \qquad {\bf \tilde \lambda}_i = \Uf^T{\bf \widehat M}_{i,i+1}^T\Ua{\bf \tilde \lambda}_{i+1} + \Uf^T{\bf \widehat H}_i^T \mathbf{R}_{i}^{-1} \big({\mathbf y}_i - \H(\Uf\,{\bf \tilde x}_i)\big),\quad i=N-1,..,0, \\
\label{eqn:gradient_ARRA_reduced_model}
&& \textnormal{{\it ARRA cost function gradient} }: \\
\nonumber
&& \qquad \nabla_{{\bf \tilde x}_0}\mathcal{L}^\textsc{pod} = -\Uf^T\, {\bf B}_0^{-1}\big({\mathbf x}^{\rm b}_0-\Uf\,{\bf \tilde x}_0\big) - {\bf \tilde \lambda}_0 = 0.
\end{eqnarray}
\end{subequations}


ARRA approach leads to consistent and accurate reduced KKT conditions  \eqref{eqn:KKT_ARRA}, in the following sense.
Equations \eqref{eqn:KKT_ARRA} are the optimality system of a reduced order problem (consistency), and each of the reduced optimality conditions \eqref{eqn:KKT_ARRA} is a surrogate model that accurately represents the corresponding full order condition (accuracy).

Iterative methods such as Broyden--Fletcher-–Goldfarb–-Shanno (BFGS) \cite{BROYDEN01031970,Fletcher01011970,goldfarb1970family,shanno1970conditioning} uses feasible triplets $(\x, \la, \nabla_{\x_0\L})$ to solve \eqref{eqn:4DVar_problem}, in the following sense

\begin{definition}\label{def:feasible_pair}
The triplet $(\x,\la,\nabla_{\x_0} \L) \in \mathbb{R}^{\Ns \times N} \times \mathbb{R}^{\Ns \times N} \times \mathbb{R}^{\Ns}$ is said to be KKT-feasible
if $\x$ is the solution of the forward model \eqref{eqn:KKT_Full_forward} initiated with a given $\x_0 \in \mathbb{R}^\Ns$, $\la$ is the solution of
the adjoint model \eqref{eqn:KKT_Full_adjoint} linearized across the trajectory $\x$, and $\nabla_{\x_0} \L$ is the gradient of the 
Lagrangian function computed from \eqref{eqn:KKT_Full_gradient}.
\end{definition}

Note that $\nabla_{\x_0}  \L$ does not have to be zero, i.e., Definition \ref{def:feasible_pair} applies away from the optimum as well. The ARRA framework selects the reduced order bases such that high-fidelity feasible triplets are well approximated by reduced order feasible triplets. This introduces the notion of accurate reduced KKT conditions away from optimality.

\begin{definition}\label{def:accurate_reduced_feasible_KKT}
Let $(\x,\la,\nabla_{\x_0} \L) \in \mathbb{R}^{\Ns \times N} \times \mathbb{R}^{\Ns \times N} \times \mathbb{R}^{\Ns}$ be a KKT-feasible triplet of the full order
optimization problem \eqref{eqn:4DVar_problem}.  If  for any positive ${\varepsilon}_f,~{\varepsilon}_a$ and $\varepsilon_g$ there exists $k\leq \Ns$ and
three bases ${\bar U},~{\bar V}$ and ${\bar W} \in 
\mathbb{R}^{\Ns \times k}$ such that the reduced KKT-feasible triplet $(\xr,\lar,\nabla_{\xr_0} \Lpod ) \in \mathbb{R}^{k \times N} \times \mathbb{R}^{k \times N} \times \mathbb{R}^k$  of the  reduced optimization problem \eqref{eqn:reduced_optimization}  satisfies: 
\begin{subequations}
\label{eqn:accurate_feasible_kkt}
\begin{eqnarray}
\label{eqn:accurate_feasible_forward_kkt}
\| {\bf x}_i - {\bar U} {\bf \tilde x}_i  \|_2 &\le& \varepsilon_\textnormal{f}, \quad  i=0,..,N, \\
\label{eqn:accurate_feasible_adjoint_kkt}
\| {\mathbf \lambda}_i - {\bar V} {\bf \tilde \lambda}_i \|_2 &\le& \varepsilon_\textnormal{a},\quad  i=0,..,N, \\
\label{eqn:accurate_feasible_gradient_kkt}
\| \nabla_{{\bf x}_0}\mathcal{L} -{\bar W}\nabla_{{\bf \tilde x}_0}\mathcal{L}^\textsc{pod} \|_2 &\le& \varepsilon_\textnormal{g}, 
\end{eqnarray}
\end{subequations}
then the reduced order KKT system \eqref{eqn:KKT_AR}  built using  $\Uf = {\bar U}$, $\Wf={\bar V}$, and ${\bar W}$ that generated $(\xr,\lar,\nabla \Lpod_{\xr_0})$ is said to be accurate with respect to the full order KKT system \eqref{eqn:KKT_Full}.
\end{definition}

For efficiency we are interested to generate reduced bases whose dimension $k$ increases relatively slowly as ${\varepsilon}_f,~{\varepsilon}_a$ and $\varepsilon_g$ decrease. This can be achieved if local reduced order strategies are applied \cite{Rapun_2010,Peherstorfer_2013}.

It is computationally prohibitive to require all the reduced KKT systems used during the reduced optimization to be accurate according to definition \ref{def:accurate_reduced_feasible_KKT}.  ARRA framework proposes accurate KKT systems at the beginning of the reduced optimization procedure and whenever the reduced bases are updated ($\bar U = \Uf, \bar V = \bar W = \Ua$).

For the Petrov-Galerkin projection the adjoint POD basis must have an additional property, i.e to be orthonormal to the forward POD basis. Basically we are looking in the joint space of full adjoint solution and $-{\bf B}_0^{-1}\big({\mathbf x}_0^{{\rm b}}-{\bf x}_0\big)$ for a new set of coordinates that are orthonormal with the forward POD basis. A Gram Schmidt algorithm type is employed and an updated adjoint basis is obtained. This strategy should minimally modify the basis such that as to preserve the accuracy of the adjoint and gradient reduced order versions.

The pure Petrov-Galerkin projection does not guarantee the stability of the ARRA reduced order forward model \eqref{eqn:forward_ARRA_reduced_model}. The stability of the reduced pencil  is not guaranteed, even when the pencil is stable \cite{Bui-Thanh2007}. In our numerical experiments using shallow water equation model, the ARRA approach exhibited large instabilities in the solution of the forward reduced ordered model when the initial conditions were slightly perturbed from the ones used to generate POD manifolds. As it is described in appendix, we propose a quasi-Newton approach to solve the nonlinear algebraic system of equations obtained from projecting the discrete ADI swallow water equations onto the reduced manifolds. In a quasi-Newton iteration, the forward implicit reduced Petrov-Galerkin ADI SWE discrete model requires solving linear systems with the corresponding matrices given by $\Ua^T\, {\bf \widehat M}_{i+1,i}^*\, \Uf$, $i=1,2,..,N-1$. During the second iteration of the reduced optimization algorithm, for the evaluation of the objective function, we noticed that the reduced Jacobian operator has a spectrum that contains eigenvalues with positive real part explaining the explosive numerical instability.
Future work will consider stabilization approaches such as posing the problem of selecting the reduced bases as a goal-oriented optimization problem \cite{Bui-Thanh2007} or small-scale convex optimization problem \cite{Amsallem2012}. In addition we can constrain the construction of the left basis $\Ua^T$ to minimize a norm of the residual arising at each Newton iteration to promote stability \cite{Carlberg2_2011}. It will be worth checking how this updated basis will affect the accuracy of the reduced adjoint model and the reduced gradient.

An elegant solution to avoid stability issues while maintaining the consistency and accuracy of the reduced KKT system is to employ a Galerkin POD framework where $\Wf = \Uf$, $\Wa = \Ua$ and $\Wg = \Ug$. From (\ref{eqn:POD_basis_optimal_strategy}) and \eqref{eqn:POD_basis_optimal_strategy2} we obtain that $\Uf = \Ua=\Ug$, i.e., in the Galerkin ARRA framework the POD bases for forward and adjoint reduced models, and for the optimality condition must coincide. In the proposed ARRA approach this unique POD basis is constructed from snapshots of the full forward model solution, the full adjoint model solution, as well the term $-{\bf B}_0^{-1}\big({\mathbf x}_0^{{\rm b}}-{\bf x}_0\big)$ for accurate reduced gradients. While the reduced ARRA KKT conditions \eqref{eqn:KKT_ARRA} are similar to the AR conditions  \eqref{eqn:KKT_AR}, the construction of the corresponding reduced order bases is significantly changed.     

\subsection{Estimation of AR optimization error}
%
In this section we briefly justify how the projection errors in the three optimality equations impact the optimal solution. A more rigorous argument can be made using the aposteriori error estimation  methodology developed in \cite{Alexe-PhD,alexe2014space,RBecker_BVexler_2005a,Rao_Sandu_CSL_TR_16_2014}. The full order KKT equations \eqref{eqn:KKT_Full} form a large system of nonlinear equations, written abstractly as
\begin{equation}
\label{eqn:KKT_Full_abstract}
\mathcal{F}(\zeta^a) = 0,
\end{equation}
where $\zeta^a = (\x^a,\la^a) \in \mathbb{R}^{\Ns \times (N+1)} \times \mathbb{R}^{\Ns \times (N+1)}$ is obtained by running the forward model \eqref{eqn:KKT_Full_forward} initiated with the solution $\x_0^a \in \mathbb{R}^{\Ns}$ of problem \eqref{eqn:4DVar_problem} and adjoint model \eqref{eqn:KKT_Full_adjoint} linearized across the trajectory $\x^a$.  The operator in \eqref{eqn:KKT_Full_abstract} is also defined $\mathcal{F} : \mathbb{R}^{\Ns \times (N+1)} \times \mathbb{R}^{\Ns \times (N+1)} \to \mathbb{R}^{2\times \Ns \times (N+1) } $. We assume that the model operators $\M_{i,i+1}: \mathbb{R}^\Ns \to \mathbb{R}^\Ns,~i=0,..,N-1$  and the observation operators $\H_i: \mathbb{R}^\Ns \to \mathbb{R}^{N_{\textnormal{obs}}}~i=0,..,N$ are smooth, that $\mathcal{F}'$ is nonsingular, and that its inverse is uniformly bounded in a sufficiently large neighborhood of the optimal full order solution. Under these assumptions  Newton's method applied to \eqref{eqn:KKT_Full_abstract} converges, leading to an all-at-once procedure to solve the 4D-Var problem.

The AR optimization problem \eqref{eqn:KKT_AR} has an optimal solution  $(\xr^a,\lar^a) \in \mathbb{R}^{k \times (N+1)} \times\mathbb{R}^{k \times (N+1)} $. This solution projected back onto the full space is denoted ${\widehat \zeta}^a  = ({\bf \widehat x}^a,\widehat \la^a)$.
From \eqref{eqn:KKT_Full_abstract}, and by assuming that ${\widehat \zeta}^a$ is located in a neighborhood of $\zeta^a$,  we have that
\begin{eqnarray*}
&&  \mathcal{F}({\widehat \zeta}^a)  =  \mathcal{F}({\widehat \zeta}^a) -\mathcal{F}(\zeta^a)  \approx \mathcal{F}'\left( \zeta^a  \right)\cdot \left( {\widehat \zeta}^a - \zeta^a  \right), \\
&& \left\Vert {\widehat \zeta}^a - \zeta^a  \right\Vert \le \left\Vert \mathcal{F}'\left( \zeta^a  \right)^{-1} \right\Vert \cdot
\left\Vert \mathcal{F}({\widehat \zeta}^a)  \right\Vert.
\end{eqnarray*}
Under the uniform boundedness of the inverse assumption the error in the optimal solution depends on the size of the residual $\mathcal{F}({\widehat \zeta}^a)$, obtained by inserting the projected reduced optimal solution into the full order optimality equations:
\[
\mathcal{F}({\widehat \zeta}^a) = \left[ \begin{array}{c}
\bigg[\left( \Uf\, \Wf^T - \Id\right) \,{\M}_{i,i+1}\left({\bf \widehat x}^a_i\right)\bigg]_{i=0,..,N-1} \\
\Bigl(\Wf\,\Uf^T-\Id\Bigr) {\bf \widehat H}_N^T \mathbf{R}_{N}^{-1} \bigg({\bf y}_N - \H_N({\bf \widehat x}_N)\bigg)\\
\Bigg[\Bigl(\Wf\,\Uf^T-\Id\Bigr) \,\bigg({\bf \widehat M}_{i+1,i}^*{\bf \widehat \lambda}_{i+1} + {\bf \widehat H}_i^T \mathbf{R}_{i}^{-1} \big({\bf y}_i - \H_i({\bf \widehat x}_i)\big)\bigg)\bigg]_{i=N-1,..,0}\\
(\Wf \, \Uf^T - \Id)\,{\bf B}_0^{-1}\big({\mathbf x}^{\rm b}_0-{\bf \widehat x}_0\big)\end{array} \right].
\]
The residual size depends on the projection errors $\left( \Uf\, \Wf^T - \Id \right)$ and $\left( \Wf\,\Uf^T-\Id \right)$, on how accurately is the high-fidelity forward trajectory represented by $\Uf$,  and on how accurately are the full adjoint trajectory and gradient captured by $\Wf$.  By including full adjoint solution and gradient snapshots in the construction of $\Wf$ the residual size is decreased, and so is the error in the reduced optimal solution. This is the ARRA basis construction strategy discussed in the previous section.



\section{4D-Var data assimilation with the shallow water equations}\label{SWE_ROMS_DA}

\subsection{SWE model}\label{subsec:SWE_model}

SWE has proved its capabilities in modeling propagation of Rossby and Kelvin waves in the atmosphere, rivers, lakes and oceans as well as gravity waves in a smaller domain. The alternating direction fully implicit finite difference scheme \citet{Gus1971} was considered in this paper and it is stable for large CFL condition numbers (we tested the stability of the scheme for a CFL condition number equal up to $8.9301$). We refer to \citet{FN1980,NVG1986} for other research work on this topic.

The SWE model using the $\beta$-plane approximation on a rectangular domain is introduced (see \citet{Gus1971})
\begin{equation}\label{eqn:swe-pde}
\frac{\partial w}{\partial t}=A(w)\frac{\partial w}{\partial x}+B(w)\frac{\partial w}{\partial y}+C(y)w,
\quad (x,y) \in [0,L] \times [0,D], \quad t\in(0,t_{\rm f}],
\end{equation}
where $w=(u,v,\phi)^T$ is a vector function, $u,v$ are the velocity components in the $x$ and $y$ directions, respectively, $h$ is the depth of the fluid, $g$ is the acceleration due to gravity, and $\phi = 2\sqrt{gh}$.

The matrices $A$, $B$ and $C$ are assuming the form
\[
A=-\left(\begin{array}{ccc}
           u&0&\phi/2\\
           0&u&0\\
           \phi/2&0&u \end{array}\right), \quad
B=-\left(\begin{array}{ccc}
           v&0&0\\
           0&v&\phi/2\\
           0&\phi/2&v \end{array}\right), \quad
C=\left(\begin{array}{rrr}
           0&f&0\\
           -f&0&0\\
           0&0&0 \end{array}\right),
\]
where $f$ is the Coriolis term
\[
f=\widehat f + \beta(y-D/2),\quad \beta=\frac{\partial f}{\partial y}, \quad \forall\, y,
\]
with $\widehat f$ and $\beta$ constants.

We assume periodic solutions in the $x$ direction for all three state variables
while in the $y$ direction
$$v(x,0,t)=v(x,D,t)=0,\quad x\in[0,L],\quad t\in(0,t_{\rm f}]$$
and Neumann boundary condition are considered for $u$ and $\phi$.

Initially $w(x,y,0)=\psi(x,y),\quad \psi:\mathbb{R}\times\mathbb{R}\rightarrow \mathbb{R},\quad (x,y)\in[0,L]\times[0,D]$.
Now we introduce a mesh of $n = N_x\cdot N_y$ equidistant grid points on $[0,L]\times[0,D]$, with $\Delta x=L/(N_x-1),\quad \Delta y=D/(N_y-1)$. We also discretize the time interval $[0,t_{\rm f}]$ using $N_t$ equally distributed points and $\Delta t=t_{\rm f}/(N_t-1)$. Next we define vectors of the unknown variables of dimension $n$ containing approximate solutions such as

$${\boldsymbol w}(t_N)\approx [w(x_i,y_j,t_N)]_{i=1,2,..,N_x,\quad j=1,2,..,N_y} \in \mathbb{R}^{n},\quad N=1,2,..N_t. $$

The semi-discrete equations of SWE \eqref{eqn:swe-pde} are:
\begin{eqnarray}\label{eqn:swe-sd}
 {\bf u}' & = & -F_{11}({\bf u})-F_{12}(\bm{{\phi}})-F_{13}({\bf u},{\bf v}) + {\bf F}\odot {\bf v}, \\
  {\bf v}' & = & -F_{21}({\bf u})-F_{22}({\bf v}) -F_{23}(\bm{{\phi}}) - {\bf F}\odot {\bf u}, \\
  {\bm{{\phi}}}' & = & -F_{31}({\bf u},\bm{{\phi}})-F_{32}({\bf u},\bm{{\phi}})-F_{33}({\bf v},\bm{{\phi}})-F_{34}({\bf v},\bm{{\phi}}),
\end{eqnarray}
where ${\bf u}'$, ${\bf v}'$, ${\bm{{\phi}}}'$ denote semi-discrete time derivatives, ${\bf F} \in \mathbb{R}^{n}$ stores Coriolis components,  $\odot$ is the component-wise multiplication operator, while the nonlinear terms $F_{i,j}$ are defined as follows:
\begin{equation}\label{eqn:swe-nonlinear_terms}
\begin{split}
  & F_{11},F_{12},F_{21},F_{23}: \mathbb{R}^{n} \rightarrow \mathbb{R}^{n},\quad F_{13},F_{22},F_{3i} : \mathbb{R}^{n} \times \mathbb{R}^{n} \rightarrow \mathbb{R}^{n},\quad i=1,2,3,4,\\
  & \quad F_{11}({\bf u})={\boldsymbol u}\odot A_x{\boldsymbol u},\quad F_{12}({\boldsymbol \phi})=\frac{1}{2}{\boldsymbol \phi}\odot A_x{\boldsymbol\phi},\quad F_{13}({\boldsymbol u},{\boldsymbol v})={\boldsymbol v}\odot A_y{\boldsymbol u},\\
  & F_{21}({\boldsymbol u},{ \boldsymbol v})={\boldsymbol u}\odot A_x{\boldsymbol v},\quad F_{22}({\boldsymbol v})={\boldsymbol v}\odot A_y{\boldsymbol v};\quad F_{23}({\boldsymbol\phi})=\frac{1}{2}{\boldsymbol \phi}\odot A_y{\boldsymbol\phi},\\
  &F_{31}({\boldsymbol u},{\boldsymbol \phi})=\frac{1}{2}{\boldsymbol\phi} \odot A_x {\boldsymbol u},\quad F_{32}({\boldsymbol u},{\boldsymbol \phi})={\boldsymbol u} \odot {A_x\boldsymbol \phi},\\
   & F_{33}({\boldsymbol v},{\boldsymbol \phi})=\frac{1}{2}{\boldsymbol\phi}\odot A_y{\boldsymbol v},\quad F_{34}({\boldsymbol v},{\boldsymbol \phi})={\boldsymbol v} \odot A_y{\boldsymbol\phi},
\end{split}
\end{equation}
where $A_x,A_y\in \mathbb{R}^{n\times n}$ are constant coefficient matrices for discrete first-order and second-order differential operators which incorporate the boundary conditions.

The numerical scheme was implemented in Fortran and uses a sparse matrix environment. For operations with sparse matrices we employed SPARSEKIT library \citet{Saad1994}
and the sparse linear systems obtained during the quasi-Newton iterations were solved using MGMRES library \citet{Barrett94,Kelley95,Saad2003}. Here we did not decouple the model equations as in Stefanescu and Navon \cite{Stefanescu2012} where the Jacobian is either block cyclic tridiagonal or
block tridiagonal. By keeping all discrete equations together the corresponding SWE adjoint model can be solved with the same implicit scheme used for forward model. Moreover, we employed $10$ nonlinear terms in \eqref{eqn:swe-sd} in comparison with only $6$ in \cite{Stefanescu2012,Stefanescu_etal_forwardPOD_2014} to enhance the accuracy of the forward and adjoint POD/DEIM reduced order model solutions. The discrete tangent linear and adjoint models were derived by hand and their accuracy was verified using \citet{Navon_Zou_Derber_Sela_1992} techniques. At the end of the present manuscript we provide an appendix formally describing the tangent linear and adjoint ADI SWE models.

\subsection{SWE 4D-Var data assimilation reduced order systems}\label{subsec:ROMS_SWE}

The SWE model describes the evolution of a hydrostatic homogeneous, incompressible flow. It is derived \cite{VreugdenhilSWE_1995} from depth-integrating the Navier-Stokes equations, in the case where the horizontal length scale is much greater than the vertical length scale, i.e. the effects of vertical shear of the horizontal velocity are negligible. Consequently we obtain a new set of momentum and continuity equations where the pressure variable is replaced by the height of the surface topography \cite{VreugdenhilSWE_1995}.

To the best of our knowledge, the reduced order models with a pressure component use two strategies. In the decoupled approach \cite{Caiazzo2014,Noack_Papas_Monkewitz_2005}, the velocity and pressure snapshots are considered separately. In the coupled approach \cite{Bergmann2009,Weller_etal_2010}, each snapshot and POD mode have both velocity and the corresponding pressure component. In our manuscript we followed the decoupled approach to build our reduced order models. Moreover, we also decouple the velocity components as in \cite{Stefanescu2012}, thus using separate snapshots for zonal and meridional velocities, respectively.

In our current research we construct a POD basis that captures the kinetic energy of the shallow water equation model and it is not explicitly designed to represent the Rossby and Kelvin waves. Moreover, when we truncate the spectrum and define the POD basis we neglect the lower energetic modes which may describe some of the features of these waves. However we believe that long waves such as Rossby and Kelvin are less affected since they have a more energetic content in comparison with the short gravity waves. Capturing the dynamics of short gravity waves would require additional POD modes and a dense collection of snapshots.

To reduce the computational cost of 4D-Var SWE data assimilation we propose three different POD based reduced order 4D-Var SWE systems depending on standard POD, tensorial POD and standard POD/DEIM reduced order methods discussed in \citet{Stefanescu_etal_forwardPOD_2014}. These three ROMs treat the nonlinear terms in a different manner (see equations \eqref{eqn:POD_standard_nonlinearity_general},\eqref{eqn:POD_tensorial_nonlinearity_general},\eqref{eqn:POD_DEIM_nonlinearity_general}) while reduced Jacobian computation is done analytically for all three approaches via tensorial calculus.

Tensorial POD and POD/DEIM nonlinear treatment make use of an efficient decoupling of the full spatial variables from reduced variables allowing most of the calculations to be performed during the off-line stage which is not valid in the case of standard POD. Computation of tensors such as $T$ in \eqref{eqn:POD_tensorial_nonlinearity_general} is required by all three ROMs in the off-line stage since the analytic reduced Jacobian on-line calculations depend on them. Clearly, standard POD is optimized since usually the reduced Jacobians are obtained by projecting the full Jacobians onto POD spaces during on-line stage so two computational costly operations are avoided.

Since all ROMs are using exact reduced Jacobians the corresponding adjoint models have the same computational complexity. POD/DEIM backward in time reduced model relies on approximate tensors calculated with the algorithm introduced in \cite[p.7]{Stefanescu_etal_forwardPOD_2014} while the POD adjoint models make use of exact tensors. For POD/DEIM approach this leads to slightly different reduced Jacobians in comparison with the ones calculated by standard and tensorial POD leading to different adjoint models. Still, nonlinear POD/DEIM approximation \eqref{eqn:POD_DEIM_nonlinearity_general} is accurate and agrees with standard POD \eqref{eqn:POD_standard_nonlinearity_general} and tensorial POD representations \eqref{eqn:POD_tensorial_nonlinearity_general}. In comparison with standard POD, tensorial POD moves some expensive computations from the on-line stage to the off-line stage. While these pre-computations provide a faster on-line reduced order model, the two adjoint models solutions are similar.

We are now ready to describe the ARRA algorithms corresponding to each reduced data assimilation systems using Galerkin projection. For POD/DEIM SWE data assimilation algorithm we describe only the off-line part since the on-line and decisional stages are generically presented in the tensorial and standard POD reduced data assimilation algorithms.

\begin{algorithm}
 \begin{algorithmic}[1]

 \Algphase{Off-line stage}
 \State Generate background state ${\bf u},~{\bf v}$ and $\bm{{\phi}}$. \label{alg:backround_state}
 \State Solve full forward ADI SWE model to generate state variables snapshots. \label{alg:solve_forward}
 \State Solve full adjoint ADI SWE model to generate adjoint variables snapshots. \label{alg:solve_adjoint}
 \State For each state variable compute a POD basis using snapshots describing dynamics of the forward and  its corresponding adjoint trajectories.
 \State Compute tensors as $T$ in \eqref{eqn:POD_tensorial_nonlinearity_general} required for reduced Jacobian calculations. Calculate other POD coefficients corresponding to linear terms.
 \Algphase{On-line stage  - Minimize reduced cost functional $\J^\textsc{pod}$ \eqref{eqn:reduced_cost_func}}
 \State Solve forward reduced order model \eqref{eqn:KKT_AR_forward}.
 \State Solve adjoint reduced order model \eqref{eqn:KKT_AR_adjoint}.
 \State Compute reduced gradient \eqref{eqn:KKT_AR_gradient}.
 \Algphase{Decisional stage}
\State Reconstruct the conditions in full space from the suboptimal reduced initial condition (the output of the on-line stage), and perform steps  \ref{alg:solve_forward} and \ref{alg:solve_adjoint} of the off-line stage. Using full forward information evaluate  function \eqref{eqn:cost_function_4D_Var} and its gradient. If $\|\J\|>\varepsilon_3$ and $\|\nabla \J\| > \varepsilon_4$, then continue the off-line stage from step \ref{alg:solve_adjoint}, otherwise STOP.
 \end{algorithmic}
 \caption{Standard and Tensorial POD SWE DA systems}
 \label{alg_standard_tensorial_POD}
\end{algorithm}

\begin{algorithm}
 \begin{algorithmic}[1]

 \Algphase{Off-line stage}
 \State Generate background state ${\bf u},~{\bf v}$ and $\bm{{\phi}}$.
 \State Solve full forward ADI SWE model to generate nonlinear terms and state variables snapshots.
 \State Solve full adjoint ADI SWE model to generate adjoint variables snapshots.
 \State For each state variable compute a POD basis using snapshots describing dynamics of the forward and  its corresponding adjoint trajectories. For each nonlinear term compute a POD basis using snapshots from the forward model.
 \State Compute discrete empirical interpolation points for each nonlinear term.
 \State Calculate other linear POD coefficients and POD/DEIM coefficients as $W^TV(P^TV)^{-1}$ in \eqref{eqn:POD_DEIM_nonlinearity_general}.
 \State Compute tensors such as $T$ using algorithm described in \cite[p.7]{Stefanescu_etal_forwardPOD_2014} required for reduced Jacobian calculations.
 \end{algorithmic}
 \caption{POD/DEIM SWE DA systems}
 \label{alg_POD_DEIM}
\end{algorithm}

The on-line stages of all reduced data assimilation systems correspond to minimization of the cost function $\J^\textsc{pod}$  and include steps $6-8$ of the algorithm \ref{alg_standard_tensorial_POD}. The optimization is performed on a reduced POD manifold. Thus, the on-line stage is also referred as inner phase or reduced minimization cycle. The stoping criteria are
\begin{equation}\label{eqn:Red_min_stopping_criteria}
\|\nabla \J^\textsc{pod}\|\le \varepsilon_1, \quad \| \J^\textsc{pod}_{(i+1)}-\J^\textsc{pod}_{(i)}\|\le \varepsilon_2, \quad \textnormal{No of function evaluations} \le \textnormal{MXFUN},
\end{equation}
where $J^\textsc{pod}_{(i)}$ is the cost function evaluation at inner iteration (i) and $\textnormal{MXFUN}$ is the number of function evaluations allowed during one reduced minimization cycle.

Initially, the first guess for the on-line stage is given by the projection of the background state onto the current POD space while for the later inner-optimization cycles the initial guess is computed by projecting the current full initial conditions onto the updated POD subspace.

The off-line stage is called an outer iteration even if no minimization is performed during this phase. During this phase the reduced bases are updated according to the current control  \cite{Afanasiev_Hinze_2001,Ravindran_2002,Kunisch_2008,Bergmann_Cordier_2005,Ravindran_2000,Yue_Meerbergen_2013,Zahr_Farhat_2014,Zahr_Amsallem_Farhat_2013}. It also includes a general stopping criterion for reduced data assimilation system $\|\J\|\le \varepsilon_3$ where $\J$ is computed using same formulation as the full data assimilation cost function \eqref{eqn:cost_function_4D_Var}. The gradient based criterion $\|\nabla \J\| \leq \varepsilon_4$  stops the optimization process only if $\varepsilon_4$ has the same order of magnitude as $ \varepsilon_\textnormal{f}$, $ \varepsilon_\textnormal{a}$ and $ \varepsilon_\textnormal{g}$ in definition \ref{def:accurate_reduced_feasible_KKT}. The more accurate the reduced KKT conditions are the more likely the projected reduced optimization solution will converge to the minimum of the cost function \eqref{eqn:cost_function_4D_Var}.


\section{Numerical results }\label{sec:Numerical_resuls}

This section is divided in two parts. The first one focuses on POD basis construction strategies and  tensorial POD SWE 4D-Var is used for conclusive numerical experiments while the second part measures the computational performances of the three proposed reduced order SWE data assimilation systems.

For all tests we derived the initial conditions from the initial height condition No. 1 of Grammeltvedt \cite{Gram1969} i.e.

\begin{equation*}
\hspace{-10mm}h(x,y)=H_0+H_1+\tanh\biggl(9\frac{D/2-y}{2D}\biggr)+H_2\textnormal{sech}^2\biggl(9\frac{D/2-y}{2D}\biggr)\sin\biggl(\frac{2\pi x }{L}\biggr),
\end{equation*}
$$0\leq x\leq L,~0\leq y\leq D.$$

The initial velocity fields were derived from the initial height field using the geostrophic relationship
$$u = \biggl(\frac{-g}{f}\biggr)\frac{\partial h}{\partial y},~v = \biggl(\frac{g}{f}\biggr)\frac{\partial h}{\partial x}. $$

In our numerical experiments, we apply $10\%$ uniform perturbations on the above initial conditions and generate twin-experiment observations at every grid space point location and every time step. We use the following constants $ L=6000km,~ D=4400km,~ \widehat f=10^{-4}s^{-1},~ \beta=1.5\cdot10^{-11}s^{-1}m^{-1},~ g=10 m s^{-2},~ H_0=2000m,~ H_1=220m,~ H_2=133m.$

The background state is computed using a $5\%$ uniform perturbation of the initial conditions. The background and observation error covariance matrices are taken to be identity matrices. The length of the assimilation window is selected to be $3h$. The implicit scheme allowed us to integrate in time using a larger time step and select $N_t=91$ time steps.

The Broyden-Fletcher-Goldfarb-Shanno (BFGS) optimization method option contained in the CONMIN software (\citet{Shanno_Phua1980}) is employed  for high-fidelity full SWE 4-D VAR as well as all variants of reduced SWE 4D-Var data assimilation systems. BFGS uses a line search method which is globally convergent in the sense that $lim_{k \to \infty} \Vert \nabla \J^{(k)}\Vert = 0$ and utilizes approximate Hessians to include convergence to a local minimum.

In all our reduced data assimilation experiments we use $\varepsilon_1 = 10^{-14}$ and $\varepsilon_2 = 10^{-5}$ which are important for the reduced minimization
(on-line stage) stopping criteria defined in \eqref{eqn:Red_min_stopping_criteria}. The optimization algorithms \ref{alg_standard_tensorial_POD} and
\ref{alg_POD_DEIM} stop if $\|\J\| \le \varepsilon_3$,  $\|\nabla \J\| \le \varepsilon_4$ or more than $n_{out}$ outer loop iterations are performed. 
For the full data assimilation experiment we used only $\|\nabla \J\| \le 10^{-14}$ and $\|\J\| \le \varepsilon_3$. 
Results in subsections \ref{subsec:NR_choice_POD_basis}, \ref{subsubsec:NR_reduced_POD_DEIM_DA_system}, \ref{subsubsec:NR_reduced_ad_tl_POD_DEIM_SWE models_tests}, and \ref{subsubsec:NR_reduced_HYBRID_POD_DEIM_DA_system}, are obtained with $\varepsilon_3=10^{-15}$. The values of $\varepsilon_3$ and MXFUN differ for numerical tests presented in the other sections. We select $\varepsilon_4 = 10^{-5}$ for all experiments.

All reduced data assimilation systems employ Galerkin projection to construct their low-rank approximation POD models.
In the ARRA procedure the initial training set for POD basis is obtained by running the high-fidelity forward model using the background state and full adjoint model with the final time condition described by the first equation in \eqref{eqn:KKT_Full_adjoint}. During the reduced optimization process, the updated POD basis includes information from the full forward and adjoint trajectories initiated with the current sub-optimal state (the analysis given by the on-line stage in Algorithm \ref{alg_standard_tensorial_POD} and corresponding forcing term (first equation in \eqref{eqn:KKT_Full_adjoint}) and ${\mathbf x}^{\rm b}_0$. Adding the background state to the snapshots matrix ensures the accuracy of the reduced gradient since here ${\bf B}_0 = \Id$ and a snapshot of ${\bf x}_0$ is already included.

Tables \ref{tab:experiments1} and \ref{tab:experiments2} describe the reduced order 4D-Var SWE data assimilation experiments along with the models and parameters used, the main conclusions drawn from the numerical investigations and the sections on which these conclusions are based on. We recall that $n,k,m,$ MXFUN and $n_{out}$ represent the number of space points, the dimension of POD basis, the number of DEIM points, the number of function evaluations allowed during the reduced minimization cycle and the maximum number of outer iterations allowed by algorithms \ref{alg_standard_tensorial_POD} and \ref{alg_POD_DEIM}, respectively.

\begingroup
\begin{table}[h]
\centerline{
\scalebox{0.68}{
\begin{tabular}{|c|c|c|c|c|c|c|c|}\hline \hline
Model & n & k & m  & MXFUN  & $n_{out}$ & Conclusion & Section \\ \hline
 &  &  &  &  & & adjoint and forward information must be &  \\
TPOD & $31 \times 23$ & 50 & - & 25 & 13 & included into POD basis construction.  & 6.1 \\
&  &  &  & & & for accurate sub-optimal solutions & \\ \hline
POD/DEIM & $31 \times 23$ & 50 & 50,180 & 100 & 10 & m must be selected close to n  & 6.2.1 \\
 & $17 \times 13$ & 50 & 50,135 & 20 & 10 & for accurate sub-optimal solutions.  & 6.2.1 \\
 &  & & 165 &  &  &   &  \\ \hline
 &  &  &  &  &  & adjoint and tangent linear models &  \\
POD/DEIM  & $17 \times 13$ & 50 & 50 & - & - & are accurate with respect to   & 6.2.2 \\
 &  &  &  &  &  & finite difference approximations. &  \\ \hline
 &  &  &  &  &  & replacing the POD/DEIM nonlinear terms &  \\
Hybrid POD/DEIM  & $17 \times 13$ & 50 & 50 & 20 & 10 & involving height with tensorial terms & 6.2.3 \\
 &  &  &  &  &  & accurate sub-optimal solutions &  \\
 &  &  &  &  &  & are obtained even for small no of DEIM points. &  \\ \hline
 &  & 20,30, &  &  &  & for $k\geq 50$ the sub-optimal  &  \\
Hybrid POD/DEIM  & $17 \times 13$ & 50, & 50 & 20 & 10 & solutions are as accurate as & 6.2.3 \\
 &  & 70,90 &  &  &  & full order optimal solutions. &  \\ \hline
&  &  & 10,20, &  &  & for $m\geq 50$ the sub-optimal  &  \\
Hybrid POD/DEIM  & $17 \times 13$ & 50, & 30 & 20 & 10 & solutions are as accurate as & 6.2.3 \\
 &  &  & 50,100 &  &  & full order optimal solutions. &  \\ \hline
\end{tabular}}}
\caption{\label{tab:experiments1}The main reduced order 4D-Var SWE data assimilation experiments for $\varepsilon_1=10^{-14},\varepsilon_2=10^{-5}$, $\varepsilon_3=10^{-15}$ and $\varepsilon_4 = 10^{-5}$.}
\end{table} %
\endgroup%

\begingroup
\begin{table}[h]
\centerline{
\scalebox{0.7}{
\begin{tabular}{|c|c|c|c|c|c|c|c|}\hline \hline
Model & n & k & m  & MXFUN  & $\varepsilon_3$ & Conclusion & Section \\ \hline
POD &  &  &  &  & & hybrid POD/DEIM 4D-Var data &  \\
TPOD & $61 \times 45$ & 50 & 50,120 & 10 & $10^{-7}$ & assimilation system is the fastest  & \ref{subsec:NR_reduced_DA_systems} \\
Hybrid POD/DEIM&  &  &  & & & ROM optimization approach. & \\ \hline
POD &  &  &  &  & & hybrid POD/DEIM 4D-Var data &  \\
TPOD & $151 \times 111$ & 50 & 30,50 & 15 & $10^{-1}$ & assimilation system is faster  & \ref{subsec:NR_reduced_DA_systems} \\
Hybrid POD/DEIM&  &  &  & & & with $\approx$ 4774s (by 8.86 times) & \\
FULL &  &  &  & & & than full SWE DA system. & \\ \hline
ALL & $31\times 23$ &  &  &  & $10^{-9}, 10^{-14} $ & CPU time speedup rates are directly &  \\
ALL & $61 \times 45$ &  &  &  & $10^{-4}, 10^{-7}$ & proportional to the increase of   & \\
ALL & $101\times 71$ & 30,50 & 50,120 & 15 & $10^{-1}, 10^{-4} $ & the full space resolution. & \ref{subsec:NR_reduced_DA_systems} \\
ALL & $121\times 89$ &  &  &  & $5,5\cdot 10^{-3} $ & hybrid POD/DEIM 4D-Var DA is the &  \\
ALL & $151\times 111$ &  &  &  & $10^{3},10^{-1} $ & fastest approach as n is increased.  &  \\ \hline
 &  &  &  &  & & The ROM DA systems are  &  \\
ALL& $151 \times 111$ & 50 & 30 & 15 & $10^{-1}$ & faster if $n_{out}$ is maintained  & \ref{subsec:NR_reduced_DA_systems} \\
&  &  &  & & & as low as possible. & \\ \hline
ALL & $31\times 23$ &  &  &  & $10^{-9}, 10^{-14} $ & Hybrid POD/DEIM 4D-Var DA  &  \\
ALL & $61 \times 45$ &  &  &  & $10^{-4}, 10^{-7}$ & system is the fastest approach    & \\
ALL & $101\times 71$ & 30,50 & 50,120 & 10,15,20 & $10^{-1}, 10^{-4} $ & among the ROM systems.   & \ref{subsec:NR_reduced_DA_systems} \\
ALL & $121\times 89$ &  &  &  & $5,5\cdot 10^{-3} $ & MXFUN should be increased with  &  \\
ALL & $151\times 111$ &  &  &  & $10^{3},10^{-1} $ & the decrease of $\varepsilon_3$ and increase of k. &  \\ \hline
ALL & $31\times 23$ &  &  &  &  &   &  \\
ALL & $61 \times 45$ &  &  &  &  & In terms of accuracy all the  & \\
ALL & $101\times 71$ & 30,50 & 50 & 15 & $10^{-15}$ & various ROM DA systems   & \ref{subsec:Accuracy_NR_reduced_DA_systems} \\
ALL & $121\times 89$ &  &  &  &  & deliver similar results for  &  \\
ALL & $151\times 111$ &  &  &  &  & the same $k$. &  \\ \hline
\end{tabular}}}
\caption{\label{tab:experiments2}The main 4D-Var SWE data assimilation results for $\varepsilon_1=10^{-14}$, $\varepsilon_2=10^{-5}$ and $\varepsilon_4=10^{-5}$. ALL refers to all reduced and full models and $n_{out}$ is set to $10$ for most of the numerical experiments except for the ones described in subsection \ref{subsec:Accuracy_NR_reduced_DA_systems}, where $n_{out} = 20$.}
\end{table} %
\endgroup%


\subsection{Choice of POD basis}\label{subsec:NR_choice_POD_basis}

The tensorial POD reduced SWE data assimilation is selected to test which of the POD basis snapshots selection strategies perform better with respect to suboptimal solution accuracy and cost functional decrease. The ``adjoint of reduced forward model'' approach is compared with ``adjoint of reduced forward model + reduced order adjoint model'' method discussed in section \ref{sec:POD_bases}.

\begin{figure}[h]
  \centering
  \subfigure[Forward model snapshots only] {\includegraphics[scale=0.4]{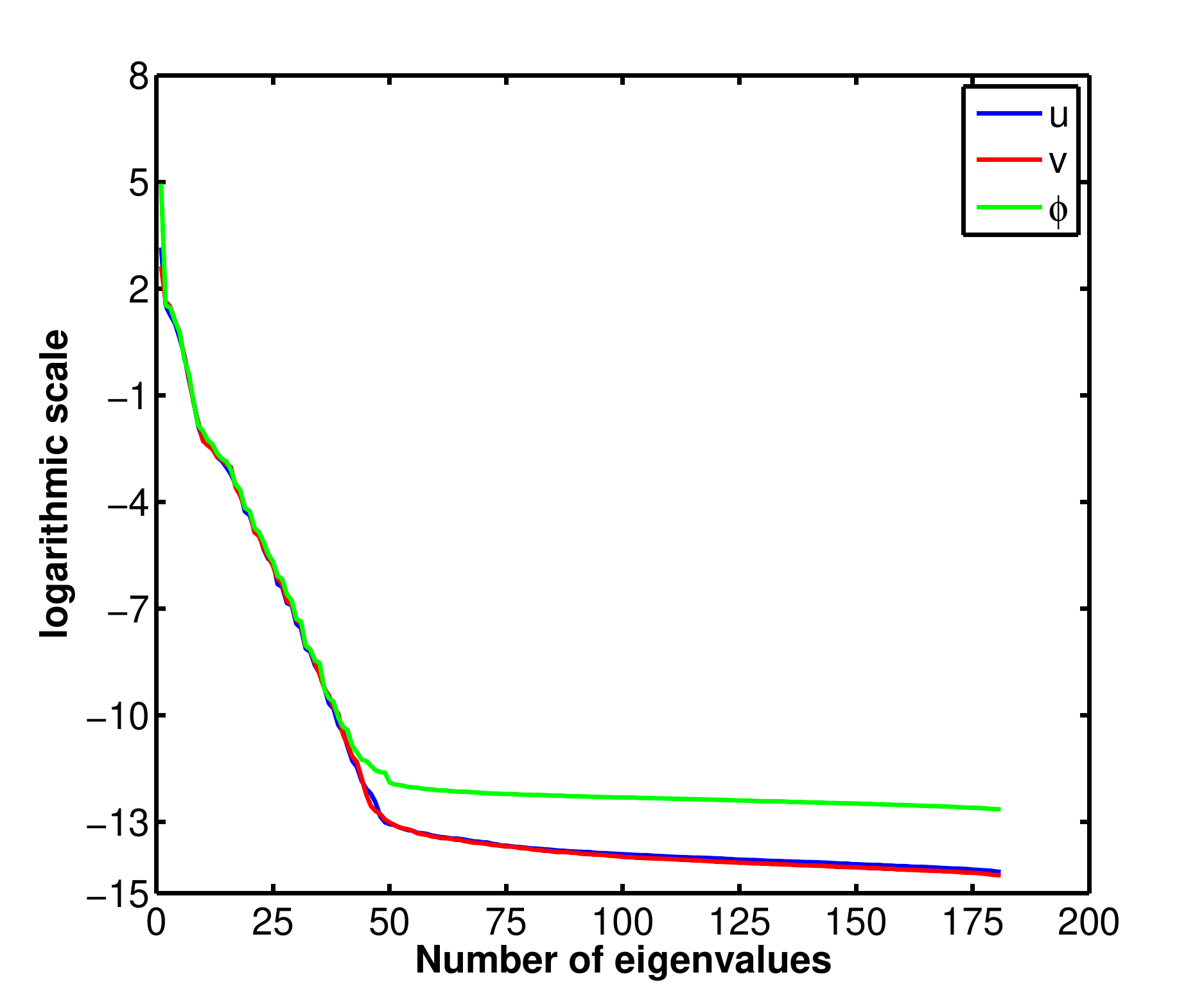}}
  \subfigure[Forward and adjoint models snapshots]{\includegraphics[scale=0.4]{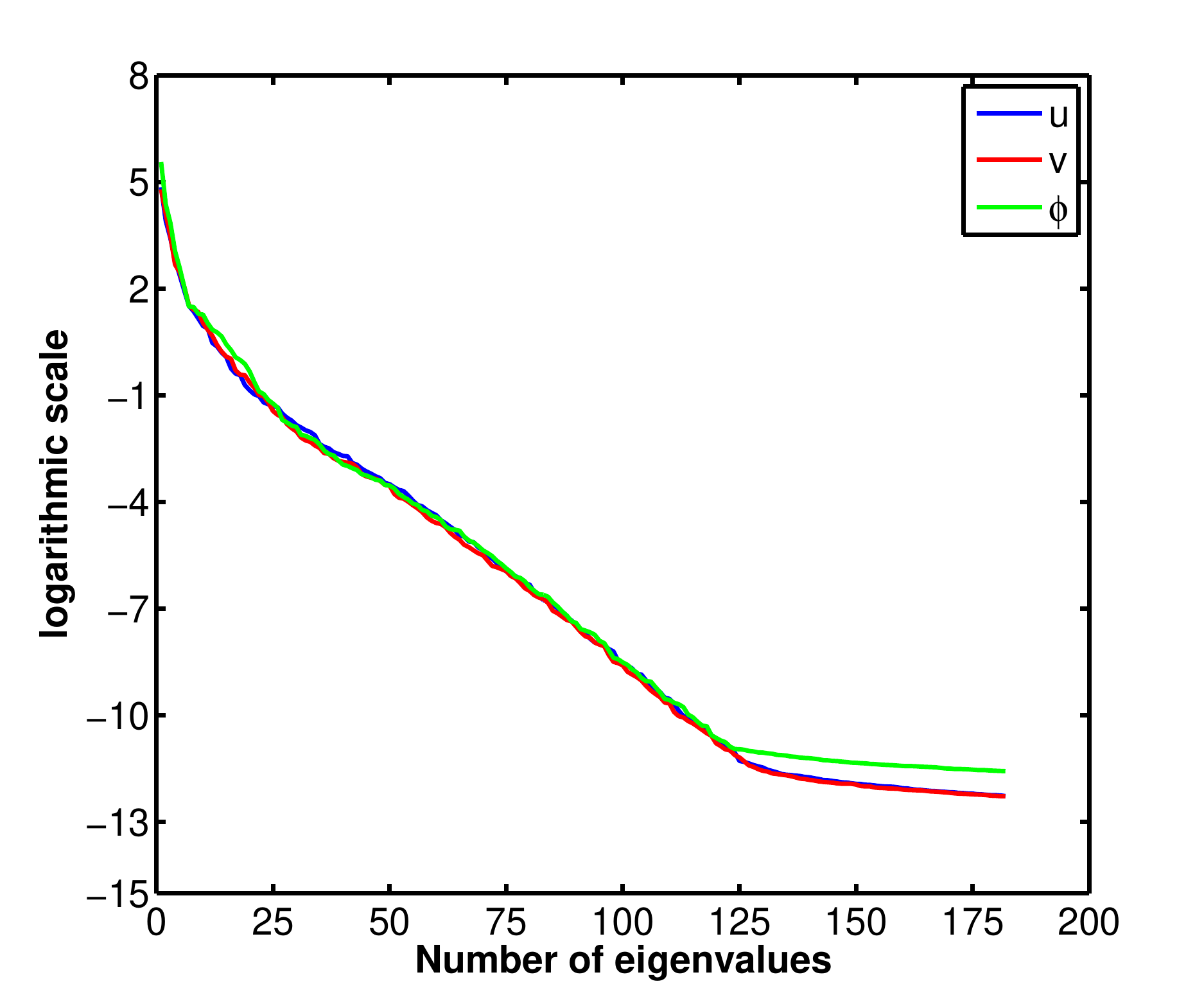}}
\caption{The decay around of the singular values of the snapshots solutions for $u, v, \phi$ for $\Delta t = 960s$ and integration time window of $3h$.
\label{Fig::Eigenvalues_AR_RA}}
\end{figure}

For AR approach there is no need for implementing the full adjoint SWE model since the POD basis relies only on forward trajectories snapshots. Consequently its off-line stage will be computationally cheaper in comparison with the similar stage of ARRA approach where adjoint model snapshots are considered inside the POD basis.  For this experiment we select $31\times 23$ mesh points and use $50$ POD basis functions. MXFUN is set to $25$. The singular values spectrums are depicted in Figure \ref{Fig::Eigenvalues_AR_RA}.

Forward snapshots consist in ``predictor'' and ``corrector'' state variables solutions ${\boldsymbol w}^{t_{i+1/2}}$, ${\boldsymbol w}^{t_i}$, $i=0,1,..,t_f-1$ and ${\boldsymbol w}^{t_f}$ obtained by solving the two steps forward ADI SWE model. The adjoint snapshots include the ``predictor'' and ``corrector'' adjoint solutions ${\bf \lambda}_{\bf w}^{t_{i-1/2}},~ {\bf \lambda}_{\bf w}^{t_{i}},~ i=1,2,.,t_f$ and ${\bf \lambda}_{\bf w}^{t_{0}}$ as well as other two additional intermediary solutions computed by the full adjoint model. An appendix is included providing details about the ADI SWE forward and adjoint models equations.


Next we compute the POD truncation relative errors for all three variables of reduced SWE models using the following norm at the initial time $t_0$
\begin{equation}\label{eqn::first_norm}
E_{\bar{\bf w}} = \frac{\Vert{\bar{\boldsymbol  w}}^{t_0}-{\bar{\boldsymbol  w}}^{t_0}_\textsc{rom}\Vert_2}{\Vert{\bar { \boldsymbol  w}}^{t_0}\Vert_2}, \quad E_{{\bar{\lambda}}_{\bf w}} = \frac{\Vert{{\bar{\lambda}}_{{\bf w}}}^{t_0}-{{\bar{\lambda}}_{{\bf w}^{ROM}}^{t_0}}\Vert_2}{\Vert{{\bar{\lambda}}_{{\bf w}}}^{t_0}\Vert_2},
\end{equation}
where ${\bar{\boldsymbol w}}$ and ${{\bar{\lambda}}_{\bf w}}$ are general variables and span the sets of SWE state variables $\{u,v,\phi\}$ and adjoint variables $\{\lambda_u,\lambda_v,\lambda_{\phi}\}$. ${\bar{\boldsymbol  w}}_\textsc{rom}$ and ${{\bar{\lambda}}_{{\bf w}^{ROM}}}$ are the full solutions reconstructed from the reduced order variables. $\|\cdot\|_2$ defines an Euclidian norm. The results are given in table \ref{table_AR_RA_POD_error1}.

\begingroup
\begin{table}[h]
\centerline{
\scalebox{1.0}{
\begin{tabular}{|c|c|c|}\hline
 & AR & ARRA \\ \hline
$E_u$ & $9.91e$-$16$  & $5.16e$-$7$ \\ \hline
$E_v$ & $1.48e$-$15$  & $1e$-$6$ \\ \hline
$E_{\phi}$ & $9.93e$-$16$  & $6.78e$-$9$ \\ \hline
\end{tabular}}
\scalebox{1.0}{
\begin{tabular}{|c|c|c|}\hline
 & AR & ARRA \\ \hline
$E_{\lambda_u}$ & $0.969$  & $5.94e$-$9$ \\ \hline
$E_{\lambda_v}$ & $0.926$  & $5.18e$-$9$ \\ \hline
$E_{\lambda_\phi}$ & $0.220$ & $1.65e$-$9$ \\ \hline
\end{tabular}}}
\caption{\label{table_AR_RA_POD_error1}Relative errors of forward (left) and adjoint (right) tensorial POD SWE initial conditions using AR and ARRA approaches.}
\end{table} %
\endgroup%

We did not scale the input snapshots. This approach seems to favor the reduced adjoint model with more accurate solutions than the reduced forward model.

\begin{figure}[ht]
\centering
\includegraphics[scale=0.48]{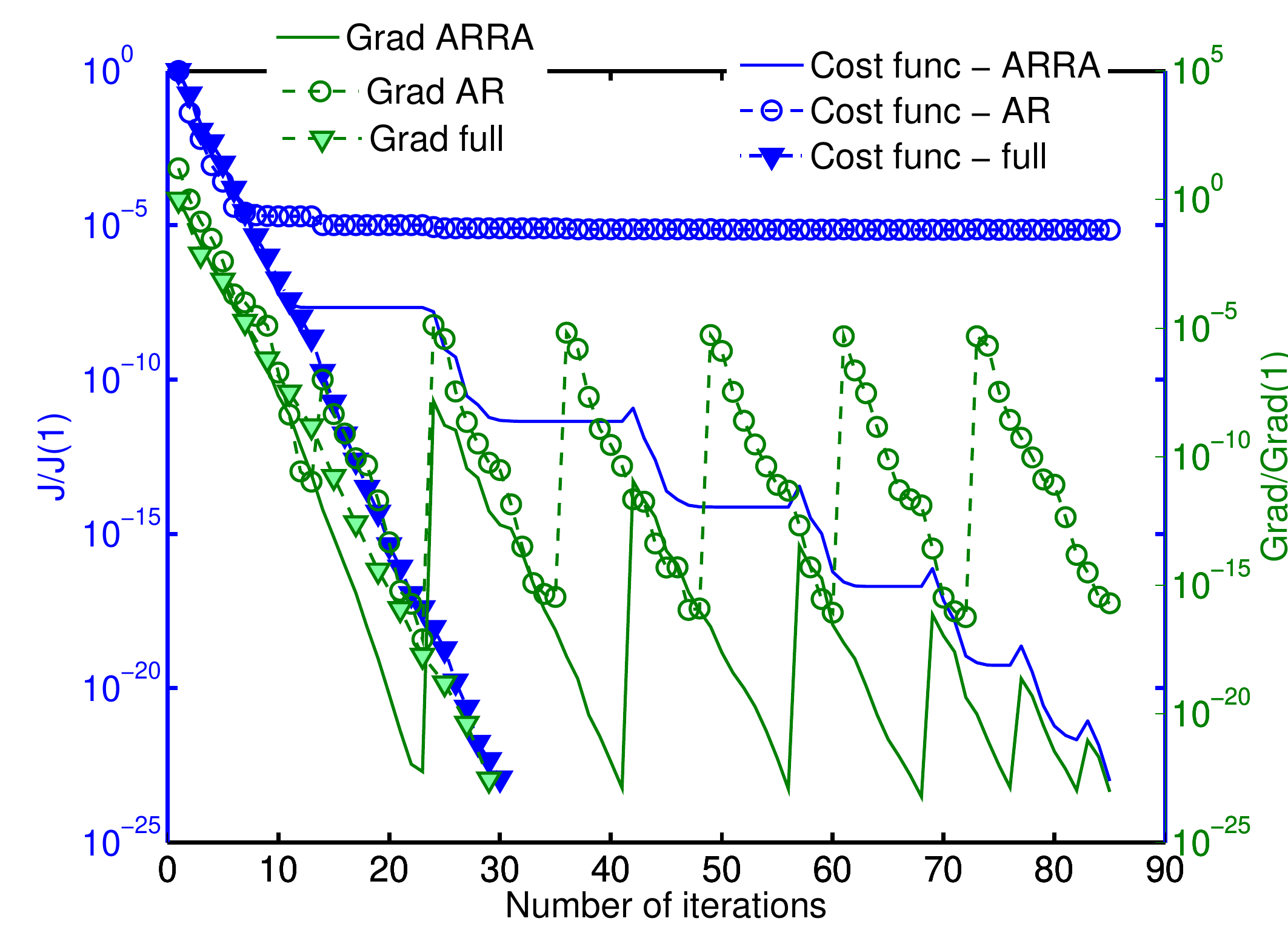}
\caption{Tensorial POD/4DVAR ADI 2D Shallow water equations -- Evolution of cost function and gradient norm as a function of the number of minimization iterations. The information from the adjoint equations has to be incorporated into POD basis.  \label{Fig::Choice_of_POD_basis_in_data_assim}}
\end{figure}

Even if AR reduced data assimilation does not require a full adjoint model we chose to display the reduced adjoint time averaged relative error as a measure of increased probability that the output local minimum is far away from the local minimum computed with the high-fidelity configuration. Figure \ref{Fig::Choice_of_POD_basis_in_data_assim} depicts the minimization performances of the tensorial POD SWE 4D-Var systems using different set of snapshots in comparison with the output of the full space ADI SWE 4D-Var system. The cost function and gradient values are normalized by dividing them with their initial values.

Clearly the best POD basis construction strategy is ``adjoint of reduced forward model + reduced order adjoint model'' approach since the corresponding tensorial POD SWE 4D-Var achieved a cost function reduction close to the one obtained by the high-fidelity ADI SWE 4D-Var. One may notice that $6$ POD bases recalculations were required to achieve the suboptimal solution since $6$ plateaus regions followed by $6$ peaks are visible on the cost function curve. If only forward trajectory snapshots are utilized for POD basis construction the cost function decay is modest achieving only $5$ orders of magnitude decrease despite of carrying out $13$ POD bases updates. This underlines that the ``adjoint of reduced forward model'' approach is not able to represent well the controlled dynamics in its reduced manifold leading to an suboptimal cost function value of $0.48e$+$02$. In the case of ``adjoint of reduced forward model + reduced order adjoint model'' strategy the suboptimal cost function value is $0.48e$-$14$ while the optimal cost function calculated by the high fidelity ADI SWE 4D-Var system is $0.65e$-$16$. Two additional measures of the minimization performances are presented in Table \ref{table_AR_RA_POD_error2} where the relative errors of tensorial POD suboptimal solutions with respect to observations and optimal solution are displayed. The data are generated using
\begin{equation}\label{eqn::second_norm}
E_{{\bar{\boldsymbol  w}}}^* = \frac{\Vert{\bar{\boldsymbol  w}}^{*}-{\bar{\boldsymbol  w}}^{*}_\textsc{rom}\Vert_2}{\Vert{\bar{\boldsymbol  w}}^{*}\Vert_2}, \quad
E_{{\bar{\boldsymbol  w}}}^o = \frac{\Vert{\bar{\boldsymbol  w}}^{\rm obs}-{\bar{\boldsymbol  w}}^{*}_\textsc{rom}\Vert_2}{\Vert{\bar{\boldsymbol  w}}^{\rm obs}\Vert_2},
\end{equation}
${{\bar{\boldsymbol  w}}}^*$ being the optimal solution provided by the full 4D-Var data assimilation system and spans the set $\{u^*,v^*,\phi^*\}$ , ${\bar{\boldsymbol  w}}^{\rm obs}$ is the observation vector that spans the set $\{u^{\rm obs},v^{\rm obs},\phi^{\rm obs}\}$ and ${\bar{\boldsymbol  w}^{*}}_\textsc{rom}$ is the sub-optimal solution proposed by the reduced 4D-Var systems and can take each of the following $\{{u^\textsc{rom}}^*,{v^\textsc{rom}}^*,{\phi^\textsc{rom}}^*\}$.
\begingroup
\begin{table}[h]
\centerline{
\scalebox{1.0}{
\begin{tabular}{|c|c|c|}\hline
 & AR & ARRA \\ \hline
$E_{u}^*$ & $1.57e$-$2$  &  $5.19e$-$11$ \\ \hline
$E_{v}^*$ & $1.81e$-$2$  &  $6.77e$-$11$ \\ \hline
$E_{\phi}*$ & $1.6e$-$2$ &  $5.96e$-$11$ \\ \hline
\end{tabular}}
\scalebox{1.0}{
\begin{tabular}{|c|c|c|}\hline
& AR & ARRA \\ \hline
$E_{u}^{o}$ & $6.69e$-$5$  & $2.39e$-$11$ \\ \hline
$E_{v}^{o}$ & $2.29e$-$4$  & $8.71e$-$11$ \\ \hline
$E_{\phi}^{o}$ & $1.08e$-$6$ & $3.7e$-$13$ \\ \hline
\end{tabular}}}
\caption{\label{table_AR_RA_POD_error2}Relative errors of suboptimal solutions of reduced tensorial POD SWE systems using different snapshot sets and optimal solution (left) and observations (right). $^*$ denotes errors with respect to the optimal solution obtained using high-fidelity ADI SWE 4D-Var SWE system while $^{o}$ characterizes errors with respect to the observations.}
\end{table} %
\endgroup%

We conclude that information from the full forward and adjoint solutions, as well as from the background term, must be included in the snapshots set used to derive  the basis for Galerkin POD reduced order models. The smaller the error bounds $\varepsilon_\textnormal{f},~\varepsilon_{\textnormal{a}},~\varepsilon_{\textnormal{g}}$ in \eqref{eqn:accurate_feasible_kkt} are, the more accurate sub-optimal solutions are generated by the reduced order data assimilation systems. Next subsection includes experiments using only the ARRA strategy.

\subsection{Reduced order POD based SWE 4D-Var data assimilation systems} \label{subsec:NR_reduced_DA_systems}

This subsection is devoted to numerical experiments of the reduced SWE 4D-Var data assimilation systems introduced in subsection \ref{subsec:ROMS_SWE} using POD based models and discrete empirical interpolation method. In the on-line stage tensorial POD and POD/DEIM SWE forward models were shown to be faster than standard POD SWE forward model being $76\times$ and $450\times$ more efficient for more than $300,000$ variables (\citet{Stefanescu_etal_forwardPOD_2014}). Moreover, a tensorial based algorithm was developed in \cite{Stefanescu_etal_forwardPOD_2014} allowing the POD/DEIM SWE model to compute its off-line stage faster than the standard and tensorial POD approaches despite additional SVD calculations and other reduced coefficients calculations.

Consequently, one can assume that POD/DEIM SWE 4D-Var system would deliver suboptimal solutions faster than the other standard and tensorial POD data assimilation systems. The reduced Jacobians needed for solving the forward and adjoint reduced models are computed analytically for all three approaches. Since the derivatives computational complexity does not depend on full space dimension the corresponding adjoint models have similar CPU time costs. Thus, most of the CPU time differences will arise from the on-line stage of the reduced forward models and their off-line requirements.


\subsubsection{POD/DEIM ADI SWE 4D-Var data assimilation system} \label{subsubsec:NR_reduced_POD_DEIM_DA_system}

Using nonlinear POD/DEIM approximation introduced in \eqref{eqn:POD_DEIM_nonlinearity_general} we implement the reduced forward POD/DEIM ADI SWE obtained by projecting the ADI SWE equations onto the POD subspace. Then the reduced optimality conditions \eqref{eqn:KKT_AR} are computed. For this choice of POD basis, the reduced POD/DEIM ADI SWE adjoint model is just the projection of the full adjoint model \eqref{eqn:KKT_Full_adjoint} onto the reduced basis, and in consequence, they have similar algebraic structures requiring two different linear algebraic systems of equations to be solved at each time level (the algebraic form of ADI SWE adjoint model is given in appendix, see equations \eqref{eqn:AD1}-\eqref{eqn:AD2}).

\begin{figure}[h]
  \centering
  \subfigure[Number of mesh points $31\times 23$] {\includegraphics[scale=0.36]{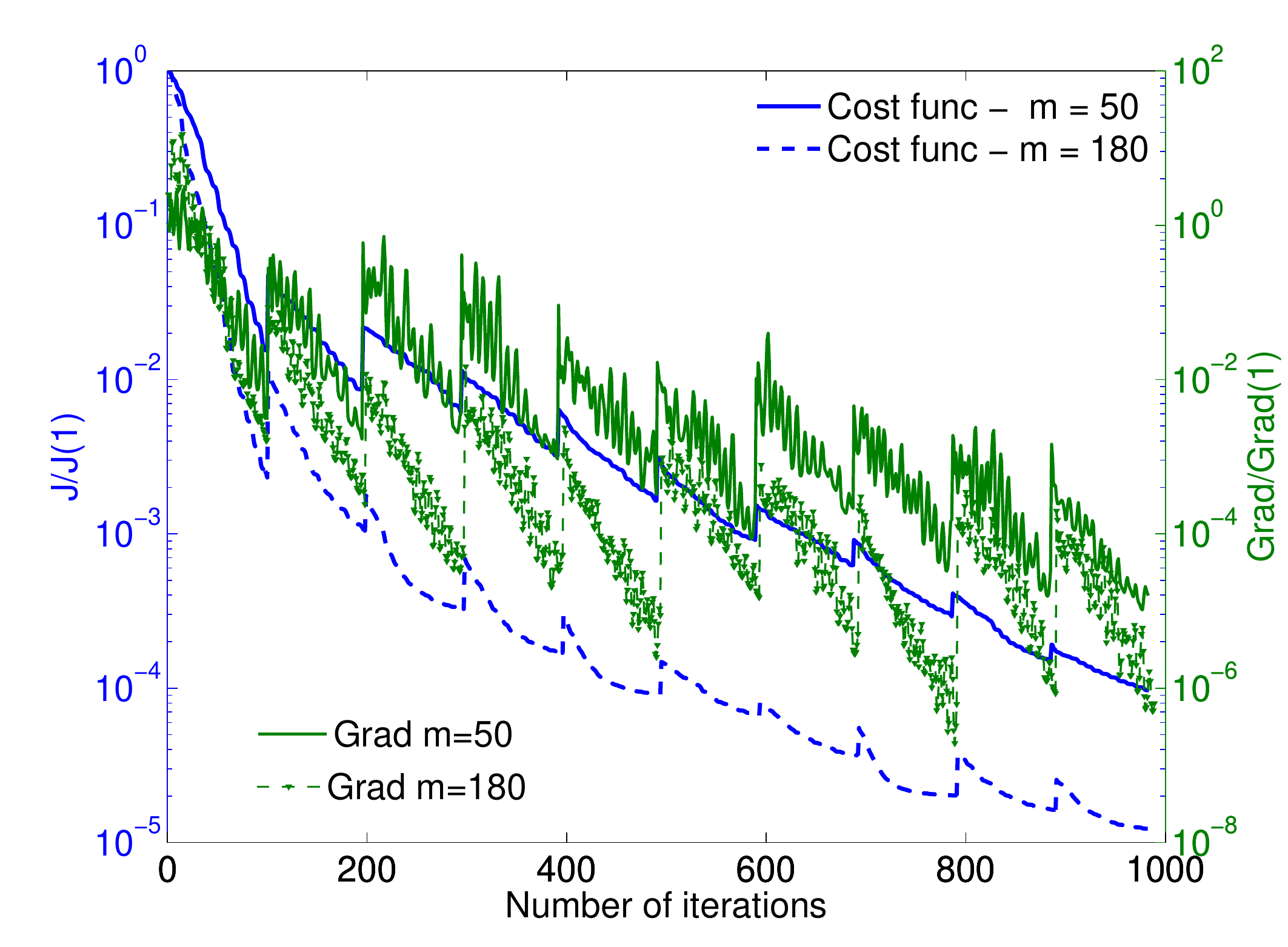}}
  \subfigure[Number of mesh points $17\times 13$]{\includegraphics[scale=0.36]{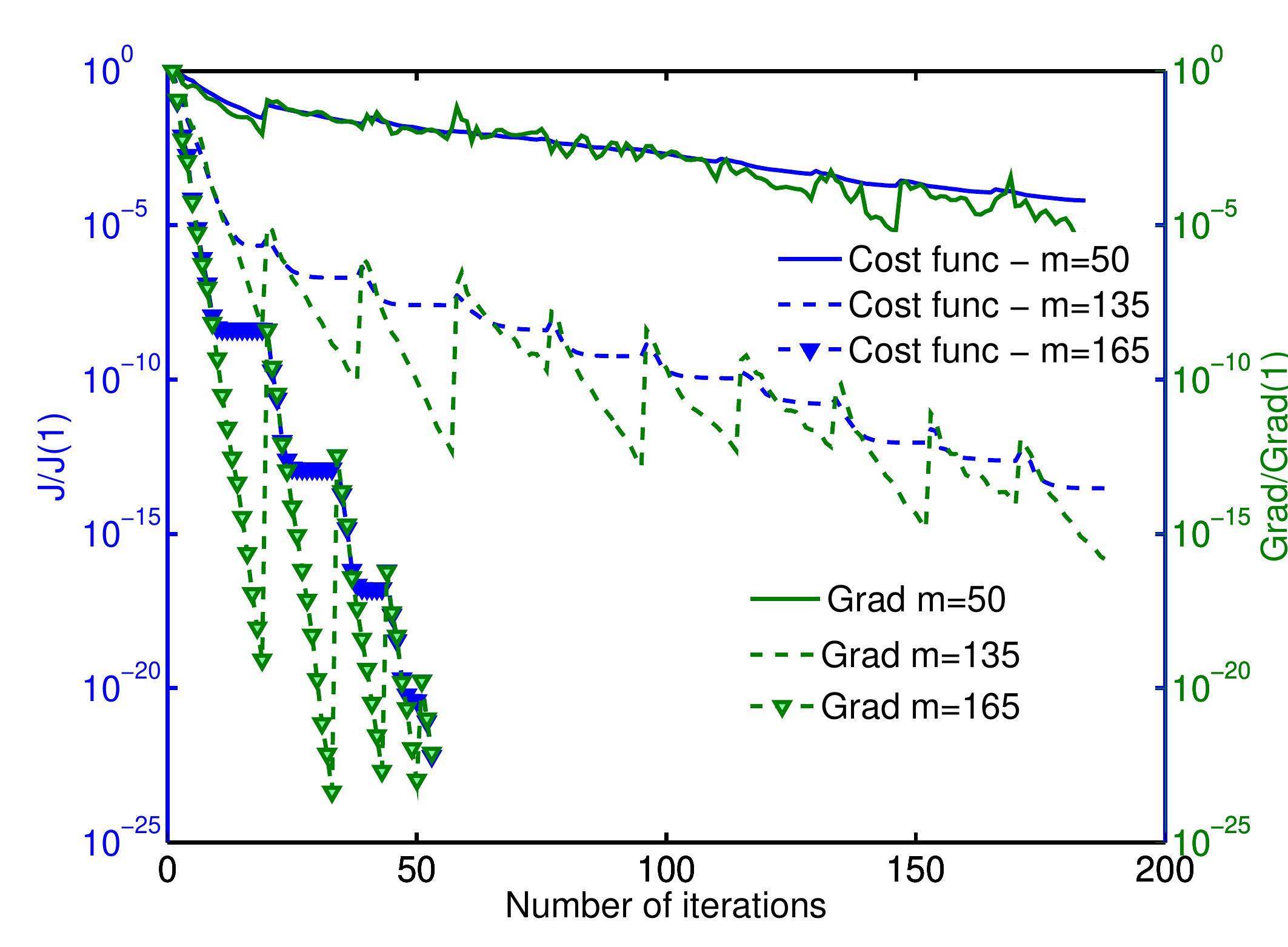}}
\caption{Standard POD/DEIM ADI SWE 4D-Var system -- Evolution of cost function and gradient norm as a function of the number of minimization iterations for different number of mesh points and various number of DEIM points. \label{Fig::POD_DEIM_4DVAR}}
\end{figure}

The first reduced optimization test is performed for a mesh of $31 \times 23$ points, a POD basis dimension of $k=50$, and $50$ and $180$ DEIM interpolation points are used. We obtain a cost function decrease of only $5$ orders of magnitude after $10$ POD bases updates and imposing a relaxed threshold of MXFUN $=100$ function evaluations per inner loop reduced optimization (see Figure \ref{Fig::POD_DEIM_4DVAR}a).

Thus we decide to incrementally increase the number of DEIM points until it reaches the number of space points and evaluate the reduced order POD/DEIM ADI SWE data assimilation system performances. However, our code is based on a truncated SVD algorithm that limits the number of POD modes of the nonlinear terms to a maximum of $2N_t-1$. This also constrains the maximum number of DEIM points to $2N_t-1$. Given this constraint, for the present space resolution $31 \times 23$ points, we can not envisage numerical experiments where the number of DEIM points is equal to the number of space points since $N_t = 91$.

In consequence we decrease the spatial resolution to $17 \times 13$ points and perform the reduced optimization with increasing number of DEIM points and MXFUN $=20$ (see Figure \ref{Fig::POD_DEIM_4DVAR}b). For $m=165$, POD/DEIM nonlinear terms approximations are identical with standard POD representations since the boundary variables are not controlled. We notice that even for $m=135$ there is an important loss of performance since the cost function decreases by only $10^{-12}$ orders of magnitude in $178$ inner iterations while for $m=165$ (standard POD) the cost functions achieves a $10^{-23}$ orders of magnitude decrease in only $52$ reduced optimization iterations.

\subsubsection{Adjoint and tangent linear POD/DEIM ADI SWE models verification tests}\label{subsubsec:NR_reduced_ad_tl_POD_DEIM_SWE models_tests}

The initial level of root mean square error (RMSE) due to by truncation of POD expansion for $k=50$ and for a number of DEIM interpolation points $m=50$ at final time $t_f=3h$ are similar for all reduced order methods (see Table \ref{table:forward_ROMS}). It means that some of  the nonlinear POD/DEIM approximations are more sensitive to changes in initial data during the optimization course while their nonlinear tensorial and standard POD counterparts proved to be more robust.

\begingroup
\begin{table}[h]
\centerline{
\scalebox{1.0}{
\begin{tabular}{|c|c|c|c|c|}\hline
& POD/DEIM & tensorial POD  & standard POD  \\ \hline
$E_{u}$ & $1.21e$-$7$& $1.2e$-$7$ & $1.2e$-$7$  \\ \hline
$E_{v}$ &$7.6e$-$8$& $7.48e$-$8$ & $7.48e$-$8$  \\ \hline
$E_{\phi}$ &$1.4e$-$7$& $1.36e$-$7$ & $1.36e$-$7$  \\ \hline
\end{tabular}}
}
\caption{\label{table:forward_ROMS} RMSE of reduced order solutions of the forward SWE ROMS with respect to the full space ADI SWE state variables at final time $t_f=3h$ for the same initial conditions used for snapshots and POD basis generations. Number of mesh points is $n = 17 \times 13$, number of POD basis functions is $k=50$ and number of DEIM points is $m=50$. }
\end{table} %
\endgroup%

 We verify the implementation of the POD/DEIM SWE 4D-Var system and the adjoint based gradient and the tangent linear model output agree with the finite difference approximations (see Figure \ref{Fig::Gr_AD_test_DEIM_hybrid}a and \cite[eq. (2.20)]{Navon_Zou_Derber_Sela_1992} for more details). The depicted values are obtained using
\begin{equation*}
\rm{adj_{test}} = \frac{(J^\textsc{pod}({\bf \tilde x}_0+\delta {\bf \tilde x}_0) - J^\textsc{pod}({\bf \tilde x}_0))}{ <\nabla J^\textsc{pod}({\bf \tilde x}_0),\delta {\bf \tilde x}_0>_2} \quad
\rm{tl_{test}} = \frac{\|\widetilde{\M}_{0,N_t}({\bf \tilde x}_0+\delta {\bf \tilde x}_0)(N_t) - \widetilde{\M}_{0,N_t}({\bf \tilde x}_0)(N_t)\|_2}{\|\widetilde{\bf M}_{0,N_t}(\delta {\bf \tilde x}_0)(N_t)\|_2},
\end{equation*}
where $\widetilde{\M}_{0,N_t}$, $\widetilde{\bf M}_{0,N_t}$ are the POD/DEIM forward and tangent linear models and $J^\textsc{pod}$ is computed using the POD/DEIM forward trajectory.

\subsubsection{Hybrid POD/DEIM ADI SWE 4D-Var data assimilation system} \label{subsubsec:NR_reduced_HYBRID_POD_DEIM_DA_system}

\begin{figure}[h]
\centering
\subfigure[Tangent Linear and Adjoint test ] {\includegraphics[scale=0.35]{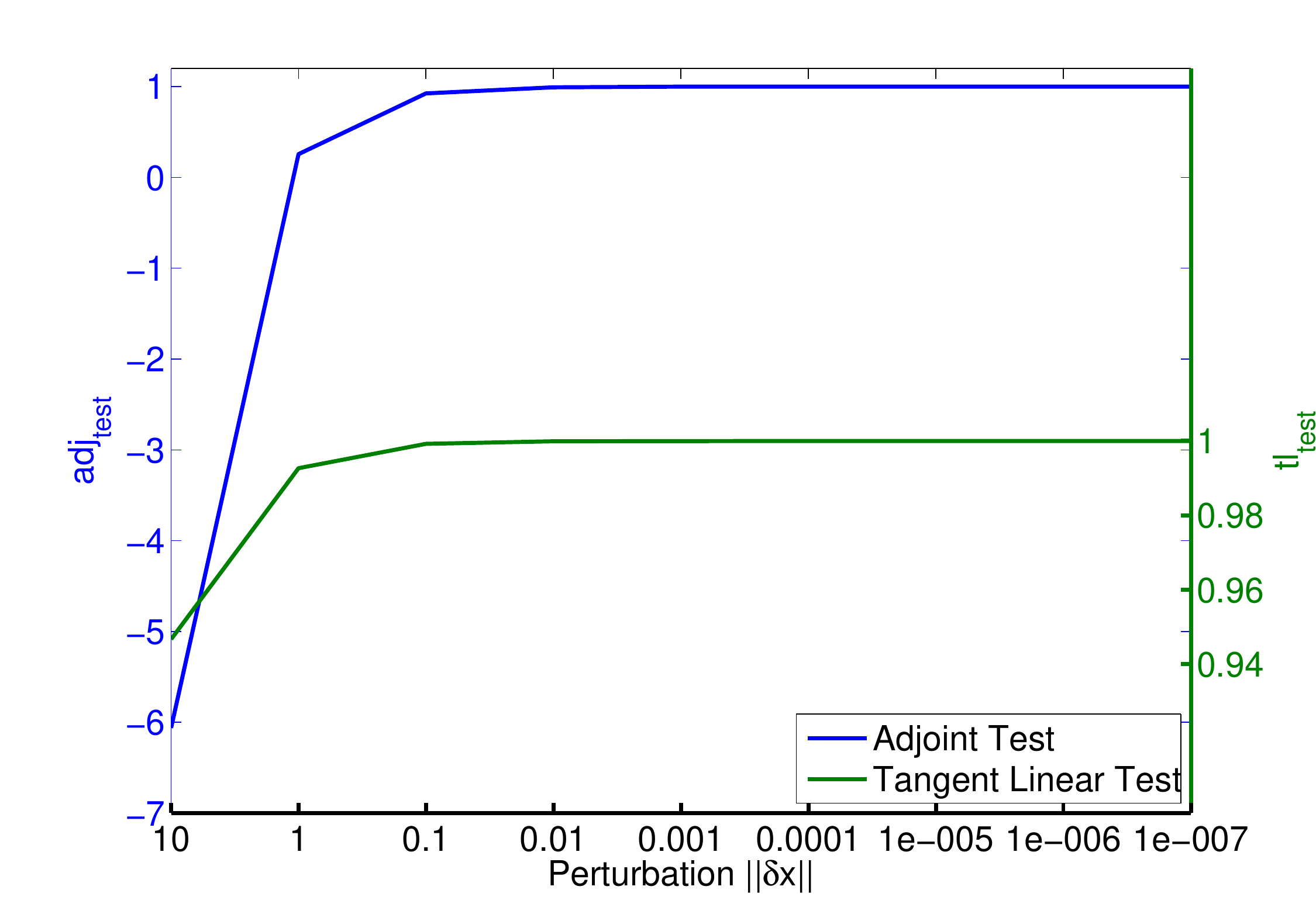}}
\subfigure[Cost function and gradient decays]{\includegraphics[scale=0.35]{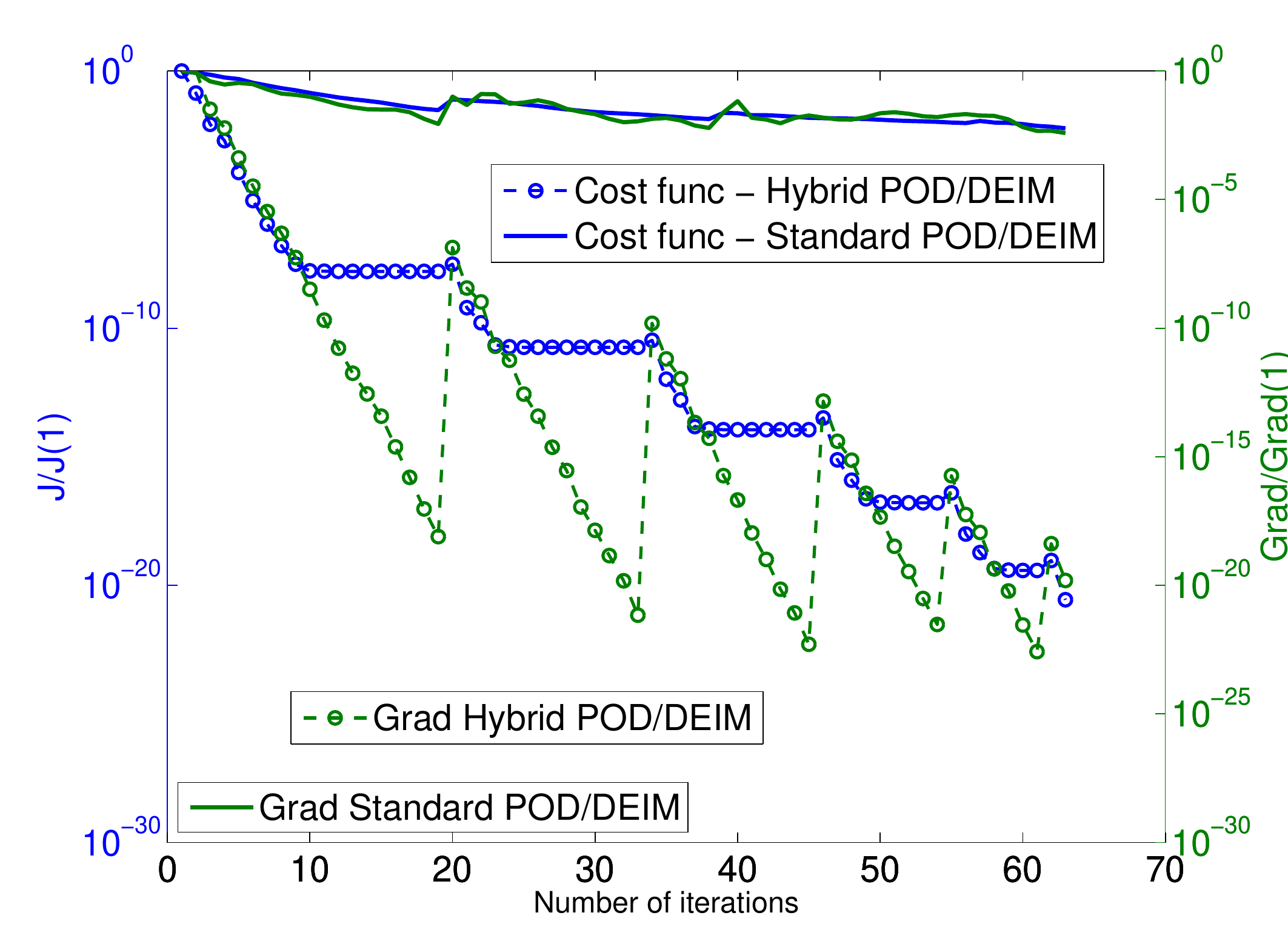}}
\caption{Tangent linear and adjoint test for standard POD/DEIM SWE 4D-Var system. Optimization performances of Standard POD/DEIM and Hybrid POD/DEIM 4D-Var ADI 2D shallow water equations for $n = 17\times 13$.}\label{Fig::Gr_AD_test_DEIM_hybrid}
\end{figure}

Next we begin checking the accuracy of the POD/DEIM nonlinear terms during the optimization and compare them with the similar tensorial POD nonlinear terms \eqref{eqn:POD_tensorial_nonlinearity_general}. We found out that POD/DEIM nonlinear terms involving height ${\boldsymbol \phi}$, i.e. ${\tilde F}_{12},~{\tilde F}_{23},~{\tilde F}_{31},~{\tilde F}_{33}$ lose $2-3$ orders accuracy in comparison with tensorial nonlinear terms. Thus we replaced only these terms by their tensorial POD representations and the new hybrid POD/DEIM SWE system using $50$ DEIM interpolation points reached the expected performances (see Figure \ref{Fig::Gr_AD_test_DEIM_hybrid}b).

\begin{figure}[h]
\centering
\subfigure[Number of DEIM points $m=50$] {\includegraphics[scale=0.36]{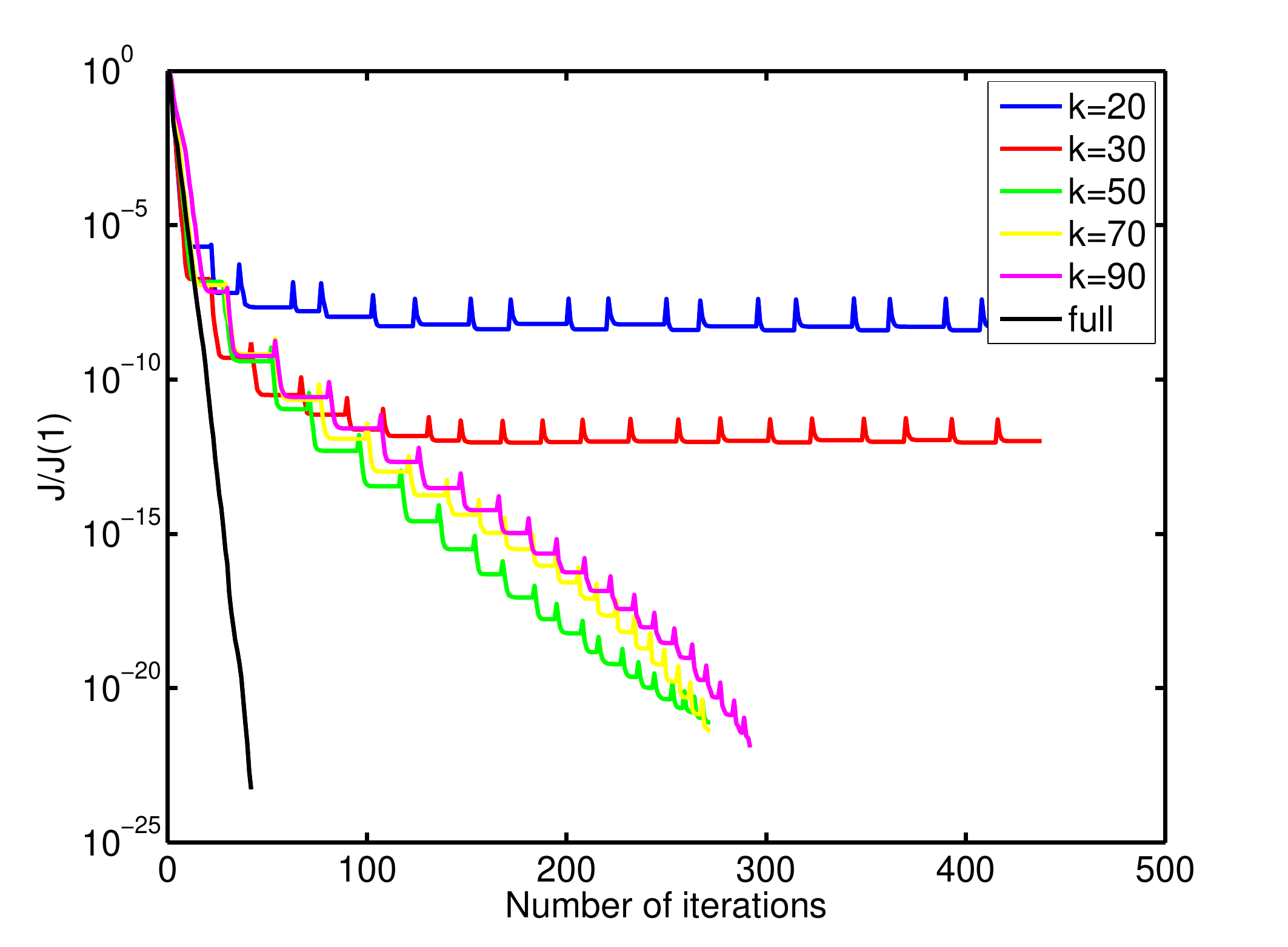}}
\subfigure[Number of POD basis dimension $k=50$] {\includegraphics[scale=0.36]{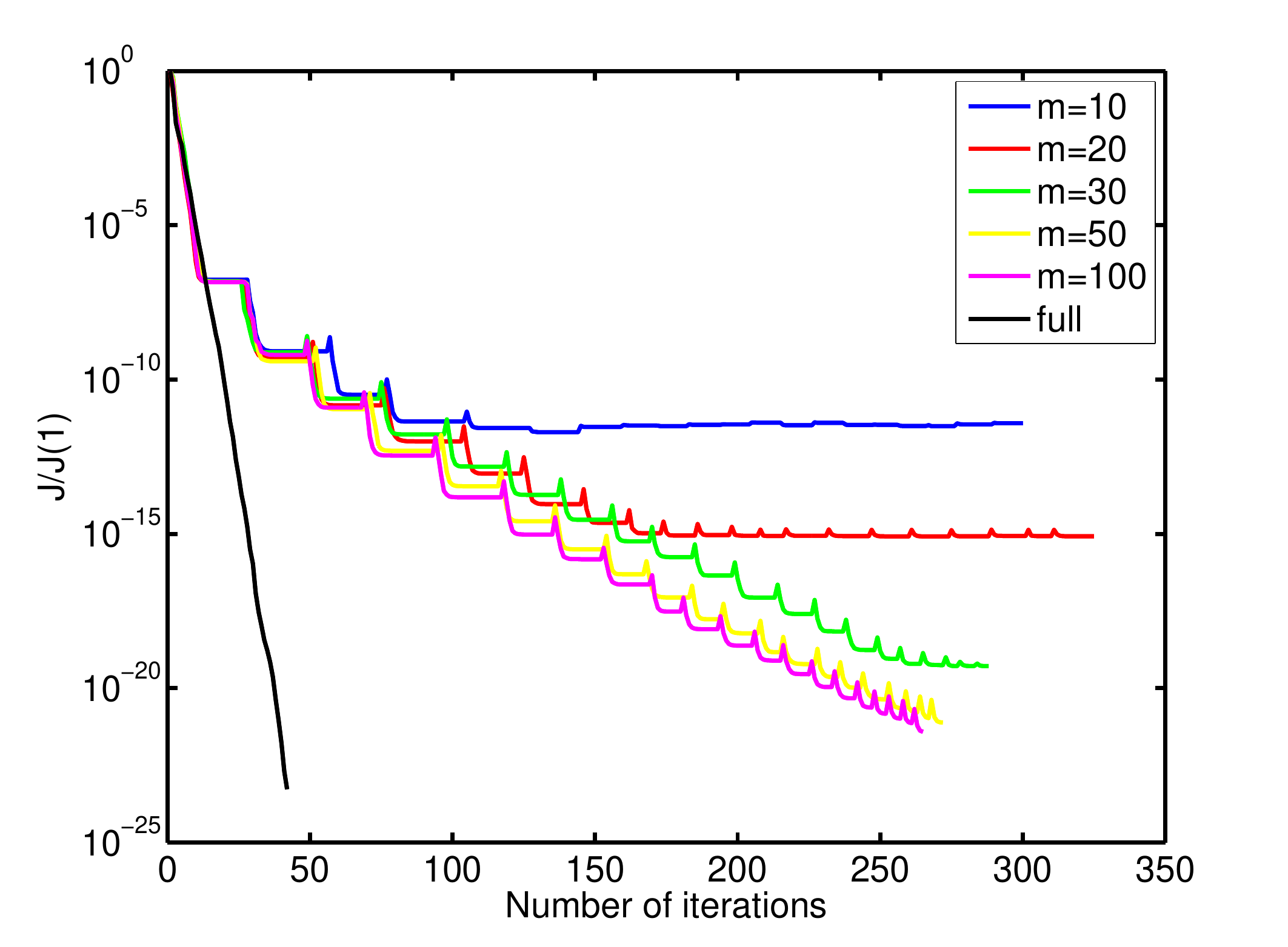}}
\caption{Performances of hybrid POD/DEIM SWE DA system with various values of POD basis dimensions (a) and different number of interpolation points (b). The spatial configuration uses $n=61\times45$ and maximum number of function evaluation per inner iteration is set MXFUN$ = 30$.\label{Fig::Hybrid_DEIM_diff_k_m}}
\end{figure}

Next we test the new hybrid POD/DEIM reduced data assimilation system using different POD basis dimensions and various numbers of DEIM points. For ROM optimization to deliver accurate suboptimal surrogate solutions similar to the output of full optimization one must increase the POD subspace dimension (see Figure \ref{Fig::Hybrid_DEIM_diff_k_m}a) for large number of mesh points configurations. Then we tested different configurations of DEIM points and for values of $m \ge 30$ the reduced optimization results are almost the same in terms of cost function decreases for $n = 17 \times 13$. Our findings were also confirmed by the relative errors accuracy of the suboptimal hybrid POD/DEIM SWE 4D-Var solutions with respect to the optimal solutions computed using high-fidelity ADI SWE 4D-Var system and observations (see Tables \ref{table:Hybrid_DEIM_diff_k},\ref{table:Hybrid_DEIM_diff_m}). We assumed that the background and observation errors are not correlated and their variances are equal to $1$.

\begingroup
\begin{table}[h]
\centerline{
\scalebox{1.0}{
\begin{tabular}{|c|c|c|c|c|c|c|}\hline
& $k=20$ & $k=30$  & $k=50$  & $k=70$ & $k=90$ & Full \\ \hline
$E_{u}^o$ & $1.03e$-$4$ & $1.57e$-$6$  & $4.1e$-$11$ & $3.75e$-$11$ & $2.1e$-$11$ & $2.77e$-$12$\\ \hline
$E_{v}^o$ & $4.04e$-$4$ & $6.29e$-$6$  & $1.72e$-$10$  & $1.16e$-$10$ & $8.55e$-$11$ & $1.27e$-$11$\\ \hline
$E_{\phi}^o$ & $2.36e$-$6$ & $3.94e$-$8$ & $1.09e$-$12$ & $6.78e$-$13$ & $2.3e$-$13$ &  $5.79e$-$14$ \\ \hline
\end{tabular}}
}
\caption{\label{table:Hybrid_DEIM_diff_k} Relative errors of suboptimal hybrid POD/DEIM SWE 4D-Var solutions with respect to observations. The number of DEIM interpolation points is held constant $m=50$ while $k$ is varied.}
\end{table} %
\endgroup%

\begingroup
\begin{table}[h]
\centerline{
\scalebox{1.0}{
\begin{tabular}{|c|c|c|c|c|c|c|}\hline
& $m=10$ & $m=20$  & $m=30$  & $m=50$ & $m=100$ & Full \\ \hline
$E_{u}^o$ & $4.28e$-$6$ & $4.91e$-$8$  & $4.15e$-$10$ & $4.1e$-$11$ & $3.08e$-$11$ & $2.77e$-$12$\\ \hline
$E_{v}^o$ & $6.88e$-$6$ & $1.889e$-$7$  & $1.54e$-$9$  & $1.72e$-$10$ & $1.25e$-$10$ & $1.27e$-$11$\\ \hline
$E_{\phi}^o$ & $7.15e$-$8$ & $1.04e$-$9$ & $7.79e$-$12$ & $1.09e$-$13$ & $7.39e$-$13$ &  $5.79e$-$14$ \\ \hline
\end{tabular}}
}
\caption{\label{table:Hybrid_DEIM_diff_m} Relative errors of suboptimal hybrid POD/DEIM SWE 4D-Var solutions with respect to observations. Different DEIM interpolation points are tested and $k$ is held constant $50$.}
\end{table} %
\endgroup%

%

\subsubsection{Computational cost comparison of reduced order 4D-Var data assimilation systems} \label{subsubsec:NR_reduced_DA_systems}

This subsection is dedicated to performance comparisons between proposed optimization systems using reduced and full space configurations. We use different numbers of mesh points resolutions $31\times 23,~61\times 45,~101\times71,~121\times189,~151\times111$ resulting in $1823,~7371,~20493,~28971$ and $48723$ control variables respectively. Various values of maximum number of function evaluations per each reduced minimization are also tested. We already proved that for increased number of POD basis dimensions the reduced data assimilation hybrid POD/DEIM ADI 4D-Var system leads to a cost function decrease almost similar with the one obtained by the full SWE 4D-Var system (see Figure \ref{Fig::Hybrid_DEIM_diff_k_m}a). Thus we are more interested to measure how fast the proposed data assimilation systems can reach the same threshold $\varepsilon_3$ in terms of cost function rate of decay.

\begin{figure}[h]
\centering
\subfigure[Iteration performance] {\includegraphics[scale=0.36]{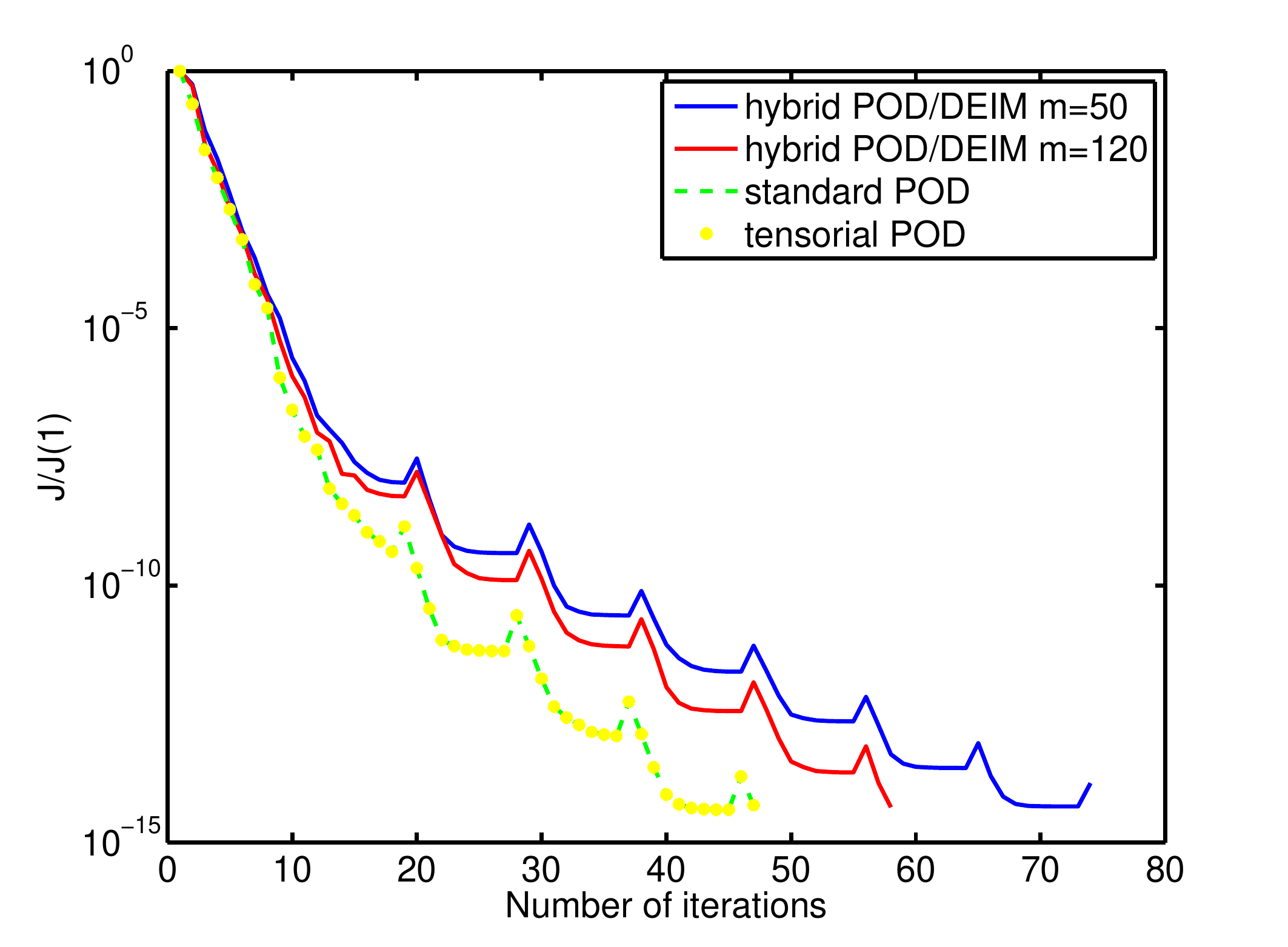}}
\subfigure[Time performance] {\includegraphics[scale=0.36]{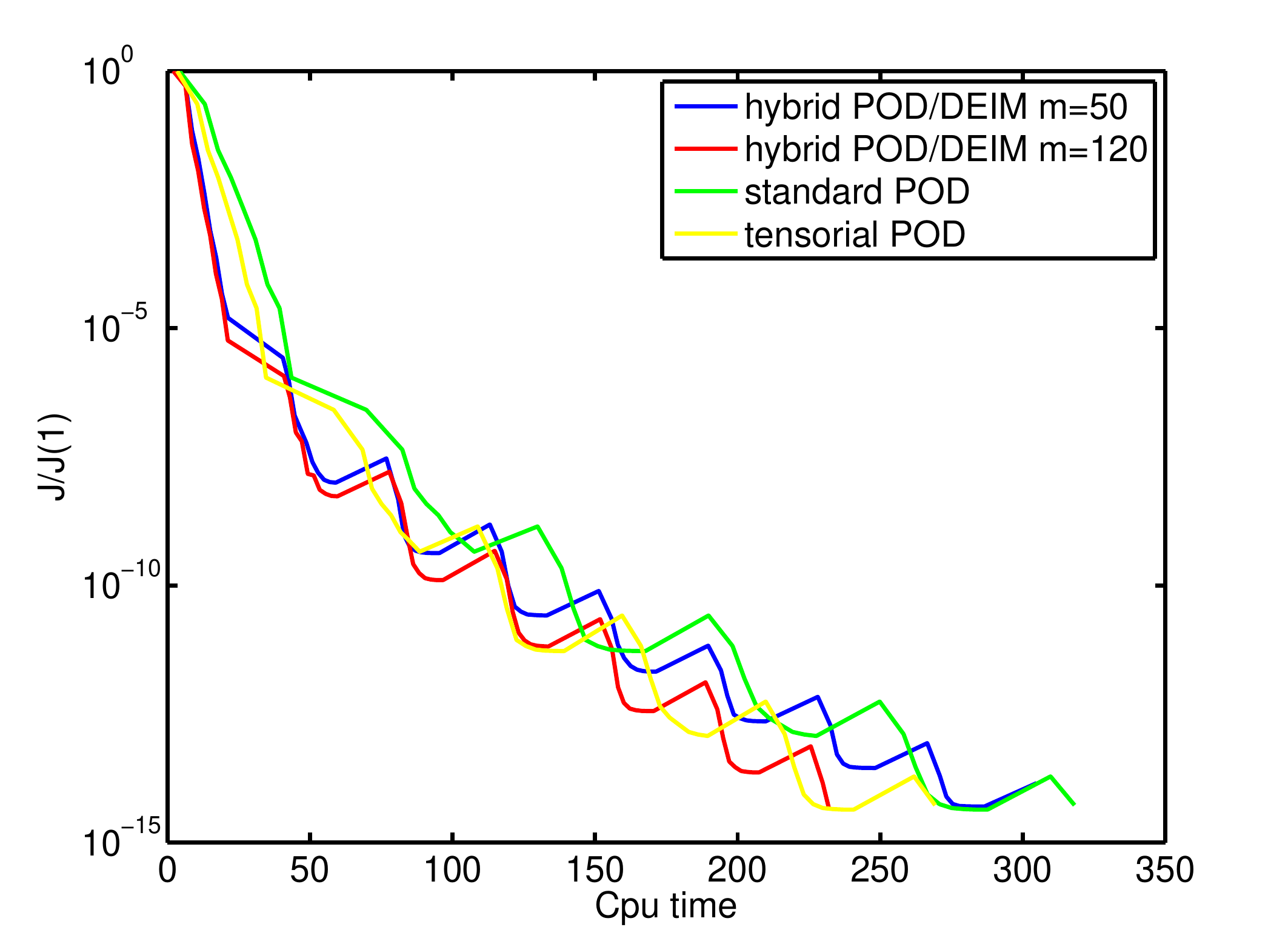}}
\caption{Number of minimization iterations and CPU time comparisons for the reduced order SWE DA systems vs. full SWE DA system. The spatial configuration uses $n=61\times45$ and maximum number of function evaluation per inner iteration is set MXFUN$ = 10$.\label{Fig::ROMs_system_compI}}
\end{figure}

\paragraph{Increasing the space resolution}
The next experiment uses the following configuration: $n = 61\times 45$ space points, number of POD basis modes $k=50$, MXFUN $= 10$ and $\varepsilon_3 = 10^{-7}$. Figure \ref{Fig::ROMs_system_compI} depicts the cost function evolution during hybrid POD/DEIM SWE 4D-Var, standard POD SWE 4D-Var and tensorial POD SWE 4D-Var minimizations versus number of iterations and CPU times. We notice that for $50$ DEIM points the hybrid POD/DEIM DA system requires $3$ additional POD basis updates to decrease the cost functional value below $10^{-7}$ in comparison with standard and tensorial POD DA systems. By increasing the number of DEIM points to $120$ the number of required POD basis recalculations is decreased by a factor of $2$ and the total number of reduced minimization iterations is reduced by $20$. The hybrid POD/DEIM SWE 4D-Var system using $m=120$ is faster with $\approx 37s$ and $\approx 86s$ than both the tensorial and standard POD SWE 4D-Var systems.

\begin{figure}[h]
\centering
\subfigure[Iteration performance] {\includegraphics[scale=0.36]{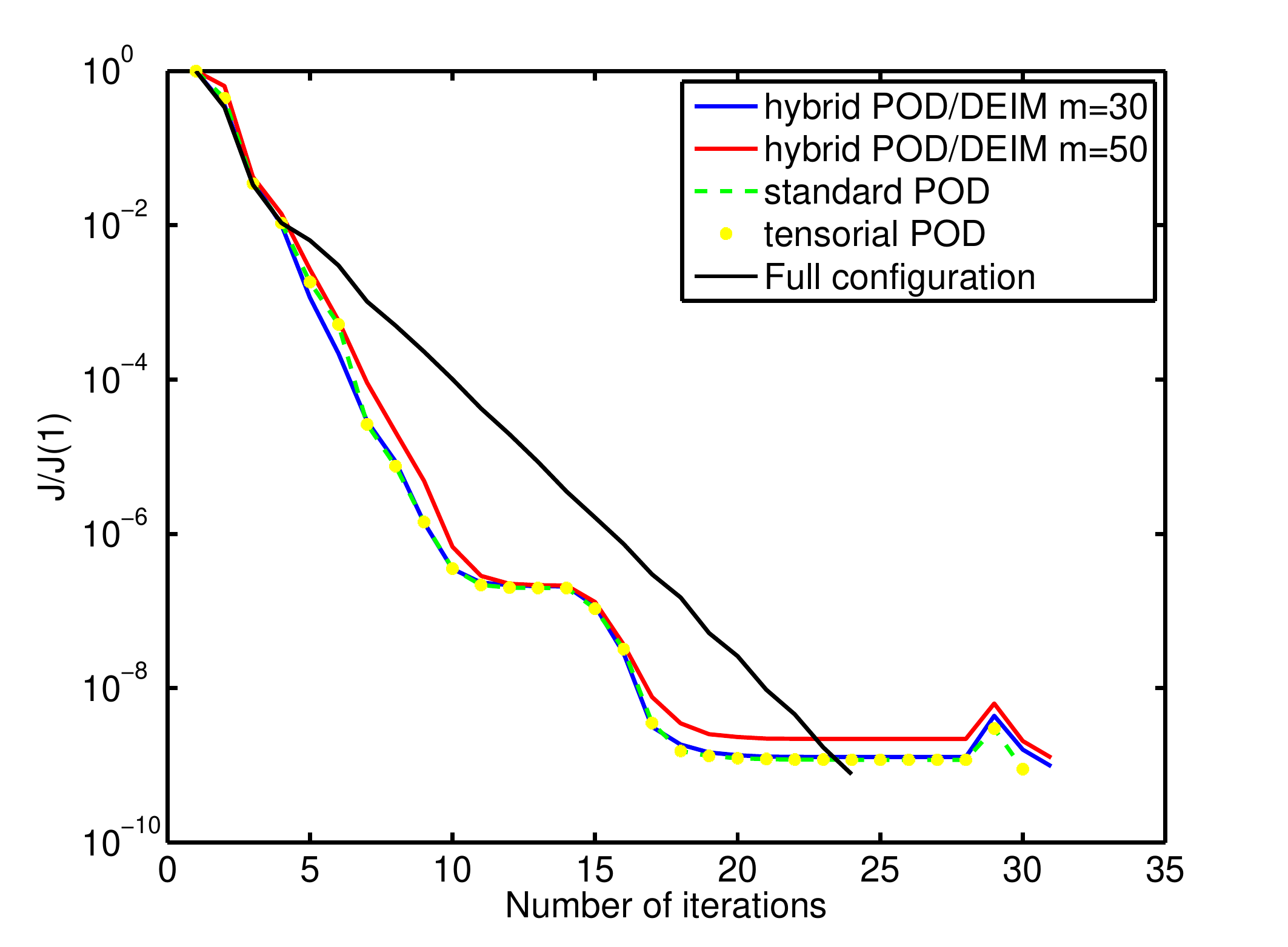}}
\subfigure[Time performance] {\includegraphics[scale=0.36]{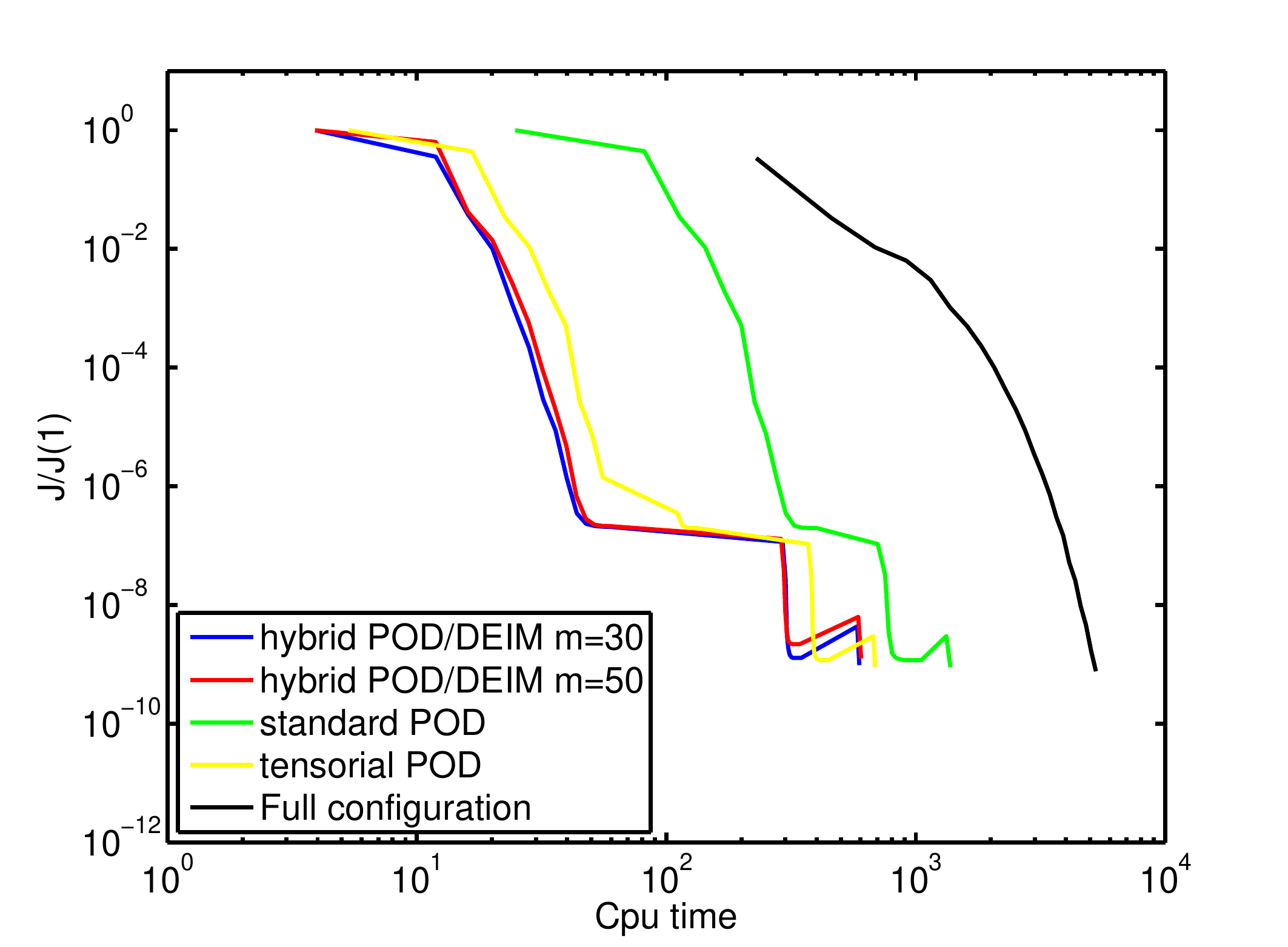}}
\caption{Number of iterations and CPU time comparisons for the reduced order SWE DA systems vs. full SWE DA system. The spatial configuration uses $n=151\times111$ and maximum number of function evaluation per inner iteration is set MXFUN$ = 15$.\label{Fig::ROMs_system_compII}}
\end{figure}

Next we increase the number of spatial points to $n = 151\times 111$ and use the same POD basis dimension $k=50$. MXFUN is set to $15$. The stopping criteria for all optimizations is $\|\J\|<\varepsilon_3=10^{-1}$. All the reduced order optimizations required two $POD$ basis recalculations and the hybrid POD/DEIM SWE 4D-Var needed one more iteration than the standard and tensorial POD systems (see Figure \ref{Fig::ROMs_system_compII}a). The hybrid POD/DEIM SWE 4D-Var system using $m=30$ is faster with $\approx 9s,92s,781s,4674s$ (by $1.01,~1.15,~2.31,~8.86$ times) than the hybrid POD/DEIM ($m=50$), tensorial POD, standard POD and full SWE 4D-Var data assimilation systems respectively (see Figure \ref{Fig::ROMs_system_compII}b).

Table \ref{tab:Cpu_ROMs_k_30_MXFUN_15} displays the CPU times required by all the optimization methods to decrease the cost function values below a specific threshold $\varepsilon_3$ specific to each space configuration (row $2$ in table \ref{tab:Cpu_ROMs_k_30_MXFUN_15}). The POD basis dimension is set to $k=30$. The bold values correspond to the best CPU time performances and some important conclusions can be drawn. There is no need for use of reduced optimization for $n=31\times 23$ or less since the full data assimilation system is faster. The hybrid POD/DEIM SWE 4D-Var using $50$ DEIM points is the most rapid optimization approach for numbers of space points larger than $61\times45$. For $n=151\times 111$ it is $1.23,~2.27,~12.329$ faster than tensorial POD, standard POD and full SWE 4D-Var systems. We also notice that the CPU time speedup rates are directly proportional with the increase of the full space resolution dimensions.

\begingroup
\begin{table}[h]
\centerline{
\scalebox{0.86}{
\begin{tabular}{|c|c|c|c|c|c|} \hline
Space points & $31\times23$ & $61\times45$ & $101\times 71$ & $121\times 89$ & $151\times 111$ \\ \hline
$\varepsilon_3$& $\|J\|<1.e$-$09$ & $\|J\|<5.e$-$04$ & $\|J\|<1.e$-$01$& $\|J\|<5$ & $\|J\|<1e$+$03$\\ \hline
hybrid DEIM 50 & 48.771 & {\bf 63.345} & {\bf 199.468} & {\bf 358.17} & {\bf 246.397}\\ \hline
hybrid DEIM 120 & 44.367 & 64.777 & 210.662& 431.460 & 286.004 \\ \hline
standard POD & 63.137 & 131.438 & 533.052 & 760.462 & 560.619\\ \hline
tensorial POD & 54.54 & 67.132 & 216.29& 391.075 & 303.95\\ \hline
FULL & {\bf 10.6441} & 117.02 & 792.929 & 1562.3425 & 3038.24 \\ \hline
\end{tabular}}}
\caption{\label{tab:Cpu_ROMs_k_30_MXFUN_15}CPU time for reduced optimization and full 4D-Var sharing the same stoping criterion $\|J\|<\varepsilon_3$. Number of POD modes is selected $30$ and MXFUN $= 15$.}
\end{table} %
\endgroup%

\paragraph{Increasing the POD basis dimension}

Next we set the POD basis dimension to $50$ and the corresponding CPU times are described in Table $7$. Notice also that $\varepsilon_3$ values are decreased. The use of reduced optimization system is justified for $n > 61\times45$ where the hybrid POD/DEIM DA system using different numbers of DEIM points proves to be the fastest choice. For $151\times111$ space points the hybrid POD/DEIM reduced optimization system is $1.15,~2.31,$ and $8.86$ times faster than tensorial POD, standard POD and full SWE 4D-Var systems respectively.

\begingroup
\begin{table}[h]
\centerline{
\scalebox{0.86}{
\begin{tabular}{|c|c|c|c|c|c|}\hline
Space points & $31\times23$ & $61\times45$ & $101\times 71$ & $121\times 89$ & $151\times 111$ \\ \hline
$\varepsilon_3$ & $\|J\|<1.e$-$14$ & $\|J\|<1.e$-$07$ & $\|J\|<1.e$-$04$& $\|J\|<5e$-$03$ & $\|J\|<1e$-$01$\\ \hline
hybrid DEIM 30 & 214.78 & 288.627 & 593.357 & {\bf 499.676} & {\bf 594.04}\\ \hline
hybrid DEIM 50 & 211.509 & 246.65 & {\bf 529.93} & 512.721 & 603.21\\ \hline
standard POD & 190.572 & 402.208 & 1243.234& 1315.573 & 1375.4 \\ \hline
tensorial POD & 269.08 & 311.106 & 585.509& 662.95 & 685.57\\ \hline
FULL & {\bf 14.1005} & {\bf 155.674} & 1057.715 & 2261.673 & 5268.7 \\ \hline
\end{tabular}}}
\caption{\label{tab:Cpu_ROMs_k_50_MXFUN_15}CPU time for reduced optimization and full 4D-Var sharing the same stoping criterion $\|J\|<\varepsilon_3$. Number of POD modes is selected $50$ and MXFUN $= 15$.}
\end{table} %
\endgroup%

\paragraph{Detailed computational cost analysis}
Now we are able to describe the computational time required by each step of the high-fidelity and reduced optimization systems. We are using $151\times111$ space points, POD basis dimension $k=50$, and number of DEIM points $m=30$. MXFUN is set to $15$ and $\varepsilon_3 = 10^{-1}$. For the full space 4D-Var system the line search and Hessian approximations are computational costly and are described separately (see table \ref{table:Full_4D-Var_CPU_time}) while for reduced data assimilation systems these costs are very small being included in the reduced adjoint model CPU time.
\begingroup
\begin{table}[h]
\centering 
\scalebox{1.0}{
\begin{tabular}{|c|c|c|c|} 
\hline
Process & Time &  \# & Total  \\ \hline
Solve full forward model & $\approx$ 80s & 26x & $\approx$ 2080s \\ \hline
Solve full adjoint model & $\approx$ 76.45s & 26x & $\approx$ 1987.7s \\ \hline
Other (Line Search, Hessian approx) & $\approx$ 46.165 & 26x & $\approx$ 1200.3s \\ \hline
\hline
Total full 4D-Var &  &  & $\approx$ 5268s \\ \hline
\end{tabular}}
\caption{The calculation times for solving the optimization problem using full 4D-Var for a number of mesh points of $151\times 111$. Stoping criterion $\Vert \J \Vert<10^{-1}$ is set.
\label{table:Full_4D-Var_CPU_time}}
\end{table}

\endgroup%

The most expensive part of the Hybrid POD/DEIM 4D-Var optimization process (see table \ref{table:Hybrid_POD_DEIM_4D-Var_CPU_time}) occurs during the off-line stage and consists in the snapshots generation stage where the full forward and adjoint models are integrated in time. This is valid also for tensorial POD 4D-Var DA system (table \ref{table:Tensorial_POD_DEIM_4D-Var_CPU_time}) while in the case of standard POD 4D-Var system (table \ref{table:Standard_POD_DEIM_4D-Var_CPU_time}) the on-line stage is far more costly since the computational complexity of the corresponding reduced forward model still depends on the full space dimension.
\begingroup
\begin{table}[h]
\centering
\scalebox{1.0}{
\begin{tabular}{|c|c|c|c|c|} \hline
Process & Time & \# & Total \\ \hline
{\bf Off-line stage} & & & \\ \hline
Solve full forward model + nonlinear snap. & $\approx$ 80.88s & 2x & $\approx$ 161.76s \\ \hline
Solve full adjoint model + nonlinear snap. & $\approx$ 76.45s& 2x & $\approx$ 152.9s \\ \hline
SVD for state variables & $\approx$ 53.8 & 2x & $\approx$ 107.6s \\ \hline
SVD for nonlinear terms &$\approx$ 11.57 & 2x & $\approx$ 23.14s \\ \hline
DEIM interpolation points &$\approx$ 0.115 & 2x & $\approx$  0.23s \\ \hline
POD/DEIM model coefficients &$\approx$ 1.06 & 2x & $\approx$ 2.12s \\ \hline
tensorial POD model coefficients &$\approx$ 8.8 & 2x & $\approx$ 17.6s \\\hline
{\bf On-line stage} & & &\\ \hline
Solve ROM forward &$\approx$ 2s & 33x & $\approx$ 66s\\ \hline
Solve ROM adjoint &$\approx$ 1.9s & 33x & $\approx$ 62.69s\\ \hline
\hline
Total Hybrid POD/DEIM 4D-Var & &  & $\approx$ 594.04s \\ \hline
\end{tabular}}
\caption{The calculation times for solving the optimization problem using Hybrid POD/DEIM 4D-Var for a number of mesh points of $151\times 111$, POD basis dimension $k=50$ and $30$ DEIM interpolation points.}
\label{table:Hybrid_POD_DEIM_4D-Var_CPU_time}

\end{table}

\endgroup%
The algorithm proposed in \citet[p.16]{Stefanescu2012} utilizes DEIM interpolation points, exploits the structure of polynomial nonlinearities and delivers fast tensorial calculations of POD/DEIM model coefficients. Consequently the hybrid POD/DEIM SWE 4D-Var systems has the fastest off-line stage among all proposed reduced data assimilation systems despite additional SVD calculations and other reduced coefficients computations.
\begingroup

\begin{table}[h]
\centering
\scalebox{1.0}{
\begin{tabular}{|c|c|c|c|} 
\hline 
Process &  Time & \# & Total \\ \hline
{\bf Off-line stage} & & &\\ \hline
Solve full forward model + nonlinear snap. & $\approx$ 80s & 2x & $\approx$ 160s \\ \hline
Solve full adjoint model + nonlinear snap. & $\approx$ 76.45s& 2x & $\approx$ 152.9s \\ \hline
SVD for state variables & $\approx$ 53.8s & 2x & $\approx$ 107.6s \\ \hline
tensorial POD model coefficients & $\approx$ 23.735s & 2x & $\approx$ 47.47s \\
\hline
{\bf On-line stage} & & & \\ \hline
Solve ROM forward &   $\approx$ 4.9s & 32x & $\approx$ 156.8s \\ \hline
Solve ROM adjoint &  $\approx$ 1.9s & 32x & $\approx$ 60.8s \\ \hline \hline
Total Tensorial POD 4D-Var &   &  & $\approx$  685.57s \\ \hline
\end{tabular}}
\caption{The calculation times for solving the optimization problem using Tensorial POD 4D-Var for a number of mesh points of $151\times 111$, and POD basis dimension $k=50$.}
\label{table:Tensorial_POD_DEIM_4D-Var_CPU_time} 

\end{table}

\endgroup
For all three reduced optimization systems the Jacobians are calculated analytically and their computations depend only on the reduced space dimension $k$. As a consequence, all the adjoint models have the same computational complexity and in the case of hybrid POD/DEIM SWE 4D-Var the on-line Jacobians computations rely partially on approximated tensors \eqref{eqn:DEIM_tensor_F11} calculated during the off-line stage  while in the other two reduced order data assimilation systems exact tensorial POD model coefficients are used.
\begingroup

\begin{table}[h]
\centering
\scalebox{1.0}{
\begin{tabular}{|c|c|c|c|} \hline
Process & Time & \# & Total \\ \hline
{\bf On-line stage} & & & \\ \hline
Solve ROM forward & $\approx$ 26.523s & 32x & $\approx$ 846.72s \\ \hline
Solve ROM adjoint & $\approx$ 1.9 & 32x & $\approx$ 60.8s \\ \hline \hline
Total Standard 4D-Var &  &  & $\approx$   1375.4s \\ 
\hline 
\end{tabular}}
\caption{The calculation times for on-line stage of Standard POD 4D-Var for a number of mesh points of $151\times 111$ and POD basis dimension $k=50$. The off-line stage is identical with the one in Tensorial POD 4D-Var system.}
\label{table:Standard_POD_DEIM_4D-Var_CPU_time}

\end{table}

\endgroup%

\paragraph{Varying the number of function evaluations per reduced minimization cycle}

The reduced optimization data assimilation systems become slow if it is repeatedly required to project back to the high fidelity model and reconstruct the reduced POD subspace. Thus, we compare the CPU times obtained by our reduced data assimilation systems using at most $10,~15,$ and $20$ function evaluations per reduced minimization cycle. The results for $k=30$ (see Table \ref{tab:Cpu_ROMs_k_30_MXFUN_diff}) shows that no more than $15$ function evaluations should be allowed for each reduced minimization cycle and hybrid POD/DEIM data assimilation system using $30$ interpolation points provides the fastest solutions. While for $101\times 71$ number of space points $15$ function evaluations are required, for other spatial configurations MXFUN $= 10$ is sufficient.

\begingroup
\begin{table}[h]
\centerline{
\scalebox{1.0}{
\begin{tabular}{|c|c|c|c|c|}\hline \hline
Space points & MXFUN & $\varepsilon_3$ & DEIM points & Method \\ \hline
$31\times 23$ & - & $1.e$-$9$ & - & Full\\ \hline
$61\times 45$ & 10 & $5.e$-$04$ & 30 & Hybrid POD/DEIM\\ \hline
$101\times 71$ & 15 & $1.e$-$01$ & 30 & Hybrid POD/DEIM \\ \hline
$121\times 89$ & 10 & $5$ & 30 & Hybrid POD/DEIM\\ \hline
$151\times 111$ & 10 & $1e$+$03$ & 30& Hybrid POD/DEIM\\ \hline
\end{tabular}}}
\caption{\label{tab:Cpu_ROMs_k_30_MXFUN_diff}The fastest optimization data assimilation systems for various number of spatial points and different MXFUN values.  Number of POD modes is $k=30$.}
\end{table} %
\endgroup%

For POD basis dimension $k=50$, we discover that more function evaluations are needed during the inner reduced minimizations in order to obtain the fastest CPU times and MXFUN should be set to $15$. More DEIM points are also required as we notice in Table \ref{tab:Cpu_ROMs_k_50_MXFUN_diff}. Thus we can conclude that MXFUN should be increased with the decrease of $\varepsilon_3$ and increase of dimension of POD basis.
\begingroup
\begin{table}[h]
\centerline{
\scalebox{1.0}{
\begin{tabular}{|c|c|c|c|c|}\hline \hline
Space points & MXFUN & $\varepsilon_3$ & DEIM points & Method \\ \hline
$31\times 23$ & - & $1.e$-$14$ & - & Full\\ \hline
$61\times 45$ & - & $1.e$-$7$ & - & Full\\ \hline
$101\times 71$ & 15 & $1.e$-$04$ & 50 & Hybrid POD/DEIM \\ \hline
$121\times 89$ & 15 & $5.e$-$3$ & 50 & Hybrid POD/DEIM\\ \hline
$151\times 111$ & 15 & $1.e$-$1$ & 50& Hybrid POD/DEIM\\ \hline
\end{tabular}}}
\caption{\label{tab:Cpu_ROMs_k_50_MXFUN_diff}The fastest optimization data assimilation systems for various number of spatial points and different MXFUN values.  Number of POD modes is $k=50$.}
\end{table} %
\endgroup%

We conclude that hybrid POD/DEIM SWE 4D-Var system delivers the fastest suboptimal solutions and is far more competitive in terms of CPU time than the full SWE data assimilation system for space resolutions larger than $61\times45$ points. Hybrid POD/DEIM SWE 4D-Var is at least two times faster than standard POD SWE 4D-Var for $n\ge 101\times 71$.

\subsubsection{Accuracy comparison of reduced 4D-Var data assimilation suboptimal solutions }
\label{subsec:Accuracy_NR_reduced_DA_systems}
In terms of suboptimal solution accuracy, the hybrid POD/DEIM delivers similar results as tensorial and standard POD SWE 4D-Var systems (see tables \ref{tab:suboptimal_error_u}, \ref{tab:suboptimal_error_v}, \ref{tab:suboptimal_error_phi}). The accuracy of the reduced order models is tested via relative norms introduced in \eqref{eqn::first_norm} at the beginning of reduced optimization algorithms and two different POD bases dimensions are tested, i.e. $k=30,~50$. To measure the suboptimal solutions accuracy we calculate the relative errors using $E_{{\bar{\boldsymbol  w}}}^*$ defined in \eqref{eqn::second_norm} and the corresponding values are depicted in the hybrid DEIM, sPOD and tPOD columns. MXFUN is set to $15$. We choose $\varepsilon_3 = 10^{-15}$ for all data assimilation systems and only $20$ outer iterations are allowed for all reduced 4D-Var optimization systems. For Hybrid POD/DEIM 4D-Var system we use $50$ DEIM interpolation points.

\begingroup
\begin{table}[h]
\centerline{
\scalebox{0.69}{
\begin{tabular}{|c|c|c|c|c|c|c|}\hline \hline
$n$ & $E_u$ & $E_{\lambda_u}$ & hybrid DEIM & sPOD & tPOD  \\ \hline
$31\times23$ & $4.51e$-$5$ & $5.09e$-$5$ & $5.86e$-$8$ & $9.11e$-$8$ & $9.11e$-$8$\\ \hline
$61\times45$ & $1.55e$-$4$ & $2.92e$-$4$ & $2.4e$-$5$ & $1.63e$-$5$ & $1.63e$-$5$ \\ \hline
$101\times71$ & $6.06e$-$4$ & $6.56e$-$4$ & $2.03e$-$4$ & $1.83e$-$4$ & $1.83e$-$4$  \\ \hline
$121\times 89$ & $9.88e$-$4$ & $1.93e$-$3$ & $1.16e$-$3$ & $8.07e$-$4$ & $8.07e$-$4$ \\ \hline
$151 \times 111$ & $2.07e$-$3$ & $2.97e$-$3$ & $4.02e$-$3$ & $4.2e$-$3$ &$4.2e$-$3$\\ \hline
\end{tabular}}
\scalebox{0.69}{
~~~~\begin{tabular}{||c|c|c|c|c|c|}\hline \hline
 $E_u$ & $E_{\lambda_u}$ & hybrid DEIM & sPOD & tPOD  \\ \hline
$1.39e$-$6$ & $1.87e$-$6$ & $1.13e$-$10$ & $2.51e$-$10$ & $2.51e$-$10$\\ \hline
$1.58e$-$5$ & $1.06e$-$5$ & $2.16e$-$7$ & $1.38e$-$7$ & $1.38e$-$7$ \\ \hline
$4.51e$-$5$ & $4.76e$-$5$ & $7.03e$-$6$ & $5.14e$-$6$ & $5.14e$-$6$  \\ \hline
$7.17e$-$5$ & $8.e$-$5$ & $5.62e$-$5$ & $3.15e$-$5$ & $3.15e$-$5$ \\ \hline
$1.02e$-$4$ & $2.77e$-$4$ & $1.59e$-$4$ & $1.05e$-$4$ &$1.05e$-$4$\\ \hline
\end{tabular}}}
\caption{\label{tab:suboptimal_error_u} Reduced order forward and adjoint model errors vs. reduced order optimal solution errors for velocity component $u$ using $30$ (left) and $50$ (right) POD modes.}
\end{table} %
\endgroup%

\begingroup
\begin{table}[h]
\centerline{
\scalebox{0.69}{
\begin{tabular}{|c|c|c|c|c|c|c|}\hline \hline
$n$ & $E_v$ & $E_{\lambda_v}$ & hybrid DEIM & sPOD & tPOD  \\ \hline
$31\times23$ & $3.20e$-$5$ & $4.06e$-$5$ & $6.03e$-$8$ & $9.95e$-$8$ & $9.95e$-$8$\\ \hline
$61\times45$ & $3.13e$-$4$ & $3.21e$-$4$ & $3.75e$-$5$ & $2.63e$-$5$ & $2.63e$-$5$ \\ \hline
$101\times71$ & $9.76e$-$4$ & $5.71e$-$4$ & $3.51e$-$4$ & $2.93e$-$4$ & $2.93e$-$4$  \\ \hline
$121\times 89$ & $1.37e$-$3$ & $1.70e$-$3$ & $1.38e$-$3$ & $1.21e$-$3$ & $1.21e$-$3$ \\ \hline
$151 \times 111$ & $1.58$-$3$ & $1.89e$-$3$ & $6.21e$-$3$ & $6.22e$-$3$ &$6.22e$-$3$\\ \hline
\end{tabular}}
~~\scalebox{0.69}{
\begin{tabular}{||c|c|c|c|c|c|}\hline \hline
$E_v$ & $E_{\lambda_v}$ & hybrid DEIM & sPOD & tPOD  \\ \hline
$7.54e$-$7 $ & $1.10e$-$6$ & $1.25e$-$10$ & $4.10e$-$10$ & $4.10e$-$10$\\ \hline
$1.09e$-$5 $ & $9.71e$-$6$ & $3.03e$-$7$ & $2.82e$-$7$ & $2.82e$-$7$ \\ \hline
$3.05e$-$5 $ & $3.55e$-$5$ & $8.32e$-$6$ & $9.45e$-$6$ & $9.45e$-$6$  \\ \hline
$7.96e$-$5 $ & $8.12e$-$5$ & $7.19e$-$5$ & $6.85e$-$5$ & $6.85e$-$5$ \\ \hline
$1.06e$-$4 $ & $1.76e$-$4$ & $2.95e$-$4$ & $1.94e$-$4$ &$1.94e$-$4$\\ \hline
\end{tabular}}}
\caption{\label{tab:suboptimal_error_v}Reduced order forward and adjoint model errors vs. reduced order optimal solution errors for velocity component $v$ using $30$ (left) and $50$ (right) POD modes.}
\end{table} %
\endgroup%

\begingroup
\begin{table}[h]
\centerline{
\scalebox{0.69}{
\begin{tabular}{|c|c|c|c|c|c|c|}\hline \hline
$n$ & $E_{\phi}$ & $E_{\lambda_{\phi}}$ & hybrid DEIM & sPOD & tPOD  \\ \hline
$31\times23$ & $4.38e$-$5$ & $4.16e$-$5$ & $1.12e$-$7$ & $1.83e$-$7$ & $1.83e$-$7$\\ \hline
$61\times45$ & $1.84e$-$4$ & $4.54e$-$4$ & $5.63e$-$5$ & $3.34e$-$5$ & $3.34e$-$5$ \\ \hline
$101\times71$ & $2.21e$-$3$ & $3.16e$-$3$ & $6.60e$-$4$ & $6.45e$-$4$ & $6.45e$-$4$  \\ \hline
$121\times 89$ & $5.60e$-$3$ & $4.38e$-$3$ & $2.88e$-$3$ & $2.54e$-$3$ & $2.54e$-$3$ \\ \hline
$151 \times 111$ & $9.04e$-$3$ & $1.2e$-$2$ & $8.54e$-$3$ & $8.46e$-$3$ &$8.46e$-$3$\\ \hline
\end{tabular}}
~~~\scalebox{0.69}{
\begin{tabular}{||c|c|c|c|c|c|}\hline \hline
$E_{\phi}$ & $E_{\lambda_{\phi}}$ & hybrid DEIM & sPOD & tPOD  \\ \hline
$1.59e$-$6$ & $1.04e$-$6$ & $1.12e$-$10$ & $2.87e$-$10$ & $2.87e$-$10$\\ \hline
$9.58e$-$6$ & $9.e$-$6$ & $1.80e$-$7$ & $2.37e$-$7$ & $2.37e$-$7$ \\ \hline
$3.52e$-$5$ & $4.49e$-$5$ & $7.60e$-$6$ & $1.04e$-$5$ & $1.04e$-$5$  \\ \hline
$7.84e$-$5$ & $1.03e$-$4$ & $5.56$e-$5$ & $5.04e$-$5$ & $5.04e$-$5$ \\ \hline
$1.8e$-$4$ & $2.77e$-$4$ & $2.76e$-$4$ & $2.75e$-$4$ &$2.75e$-$4$\\ \hline
\end{tabular}}}
\caption{\label{tab:suboptimal_error_phi} Reduced order forward and adjoint model errors vs. reduced order optimal solution errors for geopotential $\phi$ using $30$ (left) and $50$ (right) POD modes.}
\end{table} %
\endgroup%

The suboptimal errors of all reduced optimization systems $E_{{\bar{\boldsymbol  w}}}^*$ are well correlated with the relative errors of the reduced order models $E_{\bar{\bf w}}$ and $E_{{\bar{\lambda}}_{\bf w}}$ \eqref{eqn::first_norm}, having correlation coefficients higher than $0.85$. However the correlation coefficients between reduced adjoint model errors and suboptimal errors are larger than $0.9$ which confirm the a-priori error estimation results of \citet{Hinze_Volkwein_2008} developed for linear-quadratic optimal problems. It states  that error estimates for the adjoint state yield error estimates of the control. Extension to nonlinear-quadratic optimal problems is desired and represents subject of future research. In addition, an a-posteriori error estimation apparatus is required by the hybrid POD/DEIM SWE system to guide the POD basis construction and to efficiently select the number of DEIM interpolation points.

The suboptimal solutions delivered by the ROM DA systems equipped with BFGS algorithm are accurate and comparable with the optimal solution computed by the full DA system. In the future we plan to enrich the reduced data assimilation systems by implementing a trust region algorithm (see \citet{arian2000trust}). It has an efficient strategy for updating the POD basis and it is well known for its global convergence properties.


\section{Conclusions}\label{sec:Conclusions}

This work studies the use of reduced order modeling to speed up the solution of variational data assimilation problems with  nonlinear dynamical models. The novel ARRA framework proposed herein guarantees that the Karush-Kuhn-Tucker conditions of the reduced order optimization problem accurately approximate the corresponding first order optimality conditions of the full order problem. In particular, accurate low-rank approximations of the adjoint model  and of the gradient equation are obtained in addition to the accurate low-rank representation of the forward model. The construction is validated by an error estimation result.

The choice of the reduced basis in the ARRA approach depends on the type of projection employed. For a pure Petrov-Galerkin projection the test POD basis functions of the forward model coincide  with the trial POD basis functions of the adjoint model; and similarly, the adjoint test POD basis functions coincide with the forward trial POD basis functions. Moreover the trial POD basis functions of the adjoint model should also include gradient information. 
It is well known that pure Petrov-Galerkin reduced order models can exhibit severe numerical instabilities, therefore stabilization strategies have to be included with this type of reduced data assimilation system \cite{Amsallem2012,Bui-Thanh2007}. 

In the ARRA Galerkin projection approach the same reduced order basis has to represent  accurately the full order forward solution, the full order adjoint solution, and the full order gradient. The Galerkin POD bases are constructed from the dominant eigenvectors of the correlation matrix  of the aggregated set of vectors containing snapshots of the  full order forward and adjoint models, as well as the full order background term. This reduced bases selection strategy is not limited to POD framework. It extends easily to every type of reduced optimization  involving projection-based reduced order methods such the reduced basis approach.

Numerical experiments using tensorial POD SWE 4D-Var data assimilation system based on Galerkin projection and different type of POD bases support the proposed approach. The most accurate suboptimal solutions and the fastest decrease of the cost function are obtained using full forward and adjoint trajectories and background term derivative as snapshots for POD basis generation. If only forward model information is included into the reduced manifold the cost function associated with the data assimilation problem decreases by only five orders of magnitude during the optimization process. Taking into account the adjoint and background term derivative information leads to a decrease of the cost function by twenty orders of magnitude and the results of the reduced-order data assimilation system are similar with the ones obtained with the Full order SWE 4D-Var DA system. This highlights the importance of choosing appropriate reduced-order bases.

A numerical study of how the choice of reduced order technique impacts the solution of the inverse problem is performed. We consider for comparison standard POD, tensorial POD and standard POD/DEIM.  For the first time POD/DEIM is employed to construct a reduced-order data assimilation system for a geophysical two-dimensional flow model. All reduced-order DA systems employ a Galerkin projection and the reduced-order bases use information from both forward and dual solutions and the background term derivative. The POD/DEIM approximations of several nonlinear terms involving the height field partially lose their accuracy during the optimization. It suggests that POD/DEIM reduced nonlinear terms are sensitive to input data changes and the selection of interpolation points is no longer optimal. On-going research focuses on increasing the robustness of DEIM for optimization applications. The number of DEIM points must be taken closer to the number of space points for accurate sub-optimal solutions leading to slower on-line stage. The reduced POD/DEIM approximations of the aforementioned nonlinear terms are replaced with tensorial POD representations. This new hybrid POD/DEIM SWE 4D-Var DA system is accurate and faster than other standard and tensorial POD SWE 4D-Var systems. Numerical experiments with various POD basis dimensions and numbers of DEIM points illustrate the potential of the new reduced-order data assimilation system to reduce CPU time.

For a full system spatial discretization with $151 \times 111$ grid points the hybrid POD/DEIM reduced data assimilation system is approximately ten times faster then the full space data assimilation system. This rate increases in proportion to the increase in the number of grid points used in the space discretization. Hybrid POD/DEIM SWE 4D-Var is at least two times faster than standard POD SWE 4D-Var for numbers of space points larger or equal to $101\times 71$. This illustrates the power of DEIM approach not only for reduced-order forward simulations but also for reduced-order optimization.

Our results reveal a relationship between the size of the POD basis and the magnitude of the cost function error criterion $\varepsilon_3$. For a very small $\varepsilon_3$ the reduced order data assimilation system may not able to sufficiently decrease the cost function. The optimization stops only when the maximum number of outer loops is reached or the high-fidelity gradient based optimality condition is satisfied . In consequence, one must carefully select $\varepsilon_3$ since the ROM DA machinery is more efficient when the number of outer loops is kept small.  In addition, the number of function evaluations allowed during the inner minimization phase should be increased with the decrease of $\varepsilon_3$ and increase of POD basis dimension in order to speed up the reduced optimization systems.

Future work will consider Petrov-Galerkin stabilization approaches \cite{Bui-Thanh2007,Amsallem2012}. Moreover, we will focus on a generalized DEIM framework \cite{Stefanescu_Sandu_SMDEIM_2014} to approximate operators since faster reduced Jacobian computations will further decrease the computational complexity of POD/DEIM reduced data assimilation systems. We will also address the impact of snapshots scaling in the accuracy of the sub-optimal solution. One approach would be to normalize each snapshot and to use vectors of norm one as input for the singular value decompositions.
 
We intend to extend our reduced-order data assimilation systems by implementing a trust region algorithm to guide the re-computation of the bases.  On-going work of the authors seeks to develop a-priori and a-posteriori error estimates for the reduced-order optimal solutions, and to use a-posteriori error estimation apparatus to guide the POD basis construction and to efficiently select the number of DEIM interpolation points.

{\centering
\section*{Acknowledgments}
}
The work of Dr. R\u azvan \c{S}tef\u{a}nescu and Prof. Adrian Sandu was supported by the NSF CCF--1218454, AFOSR FA9550--12--1--0293--DEF, AFOSR 12-2640-06,  and by the Computational Science Laboratory at Virginia Tech. Prof. I.M. Navon acknowledges the support of NSF grant ATM-0931198. R\u azvan \c{S}tef\u{a}nescu  thanks Prof. Traian Iliescu for his valuable suggestions on the current research topic, and Vishwas Rao for useful conversations about optimization error estimation.

\newpage
\bibliographystyle{plainnat}
\bibliography{Bib/Software,Bib/ROM_state_of_the_art,Bib/POD_bib,Bib/CDS_E_proposal,Bib/sandu,Bib/comprehensive_bibliography1,Bib/Razvan_bib,Bib/Razvan_bib_ROM_IP,Bib/NSF_KB,Bib/data_assim_weak-fdvar,Bib/reduced_models,Bib/data_assim_fdvar,Bib/optimization}

\begin{thebibliography}{101}
\providecommand{\natexlab}[1]{#1}
\providecommand{\url}[1]{\texttt{#1}}
\expandafter\ifx\csname urlstyle\endcsname\relax
  \providecommand{\doi}[1]{doi: #1}\else
  \providecommand{\doi}{doi: \begingroup \urlstyle{rm}\Url}\fi

\bibitem[A. et~al.(2014)A., Iliescu, John, and Schyschlowa]{Caiazzo2014}
Caiazzo A., T.~Iliescu, V.~John, and S.~Schyschlowa.
\newblock A numerical investigation of velocity–pressure reduced order models
  for incompressible flows.
\newblock \emph{Journal of Computational Physics}, 259\penalty0 (0):\penalty0
  598 -- 616, 2014.

\bibitem[Afanasiev and Hinze(2001)]{Afanasiev_Hinze_2001}
K.~Afanasiev and M.~Hinze.
\newblock Adaptive {C}ontrol of a {W}ake {F}low {U}sing {P}roper {O}rthogonal
  {D}ecomposition.
\newblock \emph{Lecture Notes in Pure and Applied Mathematics}, 216:\penalty0
  317--332, 2001.

\bibitem[Alexe(2011)]{Alexe-PhD}
M.~Alexe.
\newblock \emph{Adjoint-based space-time adaptive solution algorithms for
  sensitivity analysis and inverse problems}.
\newblock PhD thesis, Computer Science Department, Virginia Tech, 2011.

\bibitem[Alexe and Sandu(2014)]{alexe2014space}
Mihai Alexe and Adrian Sandu.
\newblock Space--time adaptive solution of inverse problems with the discrete
  adjoint method.
\newblock \emph{Journal of Computational Physics}, 270:\penalty0 21--39, 2014.

\bibitem[Altaf et~al.(2013)Altaf, Gharamti, Heemink, and Hoteit]{Altaf2013}
M.U. Altaf, M.E. Gharamti, A.W. Heemink, and I.~Hoteit.
\newblock A reduced adjoint approach to variational data assimilation.
\newblock \emph{Computer Methods in Applied Mechanics and Engineering},
  254:\penalty0 1--13, 2013.

\bibitem[Ambrozic(2013)]{Ambrozic2013}
M.~Ambrozic.
\newblock {A Study of Reduced Order 4D-VAR with a Finite Element Shallow Water
  Model}.
\newblock Master's thesis, Delft University of Technology, Netherlands, 2013.

\bibitem[Amsallem and Farhat(2012)]{Amsallem2012}
D.~Amsallem and C.~Farhat.
\newblock Stabilization of projection-based reduced-order models.
\newblock \emph{Int. J. Numer. Meth. Engng.}, 91:\penalty0 358--377, 2012.

\bibitem[Amsallem et~al.(2013)Amsallem, Zahr, Choi, and
  Farhat]{Amsallem_et_al2013}
D.~Amsallem, M.~Zahr, Y.~Choi, and C.~Farhat.
\newblock {Design Optimization Using Hyper-Reduced-Order Models}.
\newblock Technical report, Stanford University, 2013.

\bibitem[Arian et~al.(2000{\natexlab{a}})Arian, Fahl, and Sachs]{Arian_2000}
E.~Arian, M.~Fahl, and E.W. Sachs.
\newblock {T}rust-region proper orthogonal decomposition for flow control.
\newblock \emph{ICASE: Technical Report 2000-25}, 2000{\natexlab{a}}.

\bibitem[Arian et~al.(2000{\natexlab{b}})Arian, Fahl, and
  Sachs]{arian2000trust}
E.~Arian, M.~Fahl, and E.W. Sachs.
\newblock {Trust-region proper orthogonal decomposition for flow control}.
\newblock Institute for Computer Applications in Science and Engineering,
  Hampton VA, 2000{\natexlab{b}}.

\bibitem[Atwell and King(2001)]{Atwell_King2001}
J.A. Atwell and B.B. King.
\newblock {Proper orthogonal decomposition for reduced basis feedback
  controllers for parabolic equations}.
\newblock \emph{Mathematical and Computer Modelling}, 33\penalty0
  (1--3):\penalty0 1--19, 2001.

\bibitem[Atwell and King(2004)]{Atwell_King2004}
J.A. Atwell and B.B. King.
\newblock {Reduced Order Controllers for Spatially Distributed Systems via
  Proper Orthogonal Decomposition}.
\newblock \emph{SIAM J. Sci. Comput.}, 26\penalty0 (1):\penalty0 128--151,
  2004.

\bibitem[Barrault et~al.(2004)Barrault, Maday, Nguyen, and Patera]{BMN2004}
M.~Barrault, Y.~Maday, N.C. Nguyen, and A.T. Patera.
\newblock An 'empirical interpolation' method: application to efficient
  reduced-basis discretization of partial differential equations.
\newblock \emph{Compt. Rend. Math.}, 339\penalty0 (9):\penalty0 667--672, 2004.

\bibitem[Barrett et~al.(1994)Barrett, Berry, Chan, Demmel, Donato, Dongarra,
  Eijkhout, Pozo, Romine, and der Vorst]{Barrett94}
R.~Barrett, M.~Berry, T.~F. Chan, J.~Demmel, J.~Donato, J.~Dongarra,
  V.~Eijkhout, R.~Pozo, C.~Romine, and H.~Van der Vorst.
\newblock \emph{Templates for the Solution of Linear Systems: Building Blocks
  for Iterative Methods, 2nd Edition}.
\newblock SIAM, Philadelphia, PA, 1994.

\bibitem[Baumann(2013)]{Baumann2013}
M.M. Baumann.
\newblock {Nonlinear Model Order Reduction using POD/DEIM for Optimal Control
  of Burgers equation}.
\newblock Master's thesis, Delft University of Technology, Netherlands, 2013.

\bibitem[Becker and Vexler(2005)]{RBecker_BVexler_2005a}
R.~Becker and B.~Vexler.
\newblock Mesh refinement and numerical sensitivity analysis for parameter
  calibration of partial differential equations.
\newblock \emph{J. Comput. Phys.}, 206\penalty0 (1):\penalty0 95--110, 2005.

\bibitem[Bergmann and Cordier(2007)]{Bergmann_2007}
M.~Bergmann and L.~Cordier.
\newblock Drag minimization of the cylinder wake by trust-region proper
  orthogonal decomposition.
\newblock \emph{Notes on Numerical Fluid Mechanics and Multidisciplinary
  Design}, 95\penalty0 (16):\penalty0 309--324, 2007.

\bibitem[Bergmann and Cordier(2008)]{bergmann2008optimal}
M.~Bergmann and L.~Cordier.
\newblock {Optimal control of the cylinder wake in the laminar regime by
  trust-region methods and pod reduced-order models}.
\newblock \emph{Journal of Computational Physics}, 227\penalty0 (16):\penalty0
  7813--7840, 2008.

\bibitem[Bergmann et~al.(2005)Bergmann, Cordier, and
  Brancher]{Bergmann_Cordier_2005}
M.~Bergmann, L.~Cordier, and J.P. Brancher.
\newblock {Optimal rotary control of the cylinder wake using Proper Orthogonal
  Decomposition reduced-order model}.
\newblock \emph{Physics of Fluids}, 17\penalty0 (9):\penalty0 097101, 2005.

\bibitem[Bergmann et~al.(2009)Bergmann, Bruneau, and Iollo]{Bergmann2009}
M.~Bergmann, C.H. Bruneau, and A.~Iollo.
\newblock {Enablers for robust POD models}.
\newblock \emph{Journal of Computational Physics}, 228\penalty0 (2):\penalty0
  516 -- 538, 2009.

\bibitem[BROYDEN(1970)]{BROYDEN01031970}
C.~G. BROYDEN.
\newblock {The Convergence of a Class of Double-rank Minimization Algorithms 1.
  General Considerations}.
\newblock \emph{IMA Journal of Applied Mathematics}, 6\penalty0 (1):\penalty0
  76--90, 1970.

\bibitem[Bui-Thanh et~al.(2007)Bui-Thanh, Willcox, Ghattas, and van
  Bloemen~Waanders]{Bui-Thanh2007}
T.~Bui-Thanh, K.~Willcox, O.~Ghattas, and B.~van Bloemen~Waanders.
\newblock Goal-oriented, model-constrained optimization for reduction of
  large-scale systems.
\newblock \emph{J. Comput. Phys.}, 224\penalty0 (2):\penalty0 880--896, 2007.

\bibitem[Cao et~al.(2007)Cao, Zhu, Navon, and Luo]{Cao_Zhu_2007}
Y.~Cao, J.~Zhu, I.M. Navon, and Z.~Luo.
\newblock A reduced order approach to four-dimensional variational data
  assimilation using proper orthogonal decomposition.
\newblock \emph{Int. J. Numer. Meth. Fluids.}, 53\penalty0 (10):\penalty0
  1571--1583, 2007.

\bibitem[Carlberg and Farhat(2011)]{Carlberg_2011}
K.~Carlberg and C.~Farhat.
\newblock A low-cost, goal-oriented compact proper orthogonal decomposition
  basis for model reduction of static systems.
\newblock \emph{International Journal for Numerical Methods in Engineering},
  86\penalty0 (3):\penalty0 381--402, 2011.

\bibitem[Carlberg et~al.(2011)Carlberg, Bou-Mosleh, and Farhat]{Carlberg2_2011}
K.~Carlberg, C.~Bou-Mosleh, and C.~Farhat.
\newblock Efficient non-linear model reduction via a least-squares
  {P}etrov-–{G}alerkin projection and compressive tensor approximations.
\newblock \emph{International Journal for Numerical Methods in Engineering},
  86\penalty0 (2):\penalty0 155--181, 2011.

\bibitem[Chaturantabut(2008)]{Cha2008}
S.~Chaturantabut.
\newblock Dimension {R}eduction for {U}nsteady {N}onlinear {P}artial
  {D}ifferential {E}quations via {E}mpirical {I}nterpolation {M}ethods.
\newblock \emph{TR09-38, CAAM, Rice University}, 2008.

\bibitem[Chaturantabut and Sorensen(2010)]{ChaSor2010}
S.~Chaturantabut and D.C. Sorensen.
\newblock Nonlinear model reduction via discrete empirical interpolation.
\newblock \emph{SIAM J. Sci. Comput.}, 32\penalty0 (5):\penalty0 2737--2764,
  2010.

\bibitem[Chaturantabut and Sorensen(2012)]{ChaSor2012}
S.~Chaturantabut and D.C. Sorensen.
\newblock A state space error estimate for {POD-DEIM} nonlinear model
  reduction.
\newblock \emph{SIAM J. Numer. Anal.}, 50\penalty0 (1):\penalty0 46--63, 2012.

\bibitem[Chen et~al.(2012)Chen, Akella, and Navon]{Chen2012}
X.~Chen, S.~Akella, and I.~M. Navon.
\newblock A dual weighted trust-region adaptive {POD} 4{D}-{V}ar applied to a
  {F}inite-{V}olume shallow-water {E}quations {M}odel on the sphere.
\newblock \emph{Int. J. Numer. Meth. Fluids}, 68:\penalty0 377--402, 2012.

\bibitem[Chen et~al.(2011)Chen, Navon, and Fang]{Chen2011}
Xiao Chen, I.~M. Navon, and F.~Fang.
\newblock A dual weighted trust-region adaptive {POD} 4{D}-{V}ar applied to a
  {F}inite -{E}lement {S}hallow water {E}quations {M}odel.
\newblock \emph{Int. J. Numer. Meth. Fluids}, 68:\penalty0 520--541, 2011.

\bibitem[Cohn(1997)]{Cohn1997}
S.E. Cohn.
\newblock {An introduction to estimation theory}.
\newblock \emph{Journal of the Meteorological Society of Japan}, 75\penalty0
  (B):\penalty0 257--288, 1997.

\bibitem[\c{S}tef\u{a}nescu and Navon(2013)]{Stefanescu2012}
R.~\c{S}tef\u{a}nescu and I.M. Navon.
\newblock {POD/DEIM} {N}onlinear model order reduction of an {ADI} implicit
  shallow water equations model.
\newblock \emph{J. Comput. Phys.}, 237:\penalty0 95--114, 2013.

\bibitem[\c{S}tef\u{a}nescu and Pogan(2013)]{Stefanescu_Pogan2013}
R.~\c{S}tef\u{a}nescu and M.C. Pogan.
\newblock {Optimal Control in Chemotherapy of a Viral Infection}.
\newblock \emph{Annals of the Alexandru Ioan Cuza University - Mathematics},
  59\penalty0 (2):\penalty0 321--338, 2013.

\bibitem[\c{S}tef\u{a}nescu and Sandu(2014, submitted to International Journal
  for Numerical Methods in Engineering)]{Stefanescu_Sandu_SMDEIM_2014}
R.~\c{S}tef\u{a}nescu and A.~Sandu.
\newblock {Efficient Approximation of Sparse Jacobians for Time-Implicit
  Reduced Order Models}.
\newblock Technical Report TR 15, Virginia Polytechnic Institute and State
  University, September 2014, submitted to International Journal for Numerical
  Methods in Engineering.

\bibitem[\c{S}tef\u{a}nescu et~al.(2014)\c{S}tef\u{a}nescu, Sandu, and
  Navon]{Stefanescu_etal_forwardPOD_2014}
R.~\c{S}tef\u{a}nescu, A.~Sandu, and I.M. Navon.
\newblock {Comparison of POD Reduced Order Strategies for the Nonlinear 2D
  Shallow Water Equations}.
\newblock \emph{International Journal for Numerical Methods in Fluids},
  76\penalty0 (8):\penalty0 497--521, 2014.

\bibitem[Daescu and Navon(2007)]{Daescu_Navon2007}
D.N. Daescu and I.M. Navon.
\newblock Efficiency of a {POD}-based reduced second order adjoint model in
  4-{D} {VAR} data assimilation.
\newblock \emph{Int. J. Numer. Meth. Fluids.}, 53:\penalty0 985--1004, 2007.

\bibitem[Daescu and Navon(2008)]{Daescu_Navon2_2008}
D.N. Daescu and I.M. Navon.
\newblock A {D}ual-{W}eighted {A}pproach to {O}rder {R}eduction in 4{D-V}ar
  {D}ata {A}ssimilation.
\newblock \emph{Mon. Wea. Rev.}, 136\penalty0 (3):\penalty0 1026--1041, 2008.

\bibitem[Dihlmann and Haasdonk(2013)]{Dihlmann_2013}
M.~Dihlmann and B.~Haasdonk.
\newblock Certified {PDE}-constrained parameter optimization using reduced
  basis surrogate models for evolution problems.
\newblock \emph{Submitted to the Journal of Computational Optimization and
  Applications}, 2013.
\newblock URL
  \url{http://www.agh.ians.uni-stuttgart.de/publications/2013/DH13}.

\bibitem[Dimitriu et~al.(2010)Dimitriu, Apreutesei, and
  Stefanescu]{Dimitriu_2010}
G.~Dimitriu, N.~Apreutesei, and R.~Stefanescu.
\newblock Numerical simulations with data assimilation using an adaptive {POD}
  procedure.
\newblock \emph{LNCS}, 5910:\penalty0 165--172, 2010.

\bibitem[Diwoky and Volkwein(2001)]{Diwoky_Volkwein2001}
F.~Diwoky and S.~Volkwein.
\newblock Nonlinear boundary control for the heat equation utilizing proper
  orthogonal decomposition.
\newblock In Karl-Heinz Hoffmann, Ronald~H.W. Hoppe, and Volker Schulz,
  editors, \emph{Fast Solution of Discretized Optimization Problems}, volume
  138 of \emph{ISNM International Series of Numerical Mathematics}, pages
  73--87. Birkh\"{a}user Basel, 2001.
\newblock ISBN 978-3-0348-9484-5.

\bibitem[Du et~al.(2013)Du, Navon, Zhu, Fang, and Alekseev]{Du_parab2_2012}
J.~Du, I.M. Navon, J.~Zhu, F.~Fang, and A.K. Alekseev.
\newblock Reduced order modeling based on {POD} of a parabolized
  {N}avier-{S}tokes equations model {II}: {T}rust region {POD} 4-{D} {VAR D}ata
  {A}ssimilation.
\newblock \emph{Computers and Mathematics with Applications}, 65:\penalty0
  380--–394, 2013.

\bibitem[Fairweather and Navon(1980)]{FN1980}
G.~Fairweather and I.M. Navon.
\newblock A linear {ADI} method for the shallow water equations.
\newblock \emph{Journal of Computational Physics}, 37:\penalty0 1--18, 1980.

\bibitem[Fang et~al.(2009)Fang, Pain, Navon, Piggott, Gorman, Farrell, Allison,
  and Goddard]{Fang_Pain_Navon2009}
F.~Fang, C.C. Pain, I.M. Navon, M.D. Piggott, G.J. Gorman, P.~E. Farrell,
  P.~Allison, and A.J.H. Goddard.
\newblock A {POD} reduced order 4{D}-{V}ar adaptive mesh ocean modelling
  approach.
\newblock \emph{Int. J. Numer. Meth. Fluids.}, 60\penalty0 (7):\penalty0
  709--732, 2009.

\bibitem[Fletcher(1970)]{Fletcher01011970}
R.~Fletcher.
\newblock A new approach to variable metric algorithms.
\newblock \emph{The Computer Journal}, 13\penalty0 (3):\penalty0 317--322,
  1970.

\bibitem[Goldfarb(1970)]{goldfarb1970family}
Donald Goldfarb.
\newblock A family of variable-metric methods derived by variational means.
\newblock \emph{Mathematics of computation}, 24\penalty0 (109):\penalty0
  23--26, 1970.

\bibitem[Grammeltvedt(1969)]{Gram1969}
A.~Grammeltvedt.
\newblock {A survey of finite difference schemes for the primitive equations
  for a barotropic fluid}.
\newblock \emph{Monthly Weather Review}, 97\penalty0 (5):\penalty0 384--404,
  1969.

\bibitem[Grepl and Patera(2005)]{grepl2005posteriori}
M.A. Grepl and A.T. Patera.
\newblock A posteriori error bounds for reduced-basis approximations of
  parametrized parabolic partial differential equations.
\newblock \emph{ESAIM: Mathematical Modelling and Numerical Analysis},
  39\penalty0 (01):\penalty0 157--181, 2005.

\bibitem[Gubisch and Volkwein(2013)]{Gubisch_Volkwein2013}
M.~Gubisch and S.~Volkwein.
\newblock {Proper Orthogonal Decomposition for Linear-Quadratic Optimal
  Control}.
\newblock Technical report, University of Konstanz, 2013.

\bibitem[Gustafsson(1971)]{Gus1971}
B.~Gustafsson.
\newblock {An alternating direction implicit method for solving the shallow
  water equations}.
\newblock \emph{Journal of Computational Physics}, 7:\penalty0 239--254, 1971.

\bibitem[Hay et~al.(2009)Hay, Borggaard, and
  Pelletier]{Hay_Borggaard_Pelletier_2009}
A.~Hay, J.T. Borggaard, and D.~Pelletier.
\newblock {Local improvements to reduced-order models using sensitivity
  analysis of the proper orthogonal decomposition}.
\newblock \emph{Journal of Fluid Mechanics,}, 629:\penalty0 41--72, 2009.

\bibitem[Hinze(2011)]{Hinze_adapt_2011}
M.~Hinze.
\newblock {Adaptive concepts in reduced order modeling with emphasis on PDE
  constrained optimization}, Manchester, July 2011.
\newblock URL
  \url{http://www.maths.manchester.ac.uk/~chahlaoui/AMR11/MichaelH.pdf}.

\bibitem[Hinze and Volkwein(2005)]{Hinze_2005}
M.~Hinze and S.~Volkwein.
\newblock Proper orthogonal decomposition surrogate models for nonlinear
  dynamical systems: Error estimates and suboptimal control.
\newblock \emph{Lecture Notes in Computational Science and Engineering},
  45:\penalty0 261--306, 2005.

\bibitem[Hinze and Volkwein(2008{\natexlab{a}})]{Hinze_Volkwein_2008}
M.~Hinze and S.~Volkwein.
\newblock Error {E}stimates for {A}bstract {L}inear-{Q}uadratic {O}ptimal
  {C}ontrol {P}roblems {U}sing {P}roper {O}rthogonal {D}ecomposition.
\newblock \emph{Computational Optimization and Applications}, 39\penalty0
  (3):\penalty0 319--345, 2008{\natexlab{a}}.

\bibitem[Hinze and Volkwein(2008{\natexlab{b}})]{Hinze_Wolkwein2008}
M.~Hinze and S.~Volkwein.
\newblock Error estimates for abstract linear-quadratic optimal control
  problems using proper orthogonal decomposition.
\newblock \emph{Computational Optimization and Applications}, 39:\penalty0
  319--345, 2008{\natexlab{b}}.

\bibitem[Hotelling(1933)]{hotelling1939acs}
H.~Hotelling.
\newblock Analysis of a complex of statistical variables with principal
  components.
\newblock \emph{Journal of Educational Psychology}, 24:\penalty0 417--441,
  1933.

\bibitem[Ito and Kunisch(2006)]{Ito_Kunisch_2006}
K.~Ito and K.~Kunisch.
\newblock Reduced {O}rder {C}ontrol {B}ased on {A}pproximate {I}nertial
  {M}anifolds.
\newblock \emph{Linear Algebra and its Applications}, 415\penalty0
  (2-3):\penalty0 531--541, 2006.

\bibitem[Ito and Kunisch(2008)]{Ito_Kunisch_2008}
K.~Ito and K.~Kunisch.
\newblock Reduced-{O}rder {O}ptimal {C}ontrol {B}ased on {A}pproximate
  {I}nertial {M}anifolds for {N}onlinear {D}ynamical {S}ystems.
\newblock \emph{SIAM Journal on Numerical Analysis}, 46\penalty0 (6):\penalty0
  2867--2891, 2008.

\bibitem[Kahlbacher and Volkwein(2012)]{Kahlbacher_Volkwein_2012}
M.~Kahlbacher and S.~Volkwein.
\newblock {POD} a-{P}osteriori {E}rror {B}ased {I}nexact {SQP} {M}ethod for
  {B}ilinear {E}lliptic {O}ptimal {C}ontrol {P}roblems.
\newblock \emph{ESAIM-Mathematical Modelling and Numerical
  Analysis-Modelisation Mathematique et Analyse Numerique}, 46\penalty0
  (2):\penalty0 491--511, 2012.

\bibitem[Kammann et~al.(2013)Kammann, Tr\"{o}ltzsch, and
  Volkwein]{Kammannetal2013}
E.~Kammann, F.~Tr\"{o}ltzsch, and S.~Volkwein.
\newblock {A method of a-posteriori error estimation with application to proper
  orthogonal decomposition}.
\newblock \emph{ESAIM: Mathematical Modelling and Numerical Analysis},
  47:\penalty0 555--581, 2013.

\bibitem[Karhunen(1946)]{karhunen1946zss}
K.~Karhunen.
\newblock Zur spektraltheorie stochastischer prozesse.
\newblock \emph{Annales Academiae Scientarum Fennicae}, 37, 1946.

\bibitem[Karush(1939)]{Karush1939}
W.~Karush.
\newblock {Minima of Functions of Several Variables with Inequalities as Side
  Constraints}.
\newblock Technical Report TR 3, M.Sc. Dissertation, Dept. of Mathematics,Univ.
  of Chicago, Illinois, March 1939.

\bibitem[Kelley(1995)]{Kelley95}
C.~T. Kelley.
\newblock \emph{Iterative Methods for Linear and Nonlinear Equations}.
\newblock Number~16 in Frontiers in Applied Mathematics. SIAM, 1995.

\bibitem[Kuhn and Tucker(1951)]{KT1951}
H.W. Kuhn and A.W. Tucker.
\newblock Nonlinear programming.
\newblock In \emph{{Proc. of the Second Berkeley Symposium on Mathematical
  Statistics and Probability}}, pages 481--492, Berkeley and Los Angeles, 1951.
  University of California Press.

\bibitem[Kunisch and Volkwein(1999)]{Kunisch_Volkwein_1999}
K.~Kunisch and S.~Volkwein.
\newblock Control of the {B}urgers {E}quation by a {R}educed-{O}rder {A}pproach
  {U}sing {P}roper {O}rthogonal {D}ecomposition.
\newblock \emph{Journal of Optimization Theory and Applications}, 102\penalty0
  (2):\penalty0 345--371, 1999.

\bibitem[Kunisch and Volkwein(2008)]{Kunisch_2008}
K.~Kunisch and S.~Volkwein.
\newblock {P}roper {O}rthogonal {D}ecomposition for {O}ptimality {S}ystems.
\newblock \emph{Math. Modelling and Num. Analysis}, 42:\penalty0 1--23, 2008.

\bibitem[Kunisch and Volkwein(2010)]{Kunisch_Volkwein2010}
K.~Kunisch and S.~Volkwein.
\newblock Optimal snapshot location for computing {POD} basis functions.
\newblock \emph{ESAIM: Math. Model. Numer. Anal.}, M2AN 44\penalty0
  (3):\penalty0 509--529, 2010.

\bibitem[Kunisch and Xie(2005)]{Kunisch_Xie_2005}
K.~Kunisch and L.~Xie.
\newblock {POD}-{B}ased {F}eedback {C}ontrol of the {B}urgers {E}quation by
  {S}olving the {E}volutionary {HJB} {E}quation.
\newblock \emph{Computers and Mathematics with Applications}, 49\penalty0
  (7-8):\penalty0 1113--1126, 2005.

\bibitem[Kunisch et~al.(2004)Kunisch, Volkwein, and Xie]{Kunisch_Volkwein_2004}
K.~Kunisch, S.~Volkwein, and L.~Xie.
\newblock {HJB}-{POD}-{B}ased {F}eedback {D}esign for the {O}ptimal {C}ontrol
  of {E}volution {P}roblems.
\newblock \emph{SIAM J. Appl. Dyn. Syst}, 3\penalty0 (4):\penalty0 701--722,
  2004.

\bibitem[Lass and Volkwein(2012)]{Lass_Volkwein_adapt_2012}
O.~Lass and S.~Volkwein.
\newblock {Adaptive POD basis computation for parameterized nonlinear systems
  using optimal snapshot location}.
\newblock \emph{Konstanzer Schriften Math.}, 304:\penalty0 1--27, 2012.

\bibitem[Lassila and Rozza(2010)]{Lassila_Rozza_2010}
T.~Lassila and G.~Rozza.
\newblock {Parametric free-form shape design with pde models and reduced basis
  method}.
\newblock \emph{Computer Methods in Applied Mechanics and Engineering},
  199\penalty0 (23):\penalty0 1583--1592, 2010.

\bibitem[Leibfritz and Volkwein(2008)]{Leibfritz_Volkwein_2006}
F.~Leibfritz and S.~Volkwein.
\newblock Reduced {O}rder {O}utput {F}eedback {C}ontrol {D}esign for {PDE}
  {S}ystems {U}sing {P}roper {O}rthogonal {D}ecomposition and {N}onlinear
  {S}emidefinite {P}rogramming.
\newblock \emph{Linear Algebra and its Applications}, 415\penalty0
  (2--3):\penalty0 542--575, 2008.

\bibitem[Lo\`eve(1955)]{loeve1955pt}
M.M. Lo\`eve.
\newblock \emph{Probability Theory}.
\newblock Van Nostrand, Princeton, NJ, 1955.

\bibitem[Lorenz(1956)]{lorenz1956eof}
E.N. Lorenz.
\newblock {Empirical Orthogonal Functions and Statistical Weather Prediction}.
\newblock Technical report, Massachusetts Institute of Technology, Dept. of
  Meteorology, 1956.

\bibitem[Manzoni et~al.(2012)Manzoni, Quarteroni, and
  Rozza]{Manzoni_Quarteroni_Rozza_2012}
A.~Manzoni, A.~Quarteroni, and G.~Rozza.
\newblock {Shape optimization for viscous flows by reduced basis methods and
  free-form deformation}.
\newblock \emph{International Journal for Numerical Methods in Fluids,},
  70\penalty0 (5):\penalty0 646--670, 2012.

\bibitem[Navon and Villiers(1986)]{NVG1986}
I.~M. Navon and R.~De Villiers.
\newblock { GUSTAF: A {Q}uasi-{N}ewton Nonlinear {ADI} FORTRAN IV Program for
  Solving the Shallow-Water Equations with Augmented Lagrangians}.
\newblock \emph{Computers and Geosciences}, 12\penalty0 (2):\penalty0 151--173,
  1986.

\bibitem[Navon et~al.(1992)Navon, Zou, Derber, and
  Sela]{Navon_Zou_Derber_Sela_1992}
I.M. Navon, X.~Zou, J.~Derber, and J.~Sela.
\newblock Variational data assimilation with an adiabatic version of the nmc
  spectral model.
\newblock \emph{Mon. Wea. Rev.}, 120:\penalty0 1433–--1446, 1992.

\bibitem[Noack et~al.(2005)Noack, Papas, and
  Monkewitz]{Noack_Papas_Monkewitz_2005}
B.R. Noack, P.~Papas, and P.A. Monkewitz.
\newblock The need for a pressure-term representation in empirical galerkin
  models of incompressible shear flows.
\newblock \emph{Journal of Fluid Mechanics}, 523:\penalty0 339--365, 1 2005.
\newblock ISSN 1469-7645.

\bibitem[Patera and Rozza(2007)]{patera2007reduced}
A.T. Patera and G.~Rozza.
\newblock Reduced basis approximation and a posteriori error estimation for
  parametrized partial differential equations, 2007.

\bibitem[Peherstorfer et~al.(2013)Peherstorfer, Butnaru, Willcox, and
  Bungartz]{Peherstorfer_2013}
B.~Peherstorfer, D.~Butnaru, K.~Willcox, and H.J. Bungartz.
\newblock { Localized Discrete Empirical Interpolation Method}.
\newblock MIT Aerospace Computational Design Laboratory Technical Report
  TR-13-1, 2013.

\bibitem[Pelc et~al.(2012)Pelc, Simon, Bertino, Serafy, and Heemink]{Pelc2012}
Joanna~S. Pelc, Ehouarn Simon, Laurent Bertino, Ghada~El Serafy, and Arnold~W.
  Heemink.
\newblock {Application of model reduced 4D-Var to a 1D ecosystem model}.
\newblock \emph{Ocean Modelling}, 57--58\penalty0 (0):\penalty0 43 --58, 2012.
\newblock ISSN 1463-5003.

\bibitem[Rao and Sandu(2014)]{Rao_Sandu_CSL_TR_16_2014}
V.~Rao and A.~Sandu.
\newblock {A-posteriori error estimates for inverse problems}.
\newblock Technical Report TR 16, Virginia Polytechnic Institute and State
  University, March 2014.

\bibitem[Rap\'{u}n and Vega(2010)]{Rapun_2010}
M.L. Rap\'{u}n and J.M. Vega.
\newblock Reduced order models based on local {POD} plus {G}alerkin projection.
\newblock \emph{J. Comput. Phys.}, 229\penalty0 (8):\penalty0 3046--3063, 2010.

\bibitem[Ravindran(2000)]{Ravindran_2000}
S.S. Ravindran.
\newblock {Reduced-Order Adaptive Controllers for Fluid Flows Using POD}.
\newblock \emph{Journal of Scientific Computing}, 15\penalty0 (4):\penalty0
  457--478, 2000.

\bibitem[Ravindran(2002)]{Ravindran_2002}
S.S. Ravindran.
\newblock Adaptive {R}educed-{O}rder {C}ontrollers for a {T}hermal {F}low
  {S}ystem {U}sing {P}roper {O}rthogonal {D}ecomposition.
\newblock \emph{J. Sci. Comput.}, 23:\penalty0 1924--1942, 2002.

\bibitem[Rozza and Manzoni(2010)]{Rozza_Manzoni_2010}
G.~Rozza and A.~Manzoni.
\newblock Model order reduction by geometrical parametrization for shape
  optimization in computational fluid dynamics.
\newblock In J.C.F. Pereira and A.~Sequeira, editors, \emph{ECCOMAS CFD 2010, V
  European Conference on Computational Fluid Dynamics, Lisbon, Portugal}, 2010.

\bibitem[Rozza et~al.(2008)Rozza, Huynh, and Patera]{rozza2008reduced}
G.~Rozza, D.B.P. Huynh, and A.T. Patera.
\newblock Reduced basis approximation and a posteriori error estimation for
  affinely parametrized elliptic coercive partial differential equations.
\newblock \emph{Archives of Computational Methods in Engineering}, 15\penalty0
  (3):\penalty0 229--275, 2008.

\bibitem[Saad(1994)]{Saad1994}
Y.~Saad.
\newblock {Sparsekit: a basic tool kit for sparse matrix computations}.
\newblock {Technical Report, Computer Science Department, University of
  Minnesota}, 1994.

\bibitem[Saad(2003)]{Saad2003}
Y.~Saad.
\newblock \emph{Iterative Methods for Sparse Linear Systems}.
\newblock Society for Industrial and Applied Mathematics, Philadelphia, PA,
  USA, 2nd edition, 2003.

\bibitem[Sachs and Volkwein(2010)]{Sachs_Volkwein_2010}
E.~W. Sachs and S.~Volkwein.
\newblock {POD}-{G}alerkin {A}pproximations in {PDE}-{C}onstrained
  {O}ptimization.
\newblock \emph{GAMM-Mitteilungen}, 33\penalty0 (2):\penalty0 194--208, 2010.

\bibitem[Sava(2012)]{Sava2012}
D.~Sava.
\newblock {Model-Reduced Gradient Based Production Optimization}.
\newblock (M.S) Delft University of Technology, 2012.

\bibitem[Shanno(1970)]{shanno1970conditioning}
David~F Shanno.
\newblock Conditioning of quasi-newton methods for function minimization.
\newblock \emph{Mathematics of computation}, 24\penalty0 (111):\penalty0
  647--656, 1970.

\bibitem[Shanno and Phua(1980)]{Shanno_Phua1980}
D.F. Shanno and K.H. Phua.
\newblock {Remark on algorithm 500 - a variable method subroutine for
  unconstrained nonlinear minimization}.
\newblock \emph{ACM Transactions on Mathematical Software}, 6:\penalty0
  618--622, 1980.

\bibitem[Tonna et~al.(2010)Tonna, Urbana, and Volkwein]{Tonna_Volkwein_2010}
T.~Tonna, K.~Urbana, and S.~Volkwein.
\newblock Comparison of the {R}educed-{B}asis and {POD} a-{P}osteriori {E}rror
  {E}stimators for an {E}lliptic {L}inear-{Q}uadratic {O}ptimal {C}ontrol
  {P}roblem.
\newblock \emph{Mathematical and Computer Modelling of Dynamical Systems:
  Methods, Tools and Applications in Engineering and Related Sciences},
  17\penalty0 (4):\penalty0 355--369, 2010.

\bibitem[Tr\"{o}ltzsch and Volkwein(2009)]{Troltzsch_Volkwein_2010}
F.~Tr\"{o}ltzsch and S.~Volkwein.
\newblock {POD} a-{P}osteriori {E}rror {E}stimates for {L}inear-{Q}uadratic
  {O}ptimal {C}ontrol {P}roblems.
\newblock \emph{Computational Optimization and Applications}, 44\penalty0
  (1):\penalty0 83--115, 2009.

\bibitem[Vermeulen and Heemink(2006)]{Vermeulen2006}
P.T.M. Vermeulen and A.W. Heemink.
\newblock Model-reduced variational data assimilation.
\newblock \emph{Mon. Wea. Rev.}, 134:\penalty0 2888--2899, 2006.

\bibitem[Vreugdenhil(1995)]{VreugdenhilSWE_1995}
C.B. Vreugdenhil.
\newblock \emph{{Numerical Methods for Shallow-Water Flow}}.
\newblock Springer, 1995.

\bibitem[Weller et~al.(2010)Weller, Lombardi, Bergmann, and
  Iollo]{Weller_etal_2010}
J.~Weller, E.~Lombardi, M.~Bergmann, and A.~Iollo.
\newblock {Numerical methods for low-order modeling of fluid flows based on
  POD}.
\newblock \emph{Int. J. Numer. Meth. Fluids}, 63:\penalty0 249 -- 268, 2010.

\bibitem[Willcox and Peraire(2002)]{Willcox02balancedmodel}
K.~Willcox and J.~Peraire.
\newblock Balanced model reduction via the {Proper Orthogonal Decomposition}.
\newblock \emph{AIAA Journal}, pages 2323--2330, 2002.

\bibitem[Yue and Meerbergen(2013)]{Yue_Meerbergen_2013}
Y.~Yue and K.~Meerbergen.
\newblock {Accelerating Optimization of Parametric Linear Systems by Model
  Order Reduction}.
\newblock \emph{SIAM Journal on Optimization}, 23\penalty0 (2):\penalty0
  1344--1370, 2013.

\bibitem[Zahr and Farhat(2014)]{Zahr_Farhat_2014}
M.J. Zahr and C.~Farhat.
\newblock {Progressive construction of a parametric reduced-order model for
  PDE-constrained optimization}.
\newblock \emph{International Journal on Numerical Methods in Engineering},
  (special edition on Model Reduction), accepted for publication, 2014.

\bibitem[Zahr et~al.(2013)Zahr, Amsallem, and
  Farhat]{Zahr_Amsallem_Farhat_2013}
M.J. Zahr, D.~Amsallem, and C.~Farhat.
\newblock Construction of parametrically-robust cfdbased reduced-order models
  for pde-constrained optimization.
\newblock In \emph{{42nd AIAA Fluid Dynamics Conference and Exhibit, (San
  Diego, CA)}}, June 2013.

\end{thebibliography}

\section*{Appendix}

This appendix contains a symbolic representation of the Gustafsson's nonlinear ADI finite difference shallow water equations schemes defined in \eqref{eqn:swe-sd} and the formulas of SWE tensors required by all three studied reduced order models.

\subsection*{Gustafsson's nonlinear ADI finite difference shallow water equations models}

ADI SWE scheme requires two steps to solve for ${\boldsymbol u}^{N+1},\quad {\boldsymbol v}^{N+1}, \quad {\boldsymbol \phi}^{N+1}$.
\paragraph{First step - get solution at $t_{N+\frac{1}{2}}$ }
\begin{equation}\label{eqn:swe-sd_step1}
\begin{split}
 & D_{-t}{\bf u}^{{N+\frac{1}{2}}}  =  -F_{11}({\bf u}^{N+\frac{1}{2}})-F_{12}(\bm{{\phi}}^{N+\frac{1}{2}})-F_{13}({\bf u}^N,{\bf v}^N) + {\bf F}\odot {\bf v}^N, \\
 & D_{-t}{\bf v}^{{N+\frac{1}{2}}}  =  -F_{21}({\bf u}^{N+\frac{1}{2}})-F_{22}({\bf v}^N) -F_{23}(\bm{{\phi}}^N) - {\bf F}\odot {\bf u}^{N+\frac{1}{2}}, \\
 & D_{-t}{\bm{{\phi}}}^{{N+\frac{1}{2}}}  =  -F_{31}({\bf u}^{N+\frac{1}{2}},\bm{{\phi}}^{N+\frac{1}{2}})-F_{32}({\bf u}^{N+\frac{1}{2}},\bm{{\phi}}^{N+\frac{1}{2}})-F_{33}({\bf v}^N,\bm{{\phi}}^N)-F_{34}({\bf v}^N,\bm{{\phi}}^N),
\end{split}
\end{equation}
where $D_{-t}^{{N+\frac{1}{2}}}{\bf w}$ is the backward in time difference operator, ${\bf w} = ({\bf u},{\bf v},{\bm \phi})$, $\odot$ is the component-wise multiplication operator, ${\bf F} =[\underbrace{{\bf f},{\bf f},..,{\bf f}}_{N_x}]$ stores Coriolis components ${\bf f} = [f(y_j)]_{j=1,2,..,N_y}$ while the nonlinear terms $F_{i,j}$ are defined in \eqref{eqn:swe-nonlinear_terms}.

\paragraph{Second step - get solution at $t_{N+1}$ }
\begin{equation}\label{eqn:swe-sd_step2}
\begin{split}
& D_{-t}{\bf u}^{{N+1}}  =  -F_{13}({\bf u}^{N+1},{\bf v}^{N+1}) + {\bf F}\odot {\bf v}^{N+1} -F_{11}({\bf u}^{N+\frac{1}{2}})-F_{12}(\bm{{\phi}}^{N+\frac{1}{2}}), \\
& D_{-t}{\bf v}^{{N+1}}  =  -F_{22}({\bf v}^{N+1}) -F_{23}(\bm{{\phi}}^{N+1}) -F_{21}({\bf u}^{N+\frac{1}{2}}) - {\bf F}\odot {\bf u}^{N+\frac{1}{2}}, \\
&  D_{-t}{\bm{{\phi}}}^{{N+1}}  =  -F_{33}({\bf v}^{N+1},\bm{{\phi}}^{N+1})-F_{34}({\bf v}^{N+1},\bm{{\phi}}^{N+1}) -F_{31}({\bf u}^{N+\frac{1}{2}},\bm{{\phi}}^{N+\frac{1}{2}})-F_{32}({\bf u}^{N+\frac{1}{2}},\bm{{\phi}}^{N+\frac{1}{2}}).
\end{split}
\end{equation}

The nonlinear systems of algebraic equations \eqref{eqn:swe-sd_step1} and \eqref{eqn:swe-sd_step2} are solved using quasi-Newton method, thereby we rewrite them in the form
\begin{equation}\label{eqn:symbolic_ADI_SWE}
{\bf g}({\boldsymbol w}) =0, \quad {\bf g}({\boldsymbol w}) = \left(\begin{array}{c}
{g_1({\boldsymbol w}^{N+\frac{1}{2}},{\boldsymbol w}^{N})} \\
{g_2({\boldsymbol w}^{N+1},{\boldsymbol w}^{N+\frac{1}{2}})} \\
\end{array} \right),
\end{equation}
where ${g_1}$ and ${g_2}$ represent systems \eqref{eqn:swe-sd_step1} and \eqref{eqn:swe-sd_step2}. The corresponding iterative Newton steps are
\begin{equation}\label{eqn:quasi_Newton_steps}
\begin{split}
  & {\delta \boldsymbol w}^{N+\frac{1}{2}} = - \frac{\partial g_1}{\partial w^{N+\frac{1}{2}}}({\boldsymbol w}^{N+\frac{1}{2}},{\boldsymbol w}^{N})^{-1} g_1({\boldsymbol w}^{N+\frac{1}{2}},{\boldsymbol w}^{N}), \\
  & {\delta \boldsymbol w}^{N+1} = - \frac{\partial g_2}{\partial w^{N+1}}({\boldsymbol w}^{N+1},{\boldsymbol w}^{N+\frac{1}{2}})^{-1} g_2({\boldsymbol w}^{N+1},{\boldsymbol w}^{N+\frac{1}{2}}).
\end{split}
\end{equation}
We avoid evaluating the Jacobian matrices at every iteration by proposing a Quasi-Newton approach:
\begin{equation}\label{eqn:quasi_Newton_steps1}
\begin{split}
  & {\delta \boldsymbol w}^{N+\frac{1}{2}} = - \frac{\partial g_1}{\partial w^{N+\frac{1}{2}}}({\boldsymbol w}^{N})^{-1}g_1({\boldsymbol w}^{N+\frac{1}{2}},{\boldsymbol w}^{N}), \\
  & {\delta \boldsymbol w}^{N+1} = - \frac{\partial g_2}{\partial w^{N+1}}({\boldsymbol w}^{N+\frac{1}{2}})^{-1}g_2({\boldsymbol w}^{N+1},{\boldsymbol w}^{N+\frac{1}{2}}).
\end{split}
\end{equation}
Then, we linearize the discrete models using the total variation method \cite[eq. (3.2)]{Stefanescu_Pogan2013} and obtain the tangent linear model by subtracting the set of equations \eqref{eqn:symbolic_ADI_SWE} using different arguments $({\boldsymbol w}^{N+\frac{1}{2}},{\boldsymbol w}^{N})$ and $({\boldsymbol w}^{N+1},{\boldsymbol w}^{N+\frac{1}{2}})$ and their increments $({\boldsymbol w}^{N+\frac{1}{2}}+{\delta \boldsymbol w}^{N+\frac{1}{2}},{\boldsymbol w}^{N}+{\delta \boldsymbol w}^{N})$ and $({\boldsymbol w}^{N+1}+{\delta \boldsymbol w}^{N+1},{\boldsymbol w}^{N+\frac{1}{2}}+{\delta \boldsymbol w}^{N+\frac{1}{2}})$
\begin{equation}\label{eqn:TL}
\left\{
\begin{split}
  & \frac{\partial g_1}{\partial w^{N+\frac{1}{2}}}({\boldsymbol w}^{N}) {\delta  \boldsymbol w}^{N+\frac{1}{2}} = - \frac{\partial g_1}{\partial w^{N}}({\boldsymbol w}^{N}) {\delta  \boldsymbol w}^{N}, \\
  &  \frac{\partial g_2}{\partial w^{N+1}}({\boldsymbol w}^{N+\frac{1}{2}}) {\delta  \boldsymbol w}^{N+1} = - \frac{\partial g_2}{\partial w^{N+\frac{1}{2}}}({\boldsymbol w}^{N+\frac{1}{2}}) {\delta \boldsymbol w}^{N+\frac{1}{2}},\\
\end{split}
  \right.
\end{equation}
and ${\delta \boldsymbol  w} = ({\delta \boldsymbol  u},{\delta \boldsymbol v},{\delta \boldsymbol \phi})$ are the tangent linear unknowns.

The adjoint model is obtained by transposing \eqref{eqn:TL}
\paragraph{First step - get solution at $t_{N+\frac{1}{2}}$ }
\begin{equation}\label{eqn:AD1}
\left\{
\begin{split}
  & \big[\frac{\partial g_2}{\partial w^{N+1}}({\boldsymbol w}^{N+\frac{1}{2}})\big]^T {\boldsymbol z}^{N+1} = {\boldsymbol \lambda}^{N+1},\\
  & {\boldsymbol \lambda}^{N+\frac{1}{2}} = \big[- \frac{\partial g_2}{\partial w^{N+\frac{1}{2}}}({\boldsymbol w}^{N+\frac{1}{2}})\big]^T {\boldsymbol z}^{N+1}.
\end{split}
  \right.
\end{equation}
\paragraph{Second step - get solution at $t_{N}$ }
\begin{equation}\label{eqn:AD2}
\left\{
\begin{split}
  & \big[\frac{\partial g_1}{\partial w^{N+\frac{1}{2}}}({\boldsymbol w}^{N})\big]^T {\boldsymbol z}^{N+\frac{1}{2}} = {\boldsymbol \lambda}^{N+\frac{1}{2}},\\
  & {\boldsymbol \lambda}^{N} = \big[- \frac{\partial g_1}{\partial w^{N}}({\boldsymbol w}^{N})\big]^T {\boldsymbol z}^{N+\frac{1}{2}},
\end{split}
  \right.
\end{equation}
where ${\boldsymbol \lambda } = ({\boldsymbol \lambda_u},{\boldsymbol \lambda_v},{\boldsymbol \lambda_{\phi}})$ are the adjoint unknowns and ${\boldsymbol z}$ is an intermediary variable. By $\big[\frac{\partial g_1}{\partial w^{N+\frac{1}{2}}}({\boldsymbol w}^{N})\big]^T$ we denote the transpose of Jacobian $\frac{\partial g_1}{\partial w^{N+\frac{1}{2}}}({\boldsymbol w}^{N})$.

\subsection*{SWE tensors}

 The Jacobian matrices required by the quasi-Newton method to solve the standard POD and tensorial POD reduced models are the same and their formulations are obtained analytically. The Jacobian matrices depend on specific tensors computed based on the general formula introduced in \eqref{eqn:POD_tensorial_nonlinearity_general}. In the case of standard POD/DEIM SWE model we approximate these tensors using the algorithm introduced in \cite[p.7]{Stefanescu_etal_forwardPOD_2014}. For each nonlinear SWE term \eqref{eqn:swe-nonlinear_terms} we need to define one tensor. Before introducing their formulas we first define the POD bases and their derivatives. We recall that the discrete full space dimension is $n$ and for each state variable we compute the test and trial functions assuming the bases have the same dimension $k$.
\begin{equation}\label{eqn:SWE_reduced_basis}
\begin{split}
  & {\bf U},~{\bf V},~{\bf \Phi},~{\bf W^u},~{\bf W^v},~{\bf W^{\phi}} \in \mathbb{R}^{n\times k}\\
  & {\bf U_x} = A_x{\bf U} \in \mathbb{R}^{n\times k},~{\bf V_x} = A_x{\bf V} \in \mathbb{R}^{n\times k},~{\bf \Phi_x} = A_x {\bf \Phi} \in \mathbb{R}^{n\times k},\\
  & {\bf U_y} = A_y{\bf U} \in \mathbb{R}^{n\times k},~{\bf V_y} = A_y{\bf V} \in \mathbb{R}^{n\times k},~{\bf \Phi_y} = A_y {\bf \Phi} \in \mathbb{R}^{n\times k},\\
\end{split}
\end{equation}
where $A_x,A_y\in \mathbb{R}^{n\times n}$ are constant coefficient matrices for discrete first-order and second-order differential operators which incorporate the boundary conditions.

The SWE tensors formulas for the standard POD and tensorial POD reduced order models are
\begin{equation}\label{eqn:SWE_tensors}
\begin{split}
  & T^{f} = \left(T^{f}_{i,j,l}\right)_{i,j,l=1,..,k}\in \mathbb{R}^{k \times k \times k},~f=11,12,13,21,22,23,31,32,33,34. \\
  & T^{11}_{i,j,l} = \sum_{r=1}^n W^u_{r,i}U_{r,j}{U_x}_{r,l},\quad T^{12}_{i,j,l} = \sum_{r=1}^n W^u_{r,i}\Phi_{r,j}{\Phi_x}_{r,l},\quad T^{13}_{i,j,l} = \sum_{r=1}^n W^u_{r,i}V_{r,j}{U_y}_{r,l},\\
  & T^{21}_{i,j,l} = \sum_{r=1}^n W^v_{r,i}U_{r,j}{V_x}_{r,l},\quad T^{22}_{i,j,l} = \sum_{r=1}^n W^v_{r,i}V_{r,j}{V_y}_{r,l},\quad T^{23}_{i,j,l} = \sum_{r=1}^n W^v_{r,i}\Phi_{r,j}{\Phi_y}_{r,l},\\
  & T^{31}_{i,j,l} = \sum_{r=1}^n W^{\phi}_{r,i}\Phi_{r,j}{U_x}_{r,l},\quad T^{32}_{i,j,l} = \sum_{r=1}^n W^{\phi}_{r,i}U_{r,j}{\Phi_x}_{r,l},\quad T^{33}_{i,j,l} = \sum_{r=1}^n W^{\phi}_{r,i}\Phi_{r,j}{V_y}_{r,l},\\
  & T^{34}_{i,j,l} = \sum_{r=1}^n W^{\phi}_{r,i}V_{r,j}{\Phi_y}_{r,l}.\\
\end{split}
\end{equation}

In the case of standard POD/DEIM SWE model, the above tensors are computed by making use of the DEIM approximation of the nonlinear terms, resulting in an off-line stage that is faster \cite[Figure 8(b)]{Stefanescu_etal_forwardPOD_2014} than the versions proposed by standard POD and tensorial POD, even if additional SVD decompositions and low-rank terms are calculated.

To render the manuscript as self-contained as possible we describe the methodology used for computing one POD/DEIM tensor. For example, the POD/DEIM reduced nonlinear version of $F_{11}$ defined in \eqref{eqn:swe-nonlinear_terms} is given by
\begin{equation}\label{eqn:DEIM_SWE_F11}
{\tilde F}_{11}  \approx \underbrace{{\bf W^u}^T{\bf F}_{11}^{\rm{POD}}(P^T{\bf F}_{11}^{\rm{POD}})^{-1}}_{k\times m} \underbrace{\left(P^T{\bf U}{\bf \tilde x}\odot P^T{\bf U_x}{\bf \tilde x}\right)}_{m \times 1},
\end{equation}
where ${\bf F}_{11}^{\rm{POD}}\in \mathbb{R}^{n\times m}$ is the POD basis of dimension $m$ obtained from the snapshots of the SWE nonlinear term $F_{11}$ (for more details see also the general form of POD/DEIM expansion \eqref{eqn:POD_DEIM_nonlinearity_general}). Let us denote by $E = {{\bf W^u}^T{\bf F}_{11}^{\rm{POD}}(P^T{\bf F}_{11}^{\rm{POD}})^{-1}}\in \mathbb{R}^{k\times m}$, ${\bf U}^m = P^T{\bf U} \in \mathbb{R}^{m\times k}$ and ${\bf U_x}^m = P^T{\bf U_x} \in \mathbb{R}^{m \times k}$, than the associated tensor computed during the POD/DEIM off-line stage is
\begin{equation}\label{eqn:DEIM_tensor_F11}
{T^{11}}^{\rm{DEIM}}_{i,j,l} = \sum_{r=1}^m E_{i,r}U^m_{j,r}{U_x}^m_{r,l},\quad i,j,l=1,..,k.
\end{equation}
In comparison with the formula used by standard and tensorial POD \eqref{eqn:SWE_tensors}, we notice that the summation spans only the location of DEIM points instead of entire discrete space. For completeness, we recall that $n$ is the size of the full discrete space, $k$ is the size of reduced order model and $m$ is the number of DEIM points.
\end{document}